\documentclass[12pt,preprint]{aastex}
\def\kmskpc{{\rm\,km\,s^{-1}{kpc}^{-1}}}

\def\mathnew{\mathsurround=0pt}
\def\simov#1#2{\lower .5pt\vbox{\baselineskip0pt
    \lineskip-.5pt\ialign{$\mathnew#1\hfil##\hfil$\crcr#2\crcr\sim\crcr}}}

\def\'#1{\ifx#1i{\accent"13\i}\else{\accent"13#1}\fi}

\begin{document}

\title{Tidal radii and destruction rates of globular clusters in the
Milky Way due to bulge-bar and disk shocking}

\author{Edmundo Moreno\altaffilmark{1}, B\'arbara Pichardo\altaffilmark{1} and H\'ector Vel\'azquez\altaffilmark{2}}
\altaffiltext{1}{Instituto de Astronom\'\i a, Universidad Nacional
  Aut\'onoma de M\'exico, Apdo. Postal 70-264, 04510, M\'exico, D.~F.,
  M\'exico.} 
\altaffiltext{2}{Observatorio Astron\'omico Nacional, Universidad
  Nacional Aut\'onoma de M\'exico, Apdo. Postal 877, 22800 Ensenada,
  M\'exico.}

\begin{abstract} 

We calculate orbits, tidal radii, and bulge-bar and disk shocking
destruction rates for 63 globular clusters in our Galaxy.  Orbits are
integrated in both an axisymmetric and a non-axisymmetric Galactic
potential that includes a bar and a 3D model for the spiral arms. With
the use of a Monte Carlo scheme, we consider in our simulations
observational uncertainties in the kinematical data of the clusters.
In the analysis of destruction rates due to the bulge-bar,
we consider the rigorous treatment of using the real Galactic cluster
orbit, instead of the usual linear trajectory employed in previous
studies. We compare results in both treatments.  We find that the
theoretical tidal radius computed in the non-axisymmetric Galactic
potential compares better with the observed tidal radius than that
obtained in the axisymmetric potential.  In both Galactic potentials,
bulge-shocking destruction rates computed with a linear trajectory of
a cluster at its perigalacticons give a good approximation to the
result obtained with the real trajectory of the cluster.
Bulge-shocking destruction rates for clusters with perigalacticons
in the inner Galactic region are $smaller$ in the non-axisymmetric
potential, as compared with those in the axisymmetric potential.
For the majority of clusters with high orbital eccentricities
($e > 0.5$), their total bulge+disk destruction rates are $smaller$
in the non-axisymmetric potential.

\end{abstract}

\keywords{galaxy: halo --- galaxy: kinematics and dynamics --- globular clusters:
general}

\section{Introduction}\label{introd}

In two previous papers \citep[hereafter Papers I and II]{AMP06,AMP08}
tidal radii and destruction rates due to bulge and disk shocking were
computed for 54 globular clusters in our Galaxy, using axisymmetric
and non-axisymmetric Galactic potentials. In Paper I the
non-axisymmetric Galactic potential included the Galactic bar, and in
Paper II the additional effect of three dimensional (3D) spiral arms
was also analyzed. The models for these non-axisymmetric components
given by \citet{PMME03,PMM04} were employed in those computations.
The absolute proper motion data needed to compute the Galactic orbits
of the globular clusters were obtained from the extensive studies of
\citet{D97,D99a,D99b,DI00,D01,D03} and \citet{CD07}, who have computed
the proper motions for a good fraction of the total number of globular
clusters; for other clusters, \citet{D99b} have compiled the proper
motion data from various sources.  Lately, \citet{CD10} have given
the absolute proper motions of other nine globular clusters, and
\citet{CD13} present new absolute proper motions of NGC 6397, NGC
6626, and NGC 6656; thus now we dispose of absolute proper motion data
for a total of 63 globular clusters in our Galaxy.

For the new sample of 63 globular clusters, we compute again their
tidal radii and destruction rates due to bulge and disk shocking, now
with some improvements. We use axisymmetric and non-axisymmetric
Galactic potentials, the later including both the spiral arms and the
Galactic bar models of \citet{PMME03,PMM04}, as in Paper II.  A first
part of our improvements has to do with this Galactic potential, the
initial orbital conditions of the globular clusters, and the
uncertainties in the computed quantities: (a) the Galactic potential
is now rescaled to recent values of the galactocentric distance and
rotation velocity of the local standard of rest, as found by
\citet{BRet11}, (b) we use the solar velocity obtained by
\citet{SBD10}, (c) the late compilation of clusters properties given
by \citet{H10} is used to update other parameters employed in our
computations, and (d) we make Monte Carlo simulations to estimate the
uncertainties in the tidal radii and destruction rates, and compare
with estimates in Papers I and II.

The second part of our improvements refers to the procedure to compute
destruction rates. In Papers I, II and in previous studies of tidal
heating due to the interaction with the Galactic bulge and heating by
disk shocking \citep[e.g.,][]{AHO88,GO97,GO99}, the impulse
approximation and a straight-path cluster trajectory have been
employed.  Here we relax the straight-path approximation and follow
the more rigorous treatment given by \citet{G99a}, who employ a fit to
the tidal acceleration along the true Galactic orbit of the cluster.
In this paper this procedure is undertaken in our non-spherical
Galactic potentials, as opposed to the spherical potential used by
\citet{G99a}.  The results are compared with those obtained using the
usual straight-path aproximation.

In $\S$ \ref{datos} we give the globular cluster data employed in our
study. The Galactic potential and its parameters are presented in $\S$
\ref{gpot}. In $\S$ \ref{galorb} some properties of the Galactic
orbits in both the axisymmetric and non-axisymmetric potentials are
tabulated, and for some clusters we show their meridional orbits.  The
tidal radii are analyzed in $\S$ \ref{radmar}.  The needed formulism
of destruction rates using the real trajectories of globular clusters
is summarized in $\S$ \ref{tdestr}, and our results are presented in
$\S$ \ref{destr}. In $\S$ \ref{concl} we present our conclusions.

\section{Employed data for the globular clusters}\label{datos}

In Table \ref{tbl-1} we list the cluster parameters employed in our
study. Equatorial coordinates $(\alpha,\delta)$, are given in columns
2 and 3. The distance $r$ and radial velocity $v_r$, in columns 4 and
5, are taken from the recent compilation by \citet{H10}.  The absolute
proper motions, ${\mu}_x$ = ${\mu}_{\alpha}\cos{\delta}$, ${\mu}_y$ =
${\mu}_{\delta}$, in columns 6 and 7, are the values given by
\citet{D97,D99a,D99b,DI00,D01,D03} and \citet{CD07,CD10,CD13}, except
for 47 Tuc (NGC 104) and M4 (NGC 6121) whose values are taken from
\citet{AK03} and \citet{BPKA03}, respectively.  As in Papers I and II,
the mass of a cluster, $M_c$, given in column 8, is computed using a
mass-to-light ratio $(M/L)_V$ = 2 $M_{\odot}/L_{\odot}$. \bf In 
$\S$ \ref{radmar} we also employ for some clusters their $M_c$
computed with dynamical mass-to-light ratios given by \citet{MM05} \rm.
The observed tidal radius $r_{td}$ (we call $r_{td}$=$r_K$ if this
radius is computed with a King model \citep{K62}) is not listed by
\citet{H10}, but as he points out, it can be computed with his listed
values for the concentration, $c$, and core radius, $r_c$, only for
those clusters with a non-collapsed core. For clusters with a collapsed
core, Harris suggests to take $r_{td}$ estimated by \citet{MM05} and
\citet{PK75}. For this type of clusters, and listed in Table
\ref{tbl-1}, \citet{MM05} give $r_{td}$ for NGC 362, NGC 1904, NGC
6266, and NGC 6723; we take their $r_{td}$ for a King model, which is
the model we use in our computations. For other clusters in Table
\ref{tbl-1}, \citet{PK75} estimate $r_{td}$ in NGC 6397, NGC 6752, NGC
7078, and NGC 7099.  These values of $r_{td}$ from \citet{MM05} and
\citet{PK75} are transformed according to the distances $r$ given by
\citet{H10}.  For NGC 6284, NGC 6293, NGC 6342, and NGC 6522, we take
$r_{td}$ from Harris' previous compilation, transformed with his new
listed distances. Column 9 gives the final $r_{td}$=$r_K$ employed
values, and column 10 the half-mass radius, $r_h$, in each cluster.

\section{The Galactic potential}\label{gpot}

In our analysis we employ axisymmetric and non-axisymmetric models for
the Galactic gravitational potential. The axisymmetric model is based
on the Galactic model of \citet{AS91}, which gives a circular rotation
speed on the Galactic plane ${\Theta}_0$ $\approx$ 220 km/s at its
assumed Sun's Galactocentric distance $R_0$ = 8.5 kpc. This model is
scaled to the new Galactic parameters ${\Theta}_0$, $R_0$ given by
\citet{BRet11}: ${\Theta}_0$ = 239$\pm$7 km/s, $R_0$ = 8.3$\pm$0.23 kpc.

The non-axisymmetric Galactic model is built from the scaled
axisymmetric model. First, all the mass in the spherical bulge
component in this axisymmetric model is employed to built the Galactic
bar. In Papers I and II, where only 70$\%$ of the bulge mass was
employed to built the bar, we have mentioned some properties of the
model used for this bar component. We use the third bar model given by
\citet{PMM04} (the model of superposition of ellipsoids), which
approximates the boxy COBE/DIRBE brightness profiles shown by
\citet{F98}.

We also consider a 3D gravitational potential to represent the spiral
arms. The model used for these arms, called PERLAS, is given by
\citet{PMME03}, and has already been employed in Paper II. The total
mass of the 3D spiral arms is taken as a small fraction of the mass in
the disk component of the scaled axisymmetric model.  We take
$M_{\rm arms}/M_{\rm disk}$ = 0.04$\pm$0.01, considered by \citet{PMA12}
in their analysis of the maximum value on the Galactic plane of the
parameter $Q_T$ \citep{ST80,CS81}, which, as a function of
Galactocentric distance, is the ratio of the maximum azimuthal force
of the spiral arms to the radial axisymmetric force at a given
distance.

The mass density at the center of the spiral arms falls exponentially
with Galactocentric distance, and we take its corresponding radial
scale length equal to the one of the Galactic exponential disk modeled
by \citet{BCHet05}: $H$ = 3.9$\pm$0.6 kpc, using $R_0$ = 8.5 kpc,
scaled now with the new value of $R_0$.

Other properties of the Galactic bar and the Galactic spiral arms,
have been collected in \citet{PMA12}. We use the following parameters 
in our computations: a) the present angle between the bar's major axis
and the Sun-Galactic center line is taken as 20$^\circ$, b) the angular
velocity of the bar is in the range $\approx$ 55$\pm$5 $\kmskpc$,
c) we consider the pitch angle of the spiral arms in the range
$\approx$ 15.5$\pm$3.5$^\circ$, d) the range for the angular velocity
of the spiral arms is $\approx$ 24$\pm$6 $\kmskpc$. 

Table \ref{tbl-2} summarizes all the parameters employed in our
Galactic models, along with the Solar velocity $(U,V,W)_{\odot}$
obtained by \citet{SBD10} (here $U$ is taken negative towards the
Galactic center) and its uncertainties estimated by \citet{BRet11}.

\section{Properties of the Galactic orbits}\label{galorb}

For the computation of the Galactic orbits we have employed the
Bulirsch-Stoer algorithm given by \citet{P92}, and also the Runge-Kutta
algorithm of seventh-eight order elaborated by \citet{F68}.
In our problem both algorithms give practically the same results, and
due to the complicated mathematical forms of the gravitational
potentials of the non-axisymmetric Galactic components (bar and 3D
spiral arms), we have favored the Runge-Kutta algorithm to reduce the
computing time, specially in the Monte Carlo calculations.
The Bulirsch-Stoer algorithm is employed mainly in the computations
with the axisymmetric potential.

In Table \ref{tbl-3} we give for each cluster some orbital parameters
obtained with the non-axisymmetric (first line) and axisymmetric
(second line) potentials. Except for the data given in columns 8 and
9, whose associated time interval is commented in the next section,
the data presented in this table correspond to a backward time
integration of 5 $\times$ $10^9$ yr in the axisymmetric case, and from
$10^9$ to 3 $\times$ $10^9$ yr in the non-axisymmetric case, depending
on the cluster. The second, third, and fourth columns show the average
perigalactic distance, the average apogalactic distance, and the
average maximum distance from the Galactic plane, respectively.  The
fifth column gives the average orbital eccentricity, this eccentricity
defined as $e=(r_{max}-r_{min})/(r_{max}+r_{min})$, with $r_{min}$ and
$r_{max}$ successive perigalactic and apogalactic distances. Columns 6
and 7 give the orbital energy per unit mass, $E$, and the z-component
of angular momentum per unit mass, $h$, only in the axisymmetric
potential, where these two quantities are constants of motion. Columns
8 and 9 list tidal radii, which are discussed in the next section.

\begin{figure}
\plotone{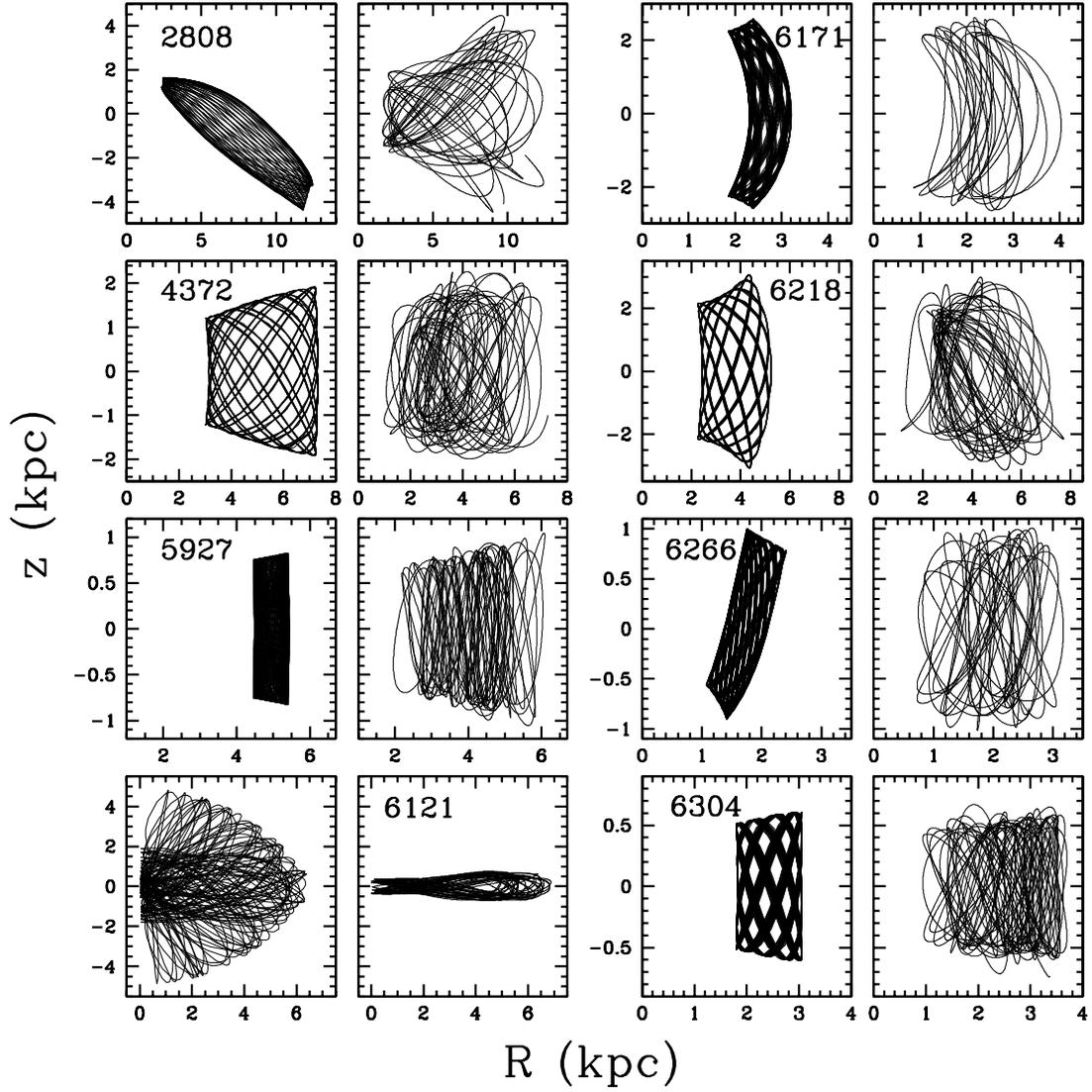}
\caption {Meridional Galactic orbits for a sample of globular clusters.
Each pair of columns shows the orbits with the axisymmetric (left)
and non-axisymmetric (right) Galactic potentials. The cluster NGC
number is given.} 
\label{fig1}
\end{figure}

\begin{figure}
\plotone{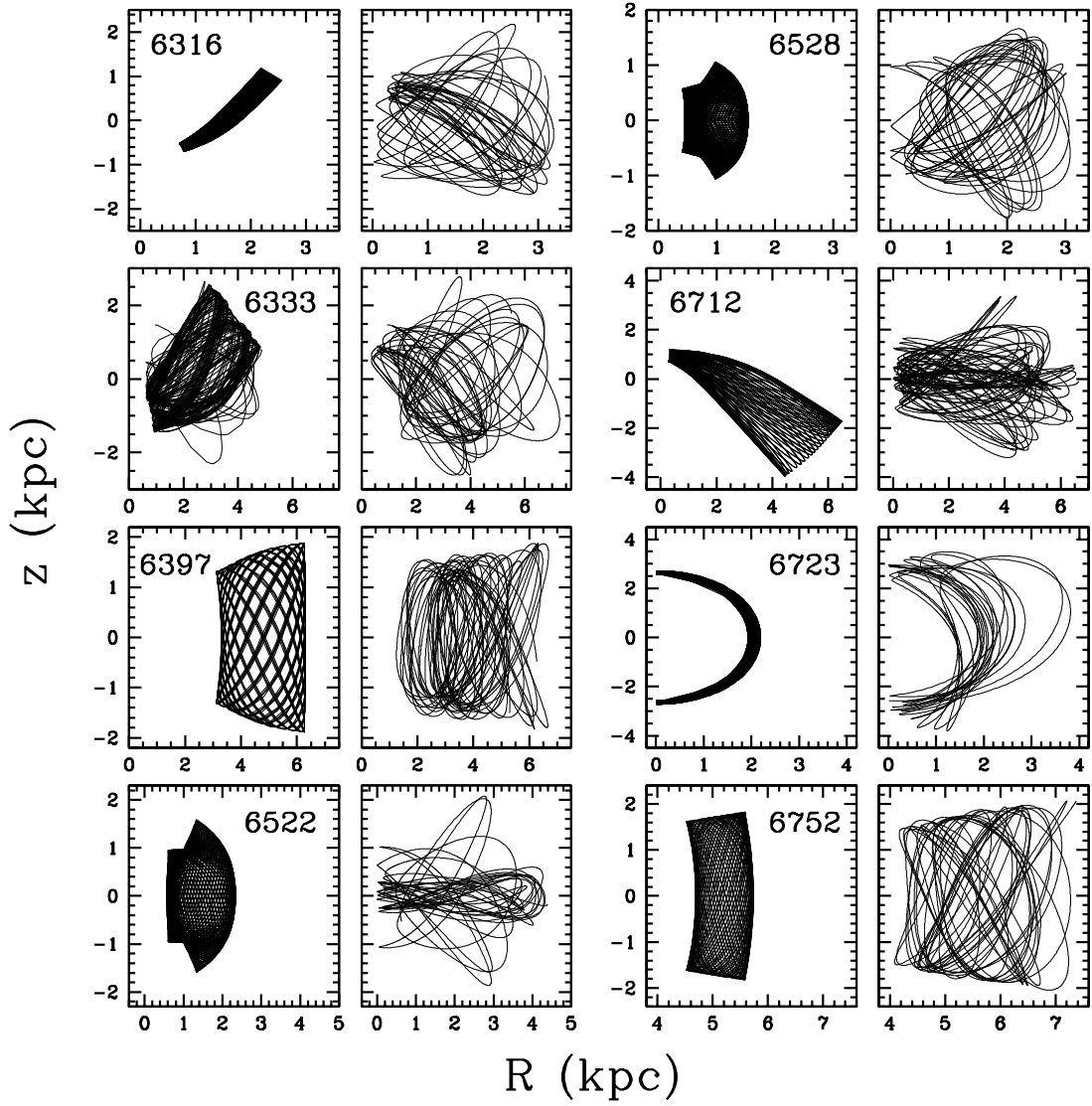}
\caption {As in Figure \ref{fig1}.}
\label{fig2}
\end{figure}

In Figures \ref{fig1} and \ref{fig2} we show meridional orbits for
some clusters, whose NGC number is given. In each pair of columns
the orbit in the axisymmetric potential is shown in the left frame,
and that in the non-axisymmetric potential in the right frame. 
We choose this sample of clusters to illustrate how strong the effect
of the non-axisymmetric Galactic components can be. The most
conspicuous difference between the computations in both potentials is
the orbital radial extent. This has important consequences in the
clusters tidal radii, as shown in the next section. 

\begin{figure}
\plotone{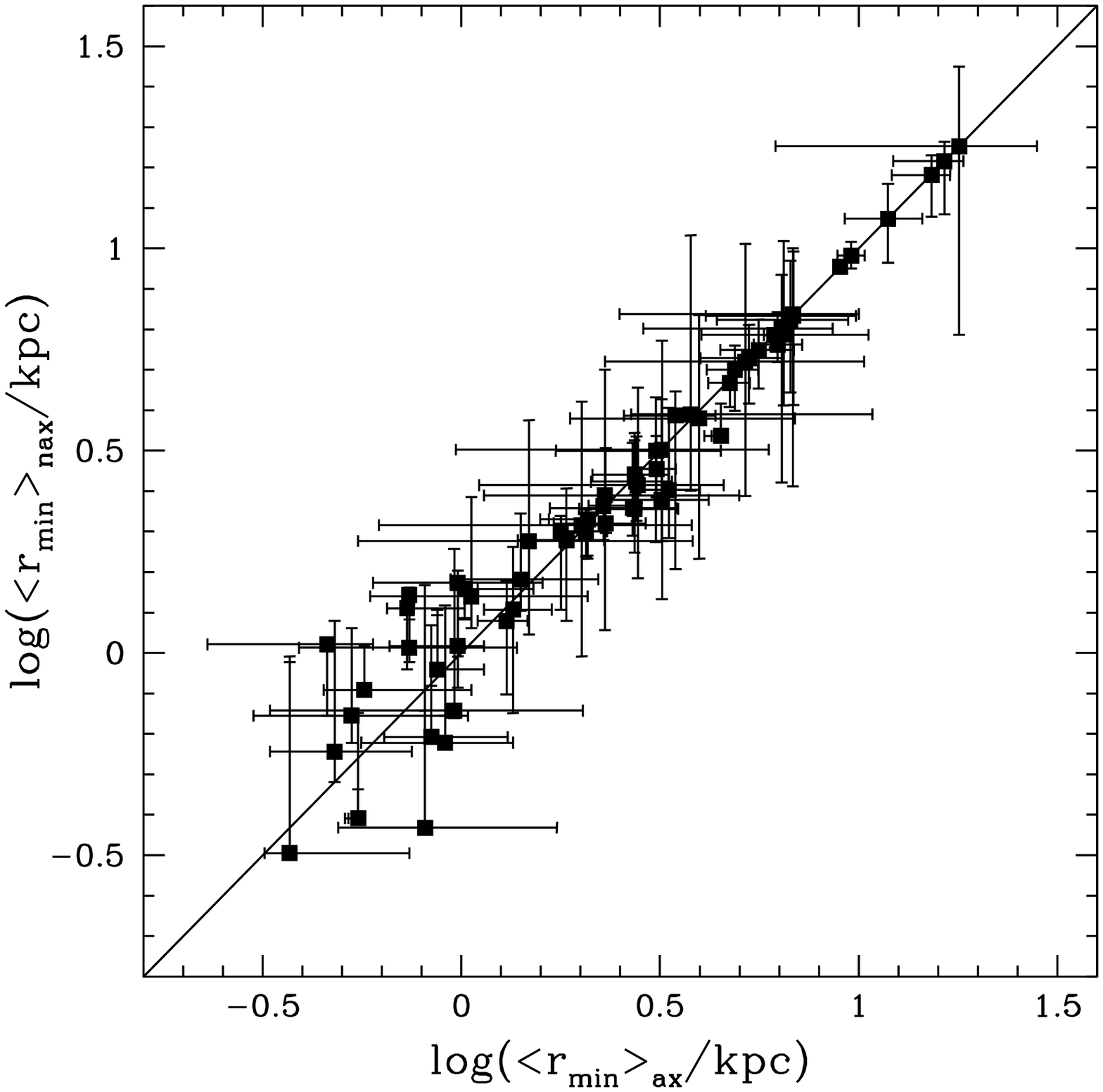}
\caption {Comparison of the cluster average perigalactic distance,
second column in Table \ref{tbl-3}, in the axisymmetric potential
(denoted with a subindex 'ax') and in the non-axisymmetric potential
(with a subindex 'nax'). The plotted line is the line of coincidence.}
\label{fig3}
\end{figure}

\begin{figure}
\plotone{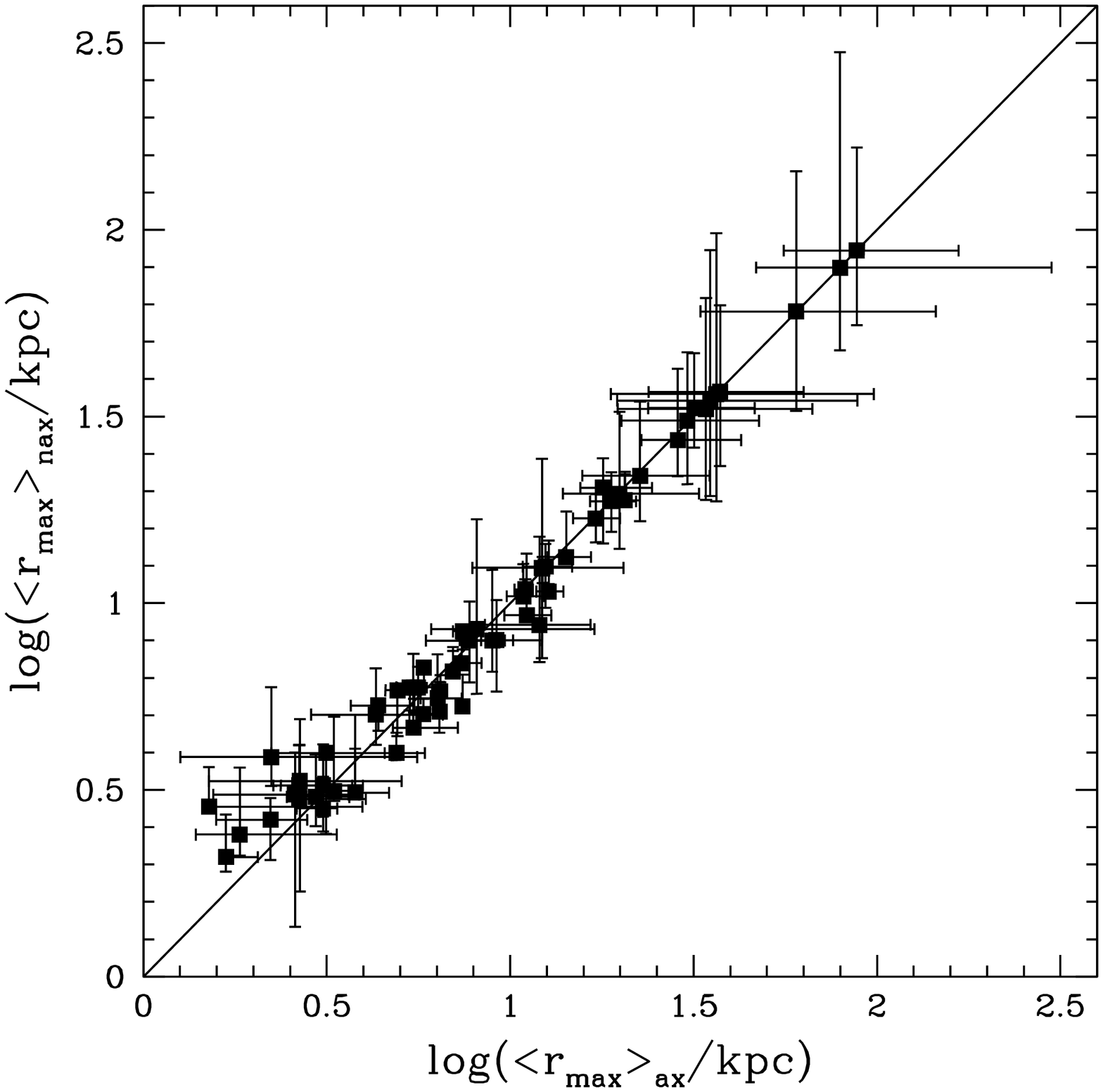}
\caption {Comparison of the cluster average apogalactic distance, 
third column in Table \ref{tbl-3}, in the axisymmetric potential
(with a subindex 'ax') and in the non-axisymmetric potential (with a
subindex 'nax'). The plotted line is the line of coincidence.}
\label{fig4}
\end{figure}

\begin{figure}
\plotone{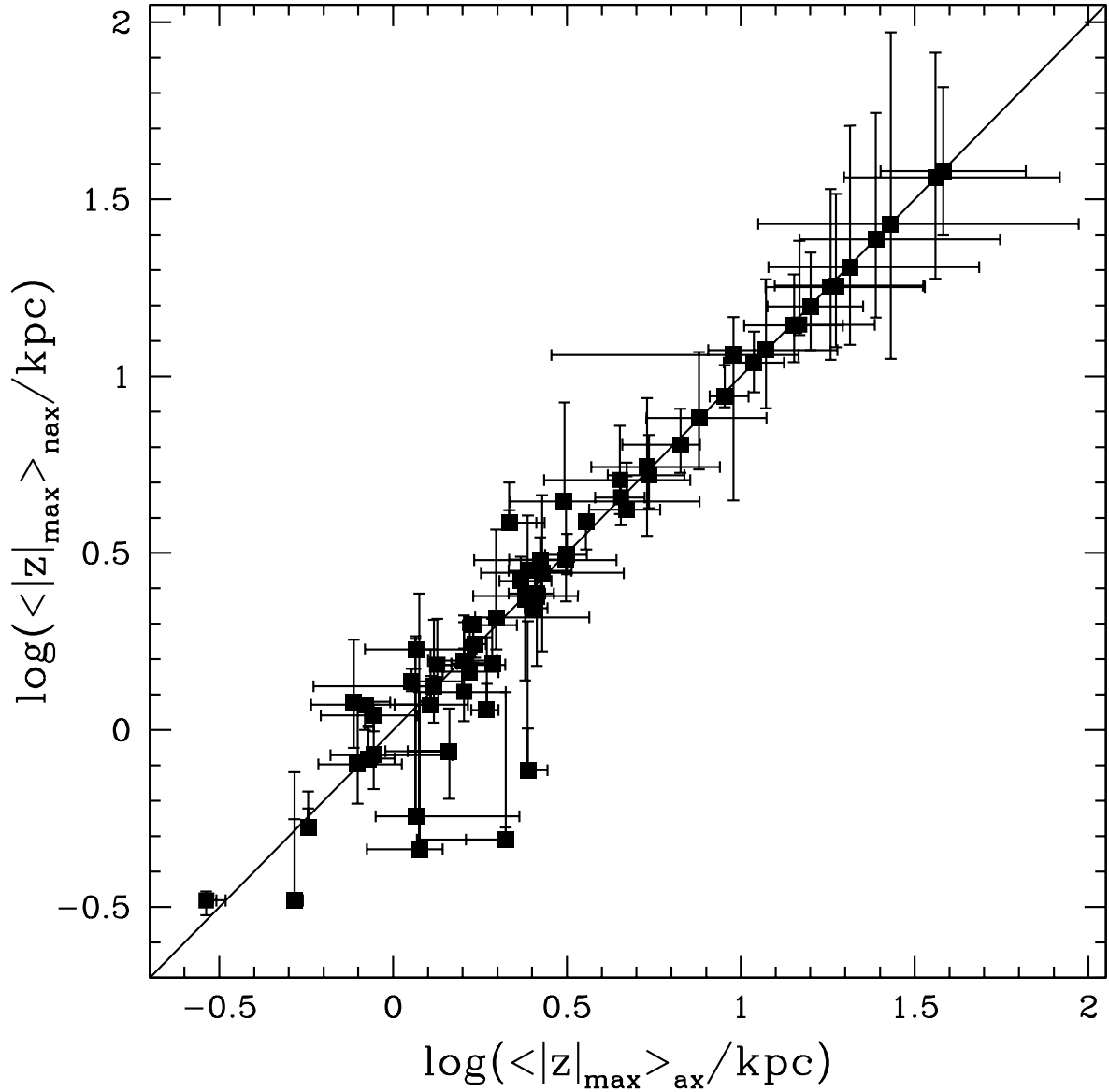}
\caption {Comparison of the cluster maximum distance from the Galactic
plane, fourth column in Table \ref{tbl-3}, in the axisymmetric potential
(with a subindex 'ax') and in the non-axisymmetric potential (with a
subindex 'nax'). The plotted line is the line of coincidence.}
\label{fig5}
\end{figure}

\bf Figures \ref{fig3}, \ref{fig4}, and \ref{fig5} show the comparison
in both Galactic potentials of the average perigalactic distance,
average apogalactic distance, and average maximum distance from the
Galactic plane, given respectively in the second, third, and fourth
columns of Table \ref{tbl-3}. Values in the axisymmetric potential 
(with a subindex 'ax') and non-axisymmetric potential (with a subindex
'nax') are given in the horizontal and vertical axes, respectively.
The uncertainties shown in these figures are estimated with the
differences from corresponding quantities obtained in the minimum and
maximum energy orbits in each cluster, according to the uncertainties
in the cluster radial velocity, distance, and proper motions. \rm

\section{Tidal radii}\label{radmar}

\subsection{Comparison of theoretical and observed tidal radii in the
axisymmetric and non-axisymmetric Galactic potentials} 
\label{compar}

\subsubsection{Comparison with observed King tidal radii $r_K$}
\label{king}

\noindent As in Papers I and II, for each globular cluster, and in both
axisymmetric and non-axisymmetric Galactic potentials employed in our
analysis, we compute a theoretical tidal radius using two expressions.
The first is King's formula \citep{K62} 

\begin{equation} r_{K_t}= \left [ \frac{M_c}{M_g(3+e)} \right ]^{1/3}r_{min}, \label{rKt} \end{equation} 

\noindent where $M_c$ is the mass of the cluster, $M_g$ is an effective
galactic mass, $e$ is the orbital eccentricity as defined in the
previous section, and $r_{min}$ is the perigalactic distance. The mass
$M_g$ is taken as the equivalent central mass point which gives an
acceleration at the given perigalactic position with a magnitude equal
to the magnitude of the actual acceleration at this point in the
corresponding Galactic potential.

The second expression is the one proposed in Paper I, computed at the
perigalactic position

\begin{equation} r_{\ast} = \left [ \frac{GM_c}{\left (\frac{\partial
F_{x'}}
{\partial x'} \right )_{{\bf r'}= 0}+ \dot{\theta}^2 + \dot{\varphi}^2\sin^2
{\theta}} \right ]^{1/3}, \label{rast} \end{equation}

\noindent with $F_{x'}$ the component of the Galactic acceleration
along the line $x'$ joining the cluster with the Galactic center, and
its partial drivative evaluated at the given perigalactic point. 
The angles ${\varphi}$ and ${\theta}$ are angular spherical coordinates
of the cluster in an inertial galactic frame. 

In Paper I these two expressions for a theoretical tidal radius gave
similar values. For a given Galactic potential, this result is
maintained in the present computations.
Columns 8 and 9 in Table \ref{tbl-3} give the average values
$<$$r_{K_t}$$>$, $<$$r_{\ast}$$>$ of $r_{K_t}$ and $r_{\ast}$, over the
last $10^9$ yr (in some clusters this time interval is extended to have
a few perigalactic points).

\begin{figure}
\plotone{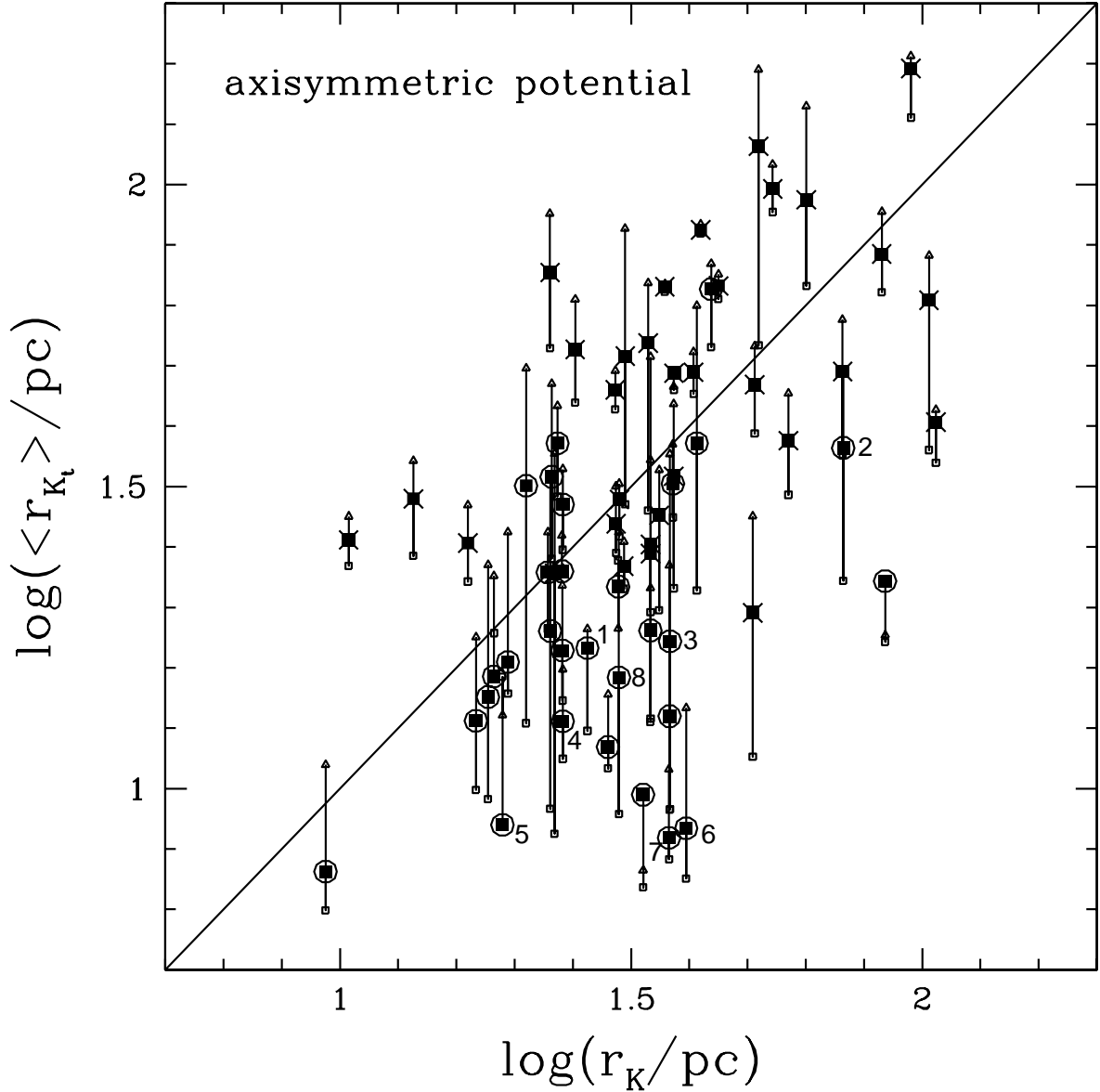}
\caption {Comparison of the theoretical tidal radius $r_{K_t}$ computed
in the axisymmetric potential, averaged over the last $10^9$ yr in
each cluster orbit, and the observed tidal radius $r_K$. Encircled and
crossed points correspond, respectively, to clusters in which
$<$$r_{K_t}$$>$ in this axisymmetric potential, is less or greater than
$<$$r_{K_t}$$>$ computed in the non-axisymmetric potential. See next
Figure \ref{fig7}. The small empty squares and triangles give
$<$$r_{K_t}$$>$ in the minimum and maximum energy orbits.
The continuous line is the line of coincidence. The numbered encircled
points are considered in Figure \ref{fig8}.}
\label{fig6}
\end{figure}

\begin{figure}
\plotone{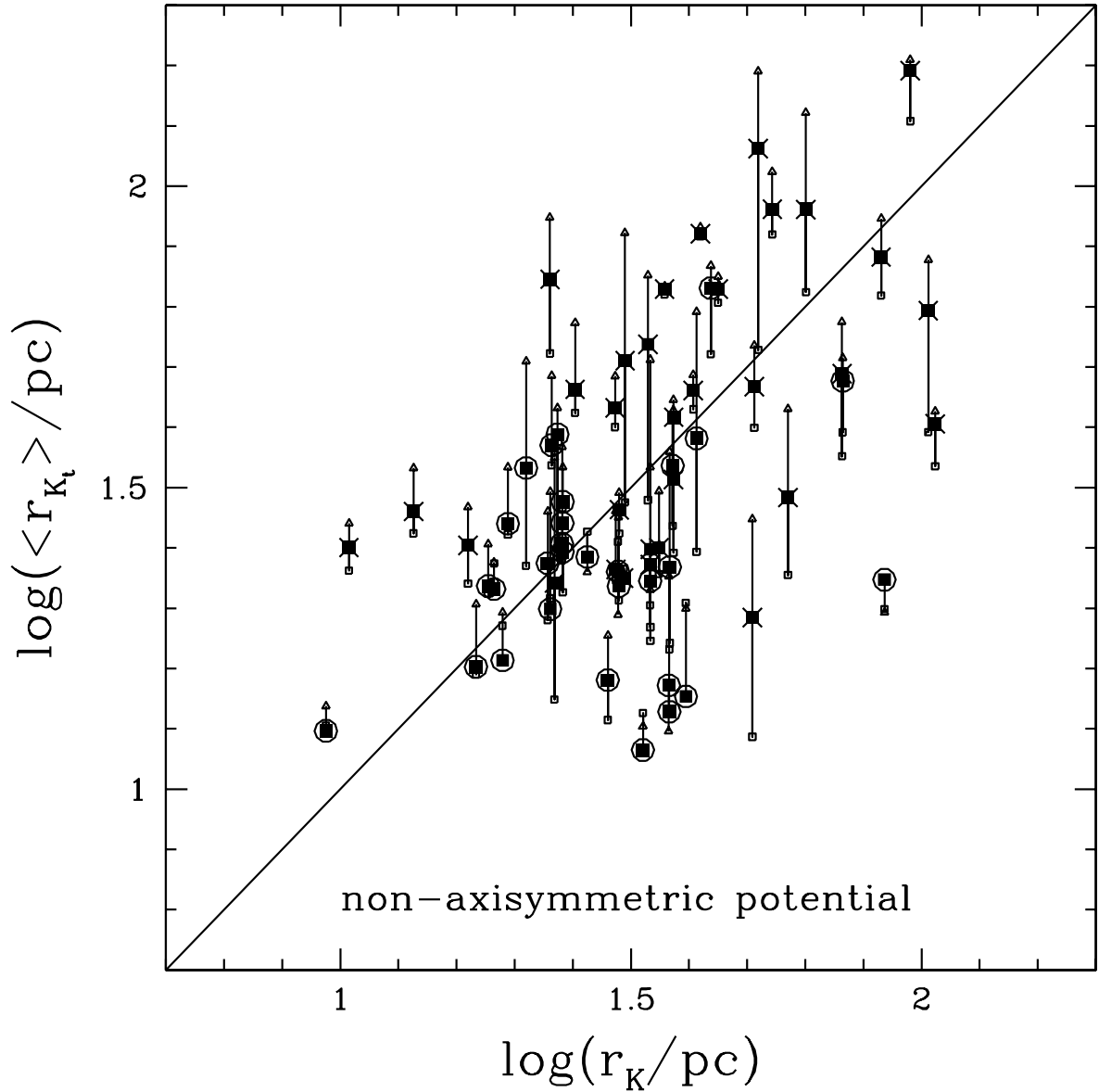}
\caption {Comparison of the averaged theoretical tidal radius $r_{K_t}$
computed in the non-axisymmetric potential and the observed tidal
radius $r_K$. The marks in the points maintain the corresponding
meaning given in Figure \ref{fig6}.}
\label{fig7}
\end{figure}

In this \bf and next sections of $\S$ \ref{radmar} we make some
comparisons using $r_{K_t}$ given by Eq. (\ref{rKt}). The first
comparison is $r_{K_t}$ with \rm the observed tidal radius (also called 
the limiting radius) $r_{td}$=$r_K$ listed in Table \ref{tbl-1},
estimated with a King model \citep{K62}. In Figures \ref{fig6} and
\ref{fig7} we show with big filled squares this comparison in the
axisymmetric and non-axisymmetric Galactic potentials. These points
have two marks: clusters in which the tidal radius $<$$r_{K_t}$$>$
computed with the non-axisymmetric potential is greater than
$<$$r_{K_t}$$>$ computed with the axisymmetric potential, are marked
with encircled points; crossed points correspond to clusters in which
$<$$r_{K_t}$$>$ computed with the non-axisymmetric potential is less
than $<$$r_{K_t}$$>$ computed with the axisymmetric potential. These
marks are shown in both figures. Thus, encircled and crossed points in
Figure \ref{fig6} will move upwards and downwards, respectively, to
give the corresponding Figure \ref{fig7}.  As in Papers I and II, the
uncertainty in $<$$r_{K_t}$$>$, or $<$$r_{\ast}$$>$, is estimated in
each cluster by computing $<$$r_{K_t}$$>$ in the minimum and maximum
energy orbits, according to the uncertainties in the cluster radial
velocity, distance, and proper motions. The small empty squares and
empty triangles in Figures \ref{fig6} and \ref{fig7} show the values
of $<$$r_{K_t}$$>$ in these minimum and maximum energy orbits,
respectively.

From these figures we note that $<$$r_{K_t}$$>$ computed in the
non-axisymmetric potential compares better with $r_K$: many clusters
whose points lie below the line of coincidence in Figure \ref{fig6},
are closer to this line in Figure \ref{fig7}; these are the encircled
points. Likewise, several clusters with points (now the crossed points)
above the line of coincidence in Figure \ref{fig6}, are closer to this
line in Figure \ref{fig7}. The rearrangement of the encircled points is
the most conspicuous.

\begin{figure}
\plotone{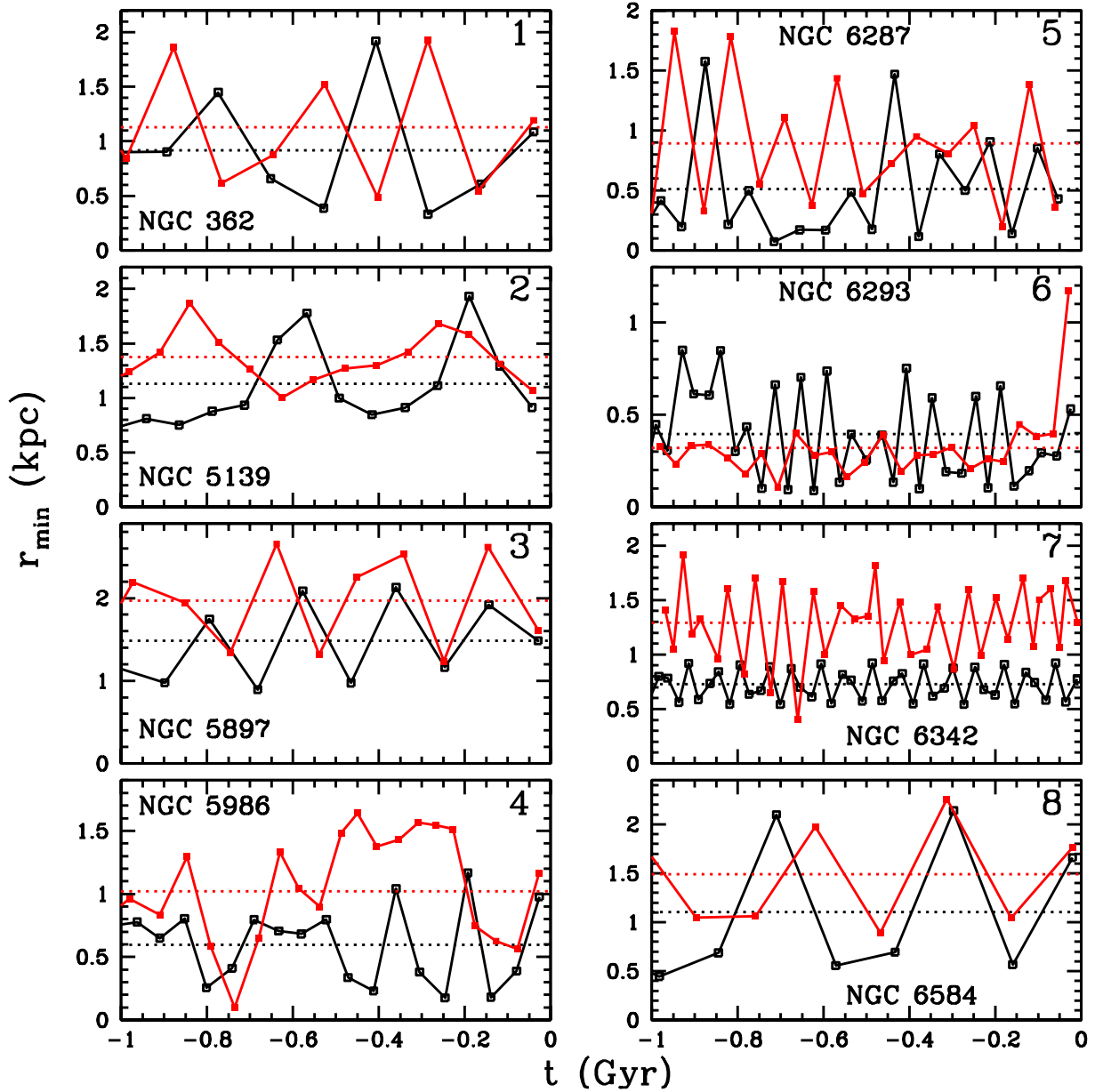}
\caption {Perigalactic distance as a function of time over the last
$10^9$ yr for the sample of globular clusters with numbered encircled
points in Figure \ref{fig6}. Black dots joined with black lines show
the values in the axisymmetric potential; those in red correspond to
the non-axisymmetric potential. The horizontal dotted lines show the
corresponding average values. In each frame the cluster name and its
identification number in Figure \ref{fig6} are given.} 
\label{fig8}
\end{figure}

In Figure \ref{fig6} a sample of eight clusters has been selected,
represented by encircled points numbered from 1 to 8, and correspond
to NGC 362, NGC 5139, NGC 5897, NGC 5986, NGC 6287, NGC 6293, NGC
6342, and NGC 6584, respectively. \bf For these clusters, in Figure
\ref{fig8} we give the values of their perigalactic distance,
$r_{min}$, as a function of time, over the last $10^9$ yr; black dots
joined by black lines show the values in the axisymmetric potential,
and the dots and lines in red correspond to the non-axisymmetric
potential. The black and red horizontal dotted lines show the
corresponding average value of $r_{min}$ in the given interval of time.
Each frame shows the cluster name and also the identification number
in Figure \ref{fig6}. Except for NGC 6293, Figure \ref{fig8} partly
explains why in the case of encircled points, $<$$r_{K_t}$$>$ increases
using the non-axisymmetric potential: in these clusters, the average
value of $r_{min}$ obtained with the non-axisymmetric potential (red
horizontal dotted lines) is greater than the corresponding average in
the axisymmetric potential (black horizontal dotted lines). 

The other factor which helps to understand the rearrangement of
encircled points from Figure \ref{fig6} to Figure \ref{fig7} (at least
those in the considered sample) is the value of the effective galactic
mass $M_g$ employed in King's formula Eq. (\ref{rKt}). As stated in
$\S$ \ref{gpot}, the original concentrated spherical bulge in the
axisymmetric potential was employed to built the bar; thus, due to the
less mass concentration of the non-axisymmetric potential in the inner
Galactic region (see upper frame in figure 7 in Paper I), the
contribution of the bar to $M_g$ computed at a given perigalactic
distance in this inner region is expected to decrease compared with the
contribution of the spherical bulge in the axisymmetric potential.
In general, the value of $M_g$ in the non-axisymmetric potential will
depend on the position of the perigalactic point relative to the axes
of the Galactic bar, because this bar generates a non-axisymmetric
force field. In addition, $M_g$ has also the effect of the spiral arms
with their relative orientation to the axes of the bar at the time of
occurrence of the perigalactic point. Thus, the value of $M_g$ depends
on the specific perigalactic point.

\begin{figure}
\plotone{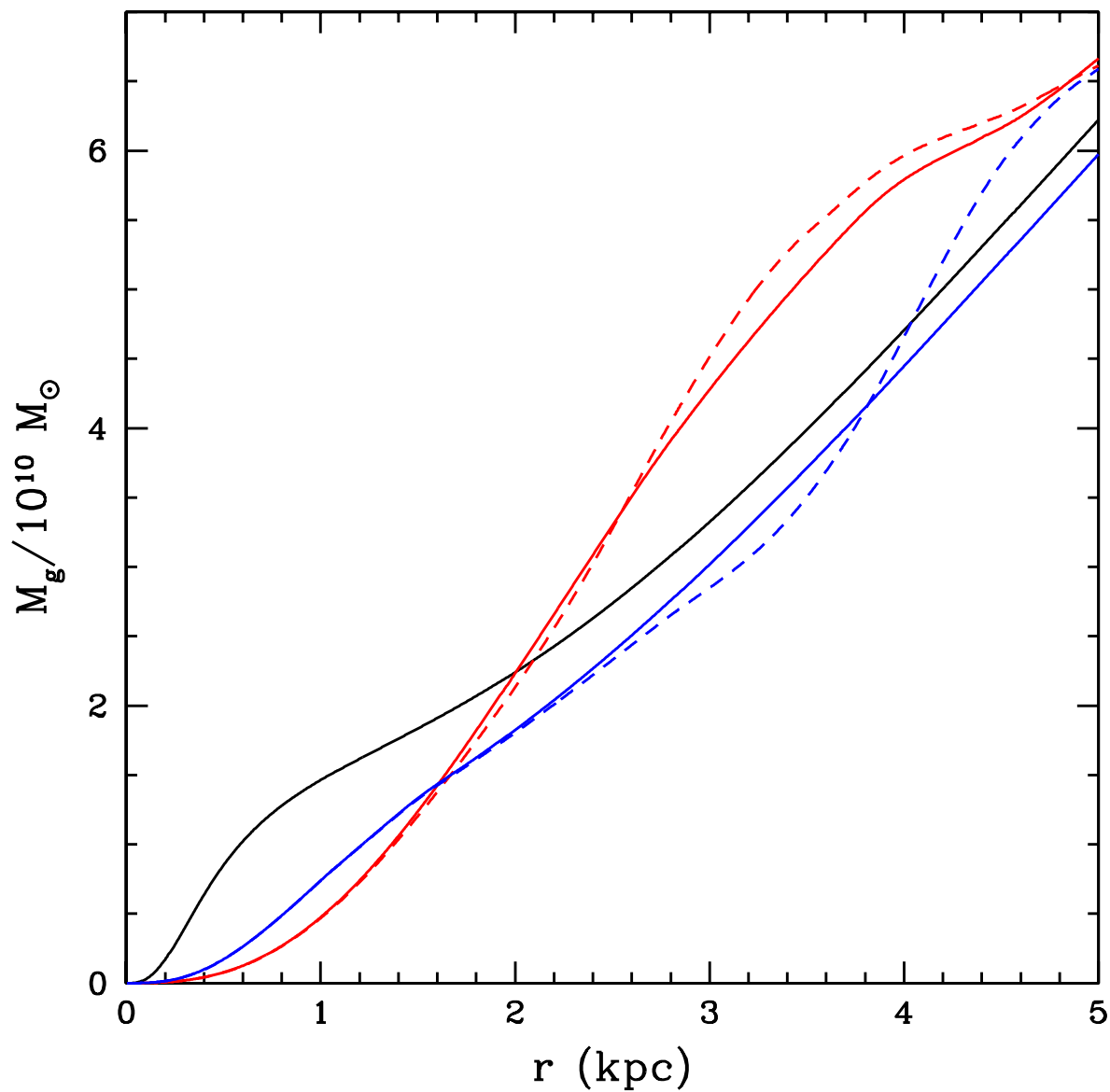}
\caption {$M_g$ computed on the Galactic plane as a function of distance
to the Galactic center. Values in the axisymmetric potential are shown
with the black line; the continuous and dashed red and blue lines show
values in the non-axisymmetric potential, along the major (red) and
minor (blue) axes of the bar. The continuous red and blue lines show the
contribution of the axisymmetric background and Galactic bar in this
potential, and the corresponding dashed lines includes the spiral arms
with a particular orientation (see main text).}
\label{fig9}
\end{figure}

To illustrate these comments, in Figure \ref{fig9} we give values
of $M_g$ computed on the Galactic plane as a function of distance
to the Galactic center. The black line corresponds to the axisymmetric
potential; the continuous red and blue lines show values of $M_g$ due
to the axisymmetric background (i.e. disk and spherical dark halo) plus
the Galactic bar, along the major and minor axes of the bar,
respectively. The dashed red and blue lines show values of $M_g$ along
these major and minor axes with the addition of the spiral arms, i.e.
considering all the mass components in the non-axisymmetric potential,
taking in particular the major axis of the bar as the line where the
spiral arms originate in the inner Galactic region. Thus, note the
smaller values that $M_g$ can take in the inner Galactic region in the
non-axisymmetric potential.

\begin{figure}
\plotone{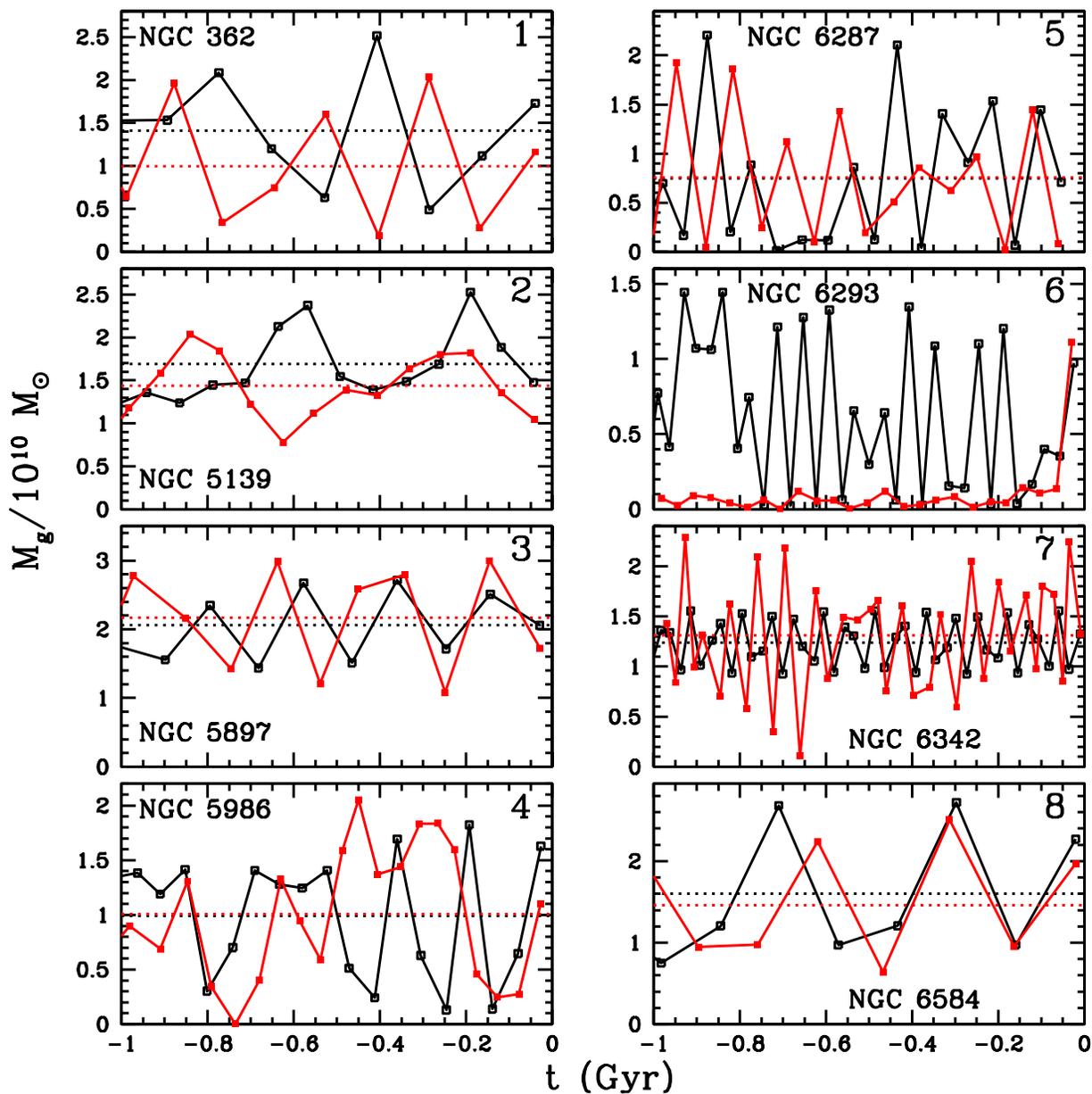}
\caption {Effective galactic mass $M_g$ employed in Eq. (\ref{rKt})
as a function of time over the last $10^9$ yr for the clusters in
Figure \ref{fig8}. The correspondence of colors is the same as in 
Figure \ref{fig8}. The horizontal dotted lines (not plotted in NGC 6293)
show the average values of $M_g$.}
\label{fig10}
\end{figure}

Figure \ref{fig10} shows the values of $M_g$ at the perigalactic points 
in Figure \ref{fig8}. The correspondence of colors in this figure is 
that given in Figure \ref{fig8}. The horizontal dotted lines show the
average values of $M_g$ over the last $10^9$ yr; these lines are not
plotted in NGC 6293 to avoid confusion in the lower continuous red
line. The average values of $M_g$ in the axisymmetric and
non-axisymmetric potentials almost coincide in the clusters NGC 5897,
NGC 5986, NGC 6287, NGC 6342, and NGC 6584; thus in these clusters the
increase of the average value of $r_{min}$ in the non-axisymmetric
potential explains the corresponding increase of $<$$r_{K_t}$$>$.
For the remaining clusters in this sample, NGC 362, NGC 5139, and
NGC 6293, the average value of $M_g$ is sensibly smaller in the
non-axisymmetric potential, specially in NGC 6293. This result 
combined with the increase of $r_{min}$, gives a net increase of
$<$$r_{K_t}$$>$ for NGC 362 and NGC 5139 in the non-axisymmetric
potential. On the other hand, in NGC 6293 the strong decrease of the
average value of $M_g$ in the non-axisymmetric potential compared with
that in the axisymmetric potential, counteracts the corresponding
slight decrease of the average value of $r_{min}$ shown in Figure
\ref{fig8}, giving a net increase of $<$$r_{K_t}$$>$ in the
non-axisymmetric potential.

\subsubsection{Comparison with improved observed limiting radii}
\label{kingwilson}

Recently \citet{MLF13} have derived the radial stellar density profiles
of 26 Galactic globular clusters from resolved star counts, using
high-resolution $Hubble Space Telescope$ observations. In particular,
they derive the limiting radius $r_l$, what we call the observed tidal
radius, employing King and Wilson \citep{W75} models. Considering the
clusters in common between our sample and their 26 clusters, we have
taken their best fits given by the least value of the reduced
${\chi}^2$ in the third column of their table 2, and compare our
theoretical tidal radii $r_{K_t}$ with their corresponding limiting
radii $r_l$.

\begin{figure}
\plotone{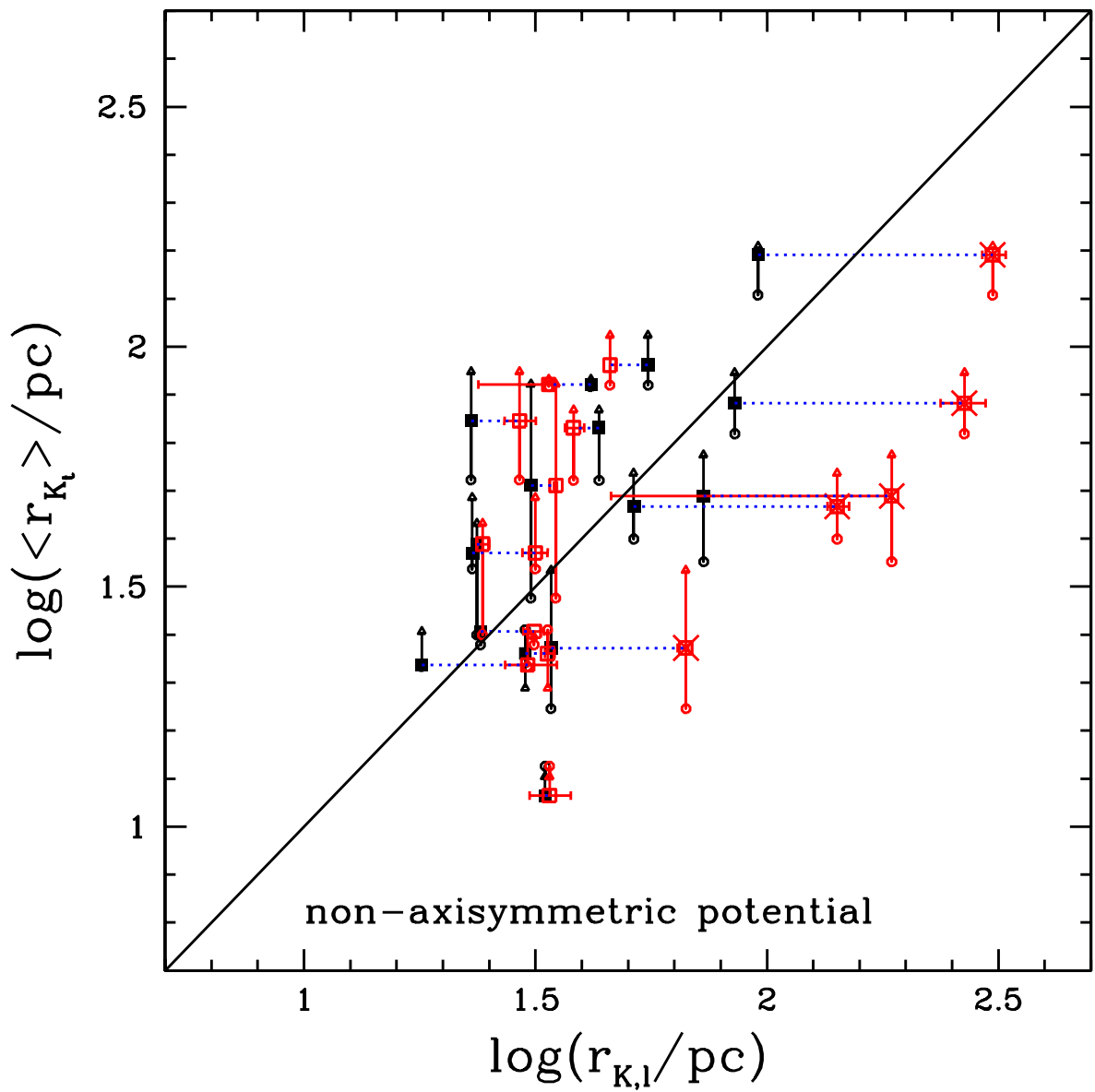}
\caption {Red points: comparison between $r_{K_t}$ and the limiting
radius $r_l$ obtained by \citet{MLF13} for some clusters in our sample.
The comparison is made in the non-axisymmetric potential. Black points
are corresponding points from $r_{K_t}$ vs $r_K$ in Figure \ref{fig7}.
Horizontal displacements between red and black points are shown with
dotted blue lines. Red crossed points correspond to clusters in which
a Wilson model gives the best fit.}
\label{fig11}
\end{figure}

We give this comparison in Figure \ref{fig11}, in particular in the
non-axisymmetric potential. The black points with their uncertainties
are points already plotted in Figure \ref{fig7} with corresponding 
King tidal radii $r_K$ employed in that figure. The red points are
the comparison between $r_{K_t}$ with $r_l$; the uncertainties in
$r_l$ are computed with data in table 2 of \citet{MLF13} using distances
given in our Table \ref{tbl-1}. Thus these points are displaced in the
horizontal axis with respect to the black points. The displacements
are shown with dotted blue lines. Red crossed points correspond to
clusters in which a Wilson model gives the best fit to the density
profile. The comparison between $r_{K_t}$ with $r_l$ is almost the same
as $r_{K_t}$ vs $r_K$ in Figure \ref{fig7}, except for the five red
crossed points, where $r_l$ is about a factor of 2-3 greater than $r_K$
and $r_{K_t}$. These five points correspond to NGC 288, NGC 5024 (M53),
NGC 5272 (M3), NGC 5466, and NGC 5904 (M5). As commented in Paper I,
the last three clusters appear to be dissolving \citep{LMC00,OG04,B06};
also, NGC 288 has extended tails \citep{GJH04}, and NGC 5024 is
possibly an accreted cluster \citep{MG04}. Thus in these five clusters
$r_l$ will be an upper bound for the tidal radius; some stars
contributing to $r_l$ may be already escaping from the cluster.

\subsubsection{Comparison with improved cluster masses}
\label{mldinamico}

In their study of structural properties of massive star clusters,
\citet{MM05} have obtained dynamical mass-to-light ratios $(M/L)_V$
for 57 Galactic globular clusters. In this part we compute $r_{K_t}$ 
for clusters in common between our sample and those listed in their
table 13, now with the cluster mass $M_c$ computed with their $(M/L)_V$,
instead of $(M/L)_V$ = 2 $M_{\odot}/L_{\odot}$ employed in our study.
With these new values of $r_{K_t}$, we compare $r_{K_t}$ vs $r_K$ in
the non-axisymmetric potential. 

\begin{figure}
\plotone{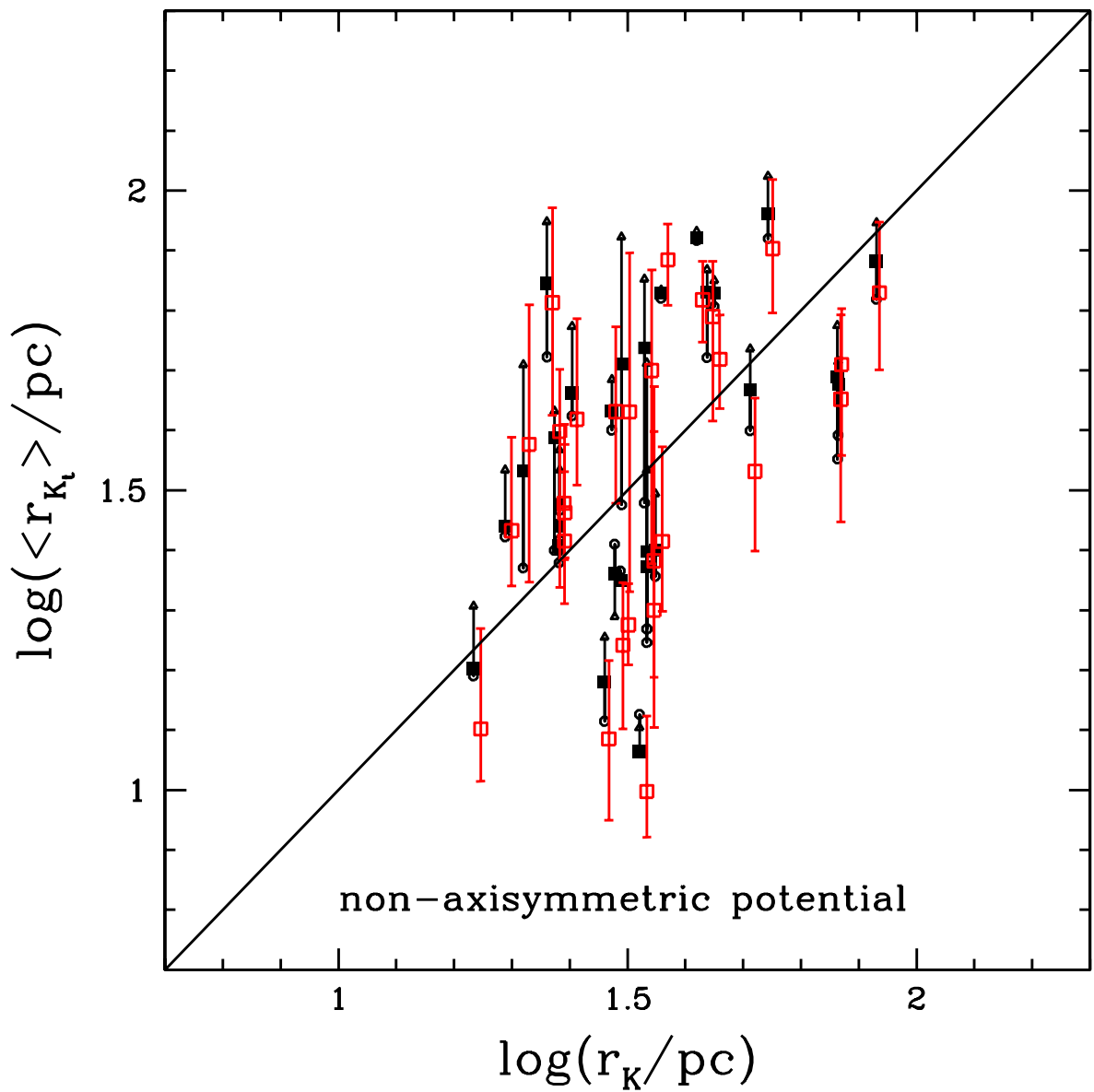}
\caption {Red points: comparison of $r_{K_t}$ vs $r_K$ in the
non-axisymmetric potential computing the cluster mass $M_c$ with
dynamical mass-to-light ratios given by \citet{MM05}. Black points:
corresponding points from Figure \ref{fig7} using $(M/L)_V$ = 2
$M_{\odot}/L_{\odot}$. For clarity, the red points are slightly
displaced to the right of the black points.}
\label{fig12}
\end{figure}

Figure \ref{fig12} shows this comparison. The black points are points
from Figure \ref{fig7}, using $(M/L)_V$ = 2 $M_{\odot}/L_{\odot}$; the
red points employ the $(M/L)_V$ values of \citet{MM05} with their
uncertainties. For clarity in the figure, these red points are slightly
displaced to the right of the black points. Thus, there is no much
difference with respect to the comparison made in Figure \ref{fig7},
and the standard $(M/L)_V$ = 2 $M_{\odot}/L_{\odot}$ is a convenient
test in our analysis. \rm

\subsection{Tidal radii with Monte Carlo computations}\label{monte}

The comparison of theoretical and observed tidal radii made in the
last section was repeated, now using Monte Carlo simulations.  The
uncertainties in the cluster distance, radial velocity, and proper
motions, listed in Table \ref{tbl-1}, plus the uncertainties of the
Galactic parameters listed in Table \ref{tbl-2}, were considered as
1$\sigma$ variations in a Gaussian Monte Carlo sampling. For each
cluster we computed a few hundreds of orbits. In each sampled cluster
orbit computed backward in time, we found the average $<$$r_{K_t}$$>$
over the last $10^9$ yr, and in turn, with all these values in a given
cluster, determined its corresponding global average, denoted by
$\langle r_{K_t} \rangle$, and its 1$\sigma$ variation.

\begin{figure}
\plotone{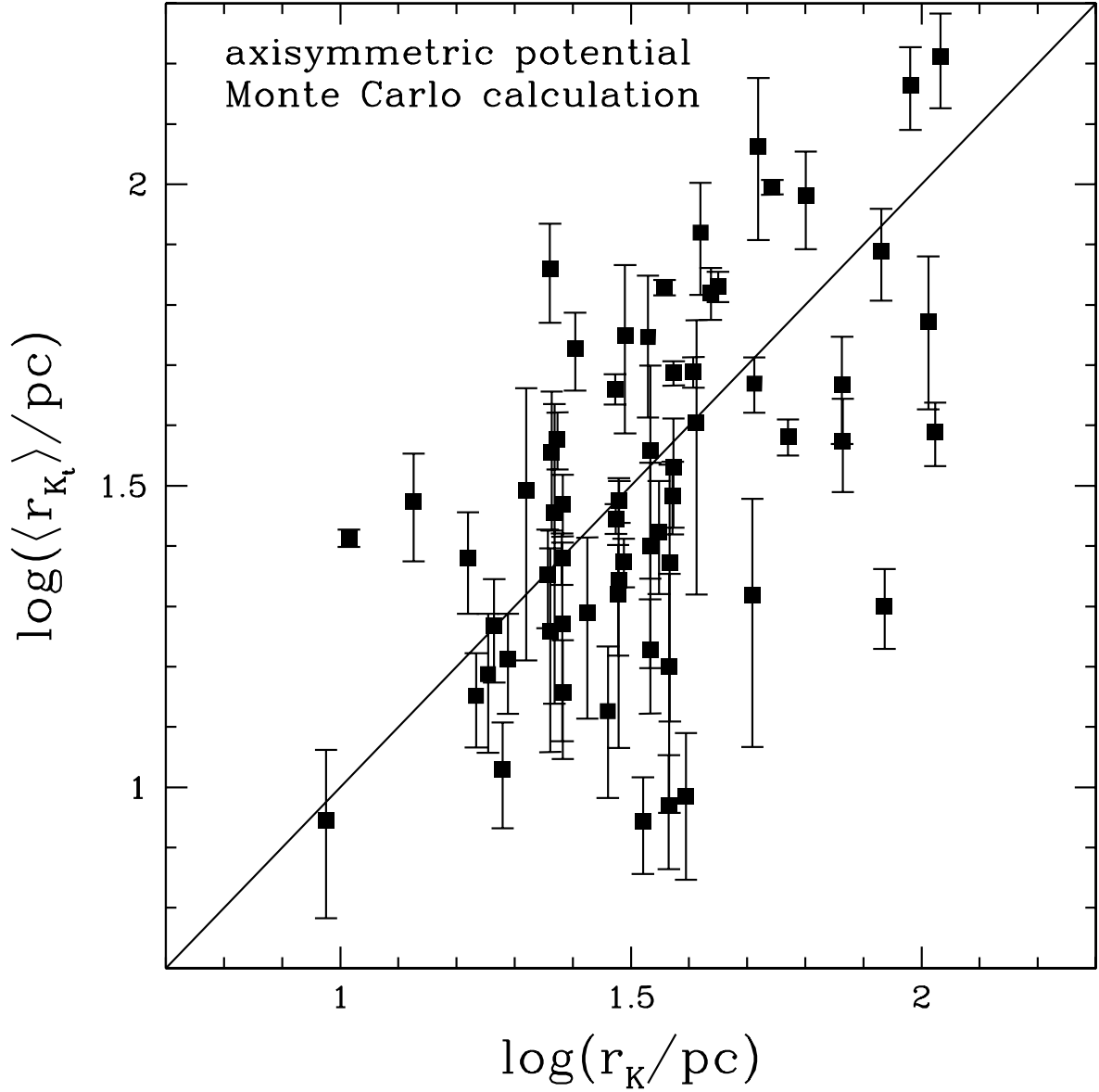}
\caption {Comparison of the Monte Carlo theoretical tidal radius
global average $\langle r_{K_t} \rangle$ computed with the
axisymmetric potential over the last $10^9$ yr, and the observed tidal
radius $r_K$. Compare this figure with Figure \ref{fig6}.}
\label{fig13}
\end{figure}

\begin{figure}
\plotone{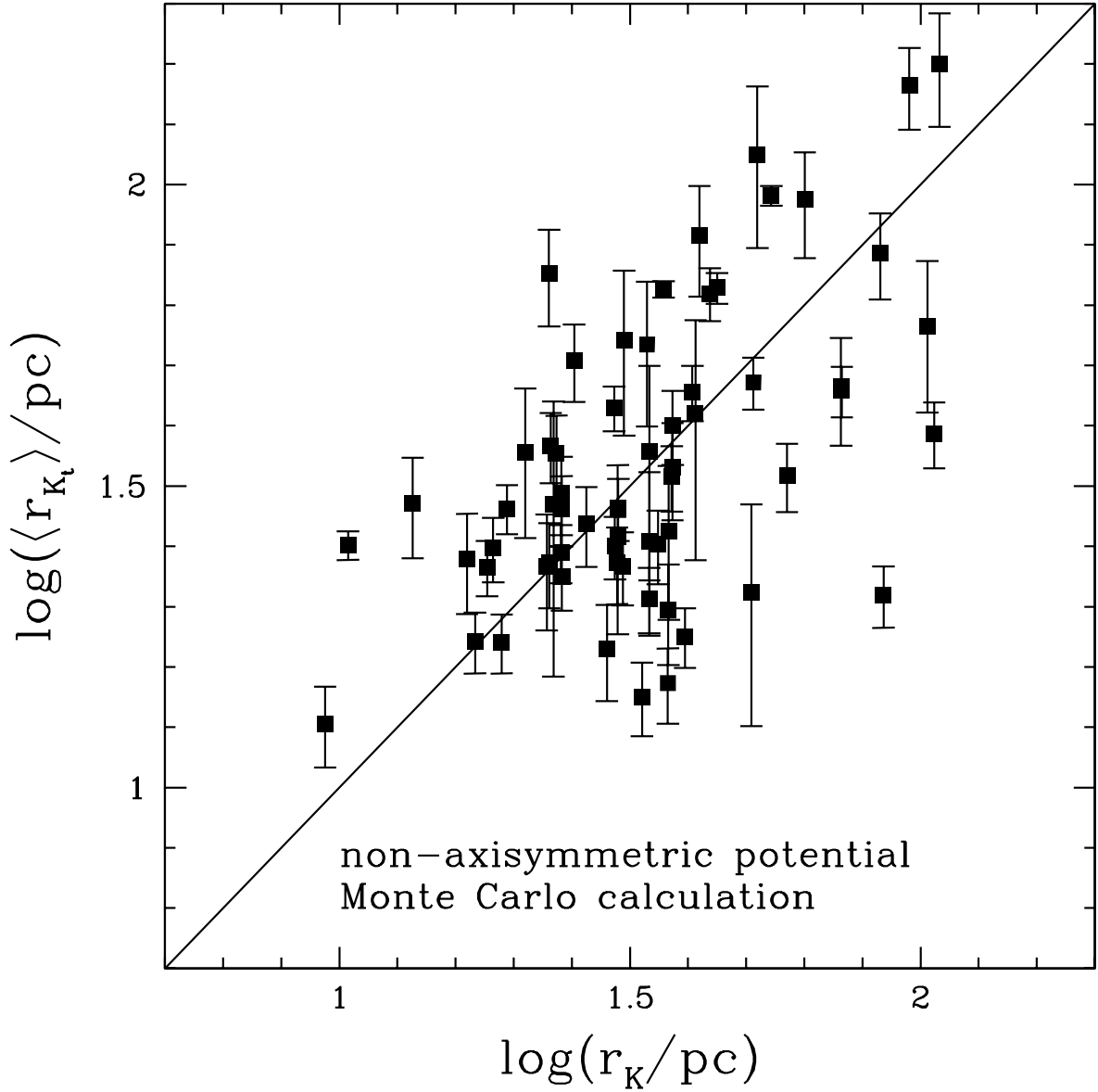}
\caption {As in Figure \ref{fig13}, here we show the results in the
non-axisymmetric potential. In this case compare with Figure
\ref{fig7}.} 
\label{fig14}
\end{figure}

Figures \ref{fig13} and \ref{fig14} show the comparison of $\langle
r_{K_t} \rangle$ with $r_K$ in the axisymmetric and non-axisymmetric
Galactic potentials. The error bars in both figures correspond to the
computed 1$\sigma$ variation. With these Monte Carlo calculations we
obtain nearly the same comparisons shown in Figures \ref{fig6} and
\ref{fig7}; thus, again our conclusion is that $\langle r_{K_t}
\rangle$ compares better with $r_K$ employing the non-axisymmetric
potential.  These Figures \ref{fig13} and \ref{fig14} also show that 
the estimate of the uncertainty in $<$$r_{K_t}$$>$ done in $\S$
\ref{compar} using the minimum and maximum energy orbits, is
acceptable.

An additional plotted point in Figures \ref{fig13} and \ref{fig14},
which does not appear in Figures \ref{fig6} and \ref{fig7}, is the one
corresponding to the cluster Pal 3, the upper point in these figures.
This cluster has an unbounded orbit computed with its mean distance,
radial velocity, and proper motions listed in Table \ref{tbl-1}, and
the mean Galactic parameters in Table \ref{tbl-2}. In the Monte Carlo
simulations we have picked out its bounded orbits.

\subsection{Overfilling excess of predicted tidal radius?}\label{exceso}

\citet{WHS13} have considered observed limiting radii of Galactic
globular clusters, given by a King model, $r_K$, and their theoretical
tidal radius, $r_t$, computed at their perigalactic distance in the
axisymmetric Galactic potential used by \citet{JSH}, but $r_t$
obtained as if the Galactic potential were spherically symmetric. They
take the ratio of the difference ($r_K$$-$$r_t$) to the average
($r_K$+$r_t$)/2, and plot this ratio against the perigalactic
distance. Clusters with this ratio greater than zero, overfill their
predicted theoretical tidal radius; underfilling occurs in clusters
which have this ratio less than zero. \citet{WHS13} show in their
figure 1 that the majority of clusters are overfilling their predicted
tidal radius.  Through $N$-body simulations of star clusters moving on
the plane of symmetry of a given axisymmetric Galactic potential, they
find an analytical correction to be applied to $r_t$ computed at
perigalacticon, to obtain a better estimate of the cluster limiting
radius $r_L$. With this value of $r_L$ employed instead of $r_t$ in
the computation of the ratio mentioned above, \citet{WHS13} find that
the overfilling excess disappears, and there is a stronger agreement
between theory and observations.

To compare with the results of \citet{WHS13}, in our axisymmetric and
non-axisymmetric Galactic potentials we compute the theoretical
tidal radius at perigalacticon with King's formula Eq. (\ref{rKt})
(or alternatively with $r_{\ast}$ in Eq. (\ref{rast})), take the
average values of $r_{K_t}$ over the last $10^9$ yr listed in column 8
of Table \ref{tbl-3}, take the ratio of the difference
($r_K$$-$$<$$r_{K_t}$$>$) to the average
($r_K$+$<$$r_{K_t}$$>$)/2 ($r_K$ is listed in Table \ref{tbl-1})
and plot this ratio against the logarithm of the average perigalactic
distance in these last $10^9$ yr. We do the same taking only the last
perigalacticon, and its distance to the Galactic center.

\begin{figure}
\plotone{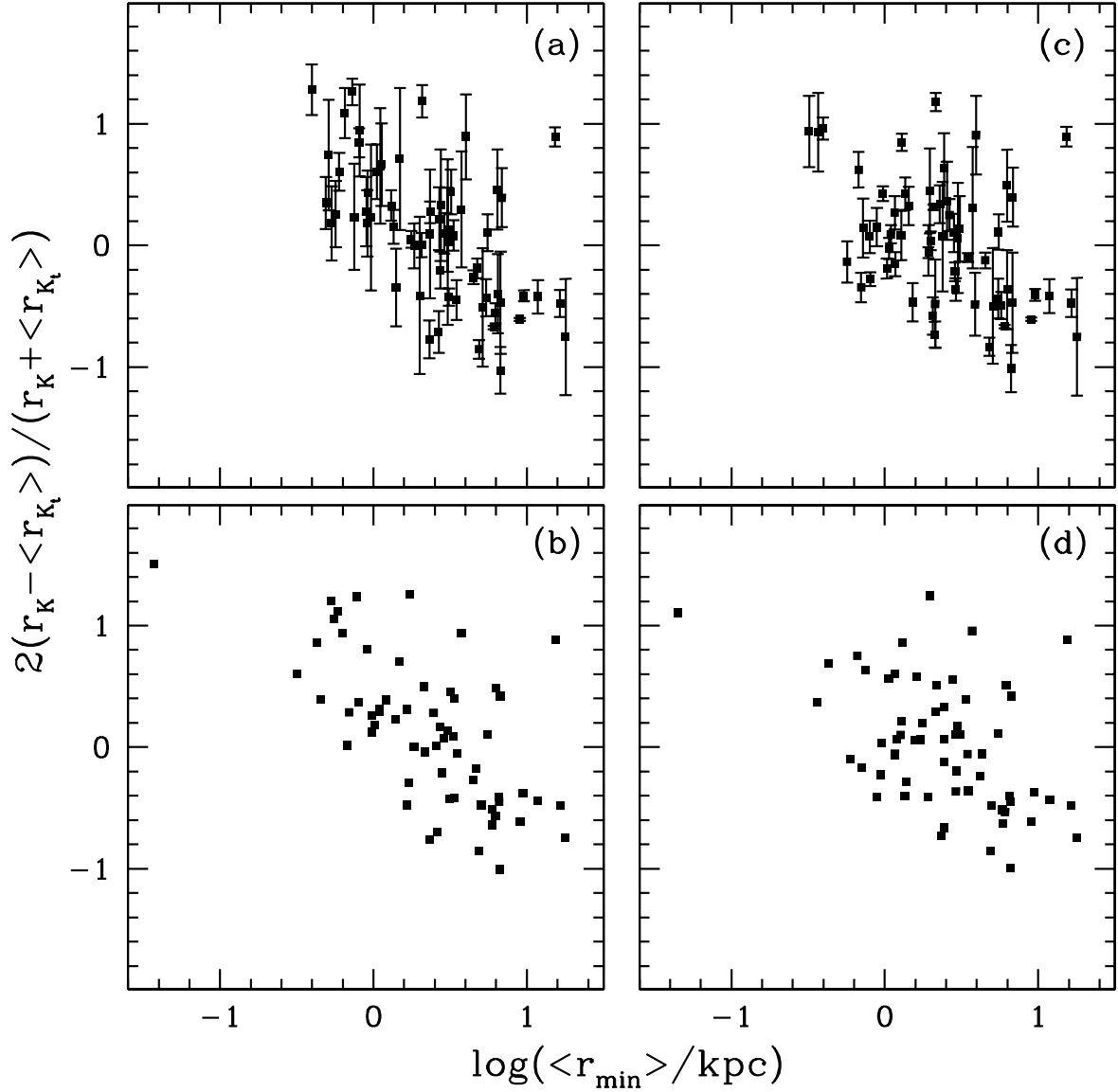}
\caption {Frames (a),(c): ratio of the difference between $r_K$ in
Table \ref{tbl-1} and $<$$r_{K_t}$$>$ in column 8 of Table
\ref{tbl-3} for the last $10^9$ yr in the orbital computation,
to their average ($r_K$+$<$$r_{K_t}$$>$)/2, plotted against
the logarithm of the average perigalactic distance in this time
interval. Frames (b),(d): the same comparison, but with $r_{K_t}$
computed only in the last perigalacticon, and the logarithm of its
corresponding distance to the Galactic center. Frames (a),(b) show
results in the axisymmetric Galactic potential, and (c),(d) in the
non-axisymmetric Galactic potential.} 
\label{fig15}
\end{figure}

Figure \ref{fig15} shows our results. The two upper frames (a),(c)
correspond to the last $10^9$ yr, and the two lower frames (b),(d) to
the last perigalacticon. The frames on the left (a),(b) give results
in the axisymmetric Galactic potential, and the frames on the right
(c),(d) in the non-axisymmetric Galactic potential. At first sight,
there is no evident overfilling nor underfilling excess of predicted
theoretical tidal radius; the points scatter approximately around zero
value in the ratio 2($r_K$$-$$<$$r_{K_t}$$>$)/($r_K$+$<$$r_{K_t}$$>$).
This holds in both the axisymmetric and non-axisymmetric Galactic
potentials, thus the main point in the discrepance of our results with
those of \citet{WHS13} seems to be the different ways in which the
theoretical tidal radius is computed at perigalactic distance. 

\begin{figure}
\plotone{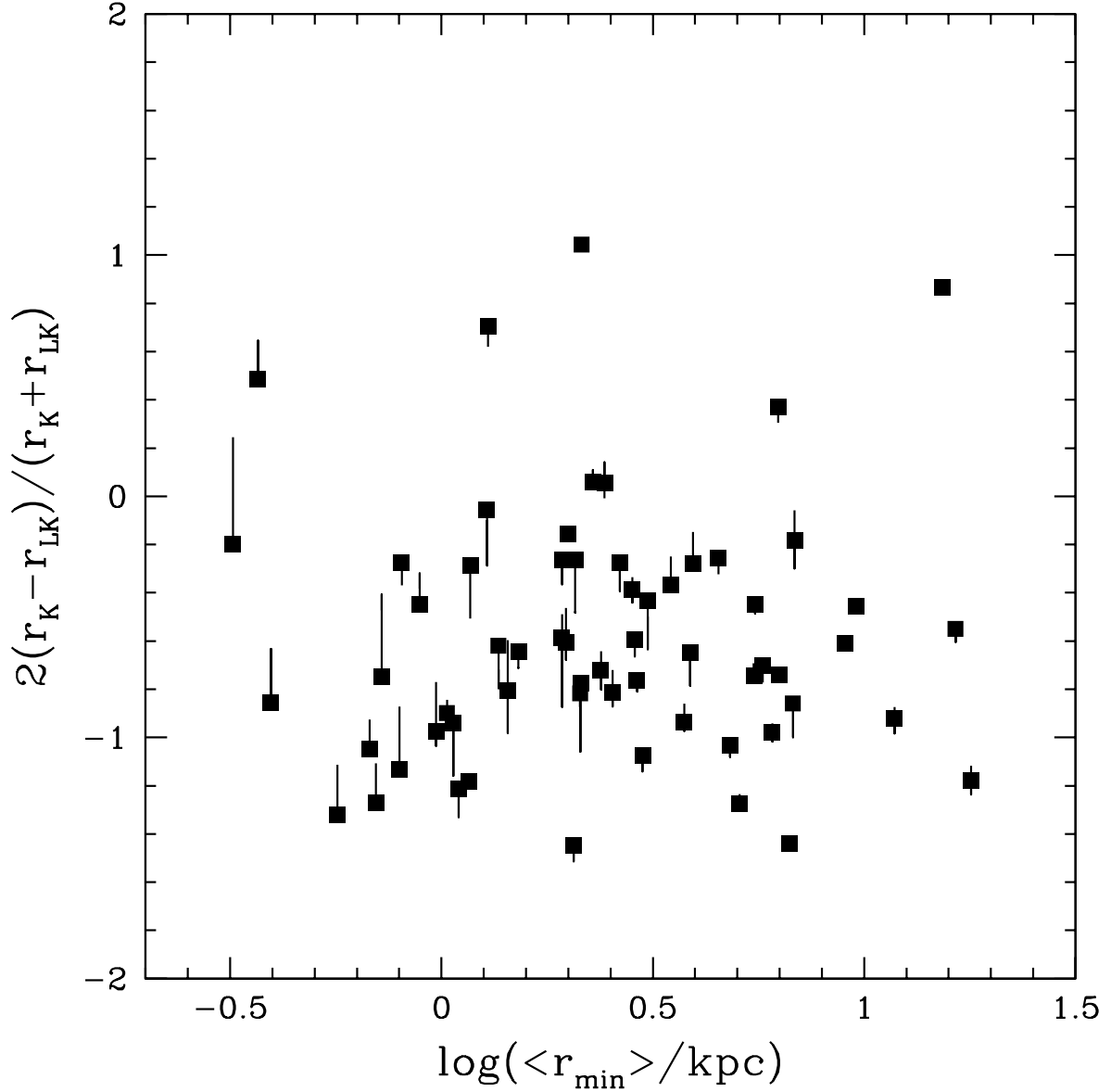}
\caption {As in Figure \ref{fig15} in the non-axisymmetric Galactic
potential over the last $10^9$ yr, but now applying to $r_{K_t}$
computed at perigalacticon the correction given by \citet{WHS13} in
their equation 8, which results in the average limiting radius 
$r_{LK}$. This new limiting radius is used instead of $<$$r_{K_t}$$>$
in the ratio plotted in the ordinate axis of Figure \ref{fig15}.}
\label{fig16}
\end{figure}

If we apply the correction given by \citet{WHS13} in their equation 8
to our $r_{K_t}$ computed at perigalacticon, this leads to a new cluster
limiting radius $r_{LK}$. To compute this limiting radius we take in
each cluster an average of $r_{K_t}$, the orbital eccentricity, and the
orbital phase, over the last $10^9$ yr. We determine the ratio of
the difference ($r_K$$-$$r_{LK}$) to the average ($r_K$+$r_{LK}$)/2,
and plot this ratio against the logarithm of the average perigalactic
distance in these last $10^9$ yr. This is shown in Figure \ref{fig16} 
only for the non-axisymmetric Galactic potential. The error bars shown
in this figure are obtained using the the minimum and maximum energy
orbits, mentioned in $\S$ \ref{compar}. We find a strong underfilling 
excess; there is a systematic shifting towards negative values in
comparison with the initial approximately zero excess found in Figure
\ref{fig15}. Then our conclusion is that in our computations we do not
need to apply the correction given by \citet{WHS13}. This issue needs
a further study.

\section{Bulge-shocking destruction rates taking the real trajectory of a cluster}\label{tdestr}

In Paper I we have listed some relations needed to compute destruction
rates of globular clusters due to bulge and disk shocking. Those
corresponding to bulge shocking employ the impulse approximation,
along with adiabatic corrections, and a linear trajectory of the
cluster when passing at a given perigalactic point. In this section we
give relations to compute destruction rates due to bulge shocking
employing the real trajectory of a globular cluster, maintaining the
impulse approximation.

Let {\boldmath $\cal F$}({\boldmath $r$}) be the Galactic
gravitational acceleration at the position {\boldmath $r$} in an
inertial reference system with origin at the Galactic center. In
particular, in the following we consider this acceleration due to the
spherical bulge component in the used axisymmetric model, and due to
the bar in the non-axisymmetric model, in which all the bulge is
represented by this bar, as mentioned in $\S$ \ref{gpot}.

With {\boldmath $r$}$_c$ the position of the center of a globular
cluster, and {\boldmath $r'$} the position of a star in the cluster
with respect to the cluster center, then 
{\boldmath $r$}={\boldmath $r$}$_c$+{\boldmath $r'$} is the position of
the star in the inertial frame, and we define
{\boldmath $F$}({\boldmath $r'$}) with the relation 
{\boldmath $\cal F$}({\boldmath $r$})$\equiv$
{\boldmath $\cal F$}({\boldmath $r$}$_c$+{\boldmath $r'$})=
{\boldmath $F$}({\boldmath $r'$}).

Then, up to linear terms in {\boldmath $r'$} and at the cluster
position {\boldmath $r$}$_c$, the tidal acceleration
{\boldmath $M$}({\boldmath $r'$}) on the star is (with a sum over a
repeated index)

\begin{equation} \mbox{\boldmath $M$} \left (\mbox{\boldmath $r'$}
\right )  =  \mbox{\boldmath $F$} \left (\mbox{\boldmath $r'$}
\right )-\mbox{\boldmath $F$} \left (\mbox{\boldmath $r'$} = 0 \right )
\simeq \mbox{\boldmath $J$} \cdot \mbox{\boldmath $r'$} =
\mbox{\boldmath $e$}_i x'_j \left (\frac{\partial
F_{x'_i}}{\partial x'_j} \right )_{{\bf r'}= 0}.
\label{acelmar}
\end{equation}

\noindent The coordinates $x'_i$, $i$ = 1, 2, 3, are Cartesian 
coordinates of {\boldmath $r'$}; this vector written in the inertial
base of unitary vectors
({\boldmath $e$}$_1$, {\boldmath $e$}$_2$, {\boldmath $e$}$_3$). 
The matrix {\boldmath $J$} is given by

\begin{equation} \mbox{\boldmath $J$}  =
\bordermatrix{ &   &   &   \cr
& \frac{\partial F_{x'}}{\partial x'} & \frac{\partial F_{x'}}{\partial y'} & \frac{\partial F_{x'}}{\partial z'} \cr
& \frac{\partial F_{y'}}{\partial x'} & \frac{\partial F_{y'}}{\partial y'} & \frac{\partial F_{y'}}{\partial z'} \cr
& \frac{\partial F_{z'}}{\partial x'} & \frac{\partial F_{z'}}{\partial y'} & \frac{\partial F_{z'}}{\partial z'} \cr}_{\bf r' = 0}.
\label{ecj}
\end{equation}
\\

To obtain the stellar velocity change due to {\boldmath $M$}, we
integrate d{\boldmath $v'$}/dt = {\boldmath $M$} in the inertial frame,
taking two successive apogalactic points in the cluster's orbit.
The stellar velocity {\boldmath $v'$} is measured in the inertial frame.
Using the impulse approximation, the change in the $ith$ component of
the stellar velocity between these two successive apogalactic points is 
(there is a sum over the index $j$)

\begin{equation} {\Delta}v'_i = x'_j \int_{t_{ap1}}^{t_{ap2}}
\left (\frac{\partial F_{x'_i}}{\partial x'_j} \right )_{{\bf r'}= 0}
dt = x'_j I_{ij},
\label{delv}
\end{equation}

\noindent with $I_{ij}$ defined by the integral, and the integration
done numerically along the real orbit of the cluster, under the whole
used Galactic potential, between successive apogalactic points
occurring at times $t_{ap1}$, $t_{ap2}$, i.e. an apogalactic period.

The change of stellar energy per unit mass is
$\Delta E$ = {\boldmath $v'$}$\cdot$$\Delta${\boldmath $v'$}+
(1/2)($\Delta${\boldmath $v'$})$^2$. Thus, with Eq. (\ref{delv}), 
assuming spherical symmetry in the cluster, and an isotropic stellar 
velocity distribution within the cluster depending only on distance
$r'$ = $|${\boldmath $r'$}$|$ from the center of the cluster, we have
the two local (i.e. averaged on a spherical surface of radius $r'$)
diffusion coefficients due to the interaction with the bulge     

\begin{equation} <\!(\Delta E)_b\!>_{loc} =
\frac{1}{2}<\!(\Delta \mbox{\boldmath $v'$})^2\!>_{loc} =
\frac{1}{6}r'^2 \sum_{i,j}^{}I^2_{ij},     
\label{delE} \end{equation}

\begin{equation} <\!(\Delta E)_b^2\!>_{loc} \approx
<\!(\mbox{\boldmath $v'$}\cdot\Delta \mbox{\boldmath $v'$})^2\!>_{loc} =
\frac{1}{9}r'^2v'^2(r')(1 + {\chi}_{r',v'}(r')) \sum_{i,j}^{}I^2_{ij},
\label{delE2} \end{equation}

\noindent with $v'(r')$ the rms velocity within the cluster at
distance $r'$, and the position-velocity correlation function
${\chi}_{r',v'}(r')$ given by \citet{GO99}. The sum in these equations
gives nine squared coefficients $I^2_{ij}$. 

To take into account the stellar motion within the cluster during its
interaction with the Galactic bulge or bar in an apogalactic period,
adiabatic correction factors ${\eta}_1 (x)$, ${\eta}_2 (x)$ are
introduced in Eqs. (\ref{delE}) and (\ref{delE2}), having the forms
\citep{GO99}

\begin{equation} {\eta}_1 (x(r')) = (1 + x^2(r'))^{-{\gamma}_1},
\label{adiab1}
\end{equation}

\begin{equation} {\eta}_2 (x(r')) = (1 + x^2(r'))^{-{\gamma}_2},
\label{adiab2}
\end{equation}

\noindent with $x(r') = {\omega}(r'){\tau}$; $\omega (r')$ is angular
velocity of stars inside the cluster at distance $r'$, and as in
Paper I, the angular velocity in circular motion at distance $r'$ is
considered to represent this $\omega (r')$.
The factor $\tau$ is an effective interaction time with the Galactic
bulge or bar in an apogalactic period. The exponents ${\gamma}_1$,
${\gamma}_2$ depend on the ratio between $\tau$ and the cluster's
inner dynamical time evaluated at the half-mass radius,
$t_{dyn,h} = ({\pi}^2{r_h}^3/2GM_c)^{1/2}$ (the half-mass radius
$r_h$ and the mass of the cluster $M_c$ are listed in Table
\ref{tbl-1}). The values of ${\gamma}_1$, ${\gamma}_2$ are those 
considered in Paper I, based on Table 2 of \citet{GO99}. 

In the usual procedure employed in the impulse approximation with a
linear trajectory of the cluster passing at a given perigalactic
point, the effective interaction time $\tau$ is estimated as
$\tau = |${\boldmath $r$}$_p|/|${\boldmath $v$}$_p|$, with
{\boldmath $r$}$_p$, {\boldmath $v$}$_p$ the position and velocity of
the cluster at this point with respect to the Galactic inertial frame.
For the real trajectory of the cluster considered in this section, we
follow the treatment of \citet{G99a} to estimate $\tau$.  \citet{G99a}
consider a potential with spherical symmetry and estimate $\tau$
making a Gaussian fit of the form $e^{-t^2/{\tau}^2}$ to the tidal
acceleration in the z-direction. In our Galactic potentials we make a
similar fit to the rms $total$ tidal acceleration.

From Eq. (\ref{acelmar}), the local averaged square tidal acceleration
is

\begin{equation}  <\!\mbox{\boldmath $M$}^2 \!>_{loc} =
\frac{1}{3}r'^2 \sum_{i,j}^{}
\left (\frac{\partial F_{x'_i}}{\partial x'_j} \right )^2_{{\bf r'}= 0}
\label{acelmar2}
\end{equation}

Taking an average over the cluster, the Gaussian fit in an apogalactic
period is made on the rms tidal acceleration given by

\begin{equation}  \left ( <\!\mbox{\boldmath $M$}^2 \!> \right )^{1/2}
 =
\left \{ \frac{1}{3} <\!r_c^2\!> \sum_{i,j}^{}
\left (\frac{\partial F_{x'_i}}{\partial x'_j} \right )^2_{{\bf r'}= 0} \right \}^{1/2},
\label{rms}
\end{equation}

\noindent where $<$$r_c^2$$>$ is the mean square cluster radius,
computed below. The resulting value of $\tau$ given by the fit is
employed in Eqs. (\ref{adiab1}) and (\ref{adiab2}).

\begin{figure}
\plotone{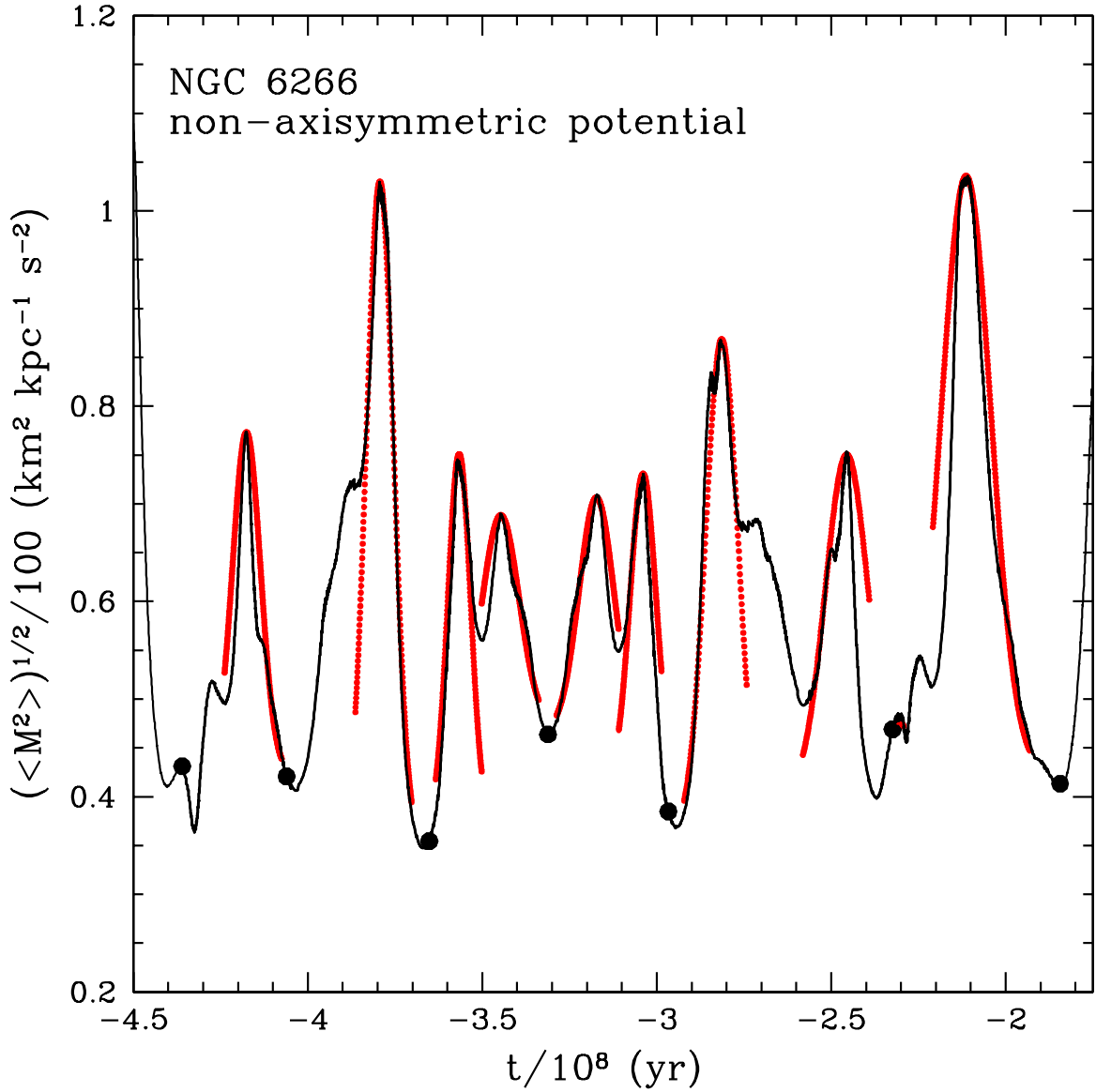}
\caption {An example in the non-axisymmetric potential of the Gaussian
fits to ($<${\boldmath $M$}$^2>$)$^{1/2}$ in some apogalactic periods,
where this tidal acceleration has more than one maximum in a given
period. The black dots show the positions of the apogalactic points, 
and the red curves the approximate fits to the main peaks in the tidal
acceleration.}
\label{fig17}
\end{figure}

In the axisymmetric potential the tidal acceleration
($<${\boldmath $M$}$^2>$)$^{1/2}$ has one maximum in every apogalactic
period. However, in the non-axisymmetric potential, and in some
clusters, this acceleration may have a complicated behavior in some
apogalactic periods, showing more than one maximum.
Figure \ref{fig17} shows as an example some apogalactic periods
in a run in the cluster NGC 6266. The black dots give the positions
in time of apogalactic points; the black curve is
($<${\boldmath $M$}$^2>$)$^{1/2}$ and the red curves show the typical
approximate fits made to the main acceleration peaks in cases like this.

Including the adiabatic correction factors, the averages of Eqs.
(\ref{delE}) and (\ref{delE2}) over the cluster are

\begin{equation} <\!(\Delta E)_b\!> =
\frac{1}{6}<\!r'^2 {\eta}_1 (x(r'))\!> \sum_{i,j}^{}I^2_{ij},
\label{delEtot} \end{equation}

\begin{equation} <\!(\Delta E)_b^2\!> \approx
\frac{1}{9}<\!r'^2v'^2(r')(1 + {\chi}_{r',v'}(r')) {\eta}_2 (x(r'))\!>
\sum_{i,j}^{}I^2_{ij},
\label{delE2tot} \end{equation}

\noindent with

\begin{equation} <\!r'^2 {\eta}_1 (x(r'))\!> = \frac{4\pi}{M_c}
\int_{0}^{r_K} {\rho}_c(r') {\eta}_1(x(r')) r'^4 dr',
\label{prom1}
\end{equation}

\begin{equation} <\!r'^2v'^2(r')(1 + {\chi}_{r',v'}(r')){\eta}_2 (x(r'))\!> =
\frac{4{\pi}G}{M_c}
\int_{0}^{r_K} {\rho}_c(r') M_c(r') {\eta}_2(x(r'))
(1+{\chi}_{r',v'}(r')) r'^3 dr',
\label{prom2}
\end{equation}

\noindent and $<$$r_c^2$$>$ in Eq. (\ref{rms})

\begin{equation} <\!r_c^2\!> = <\!r'^2 \!> = \frac{4\pi}{M_c}
\int_{0}^{r_K} {\rho}_c(r') r'^4 dr',
\label{promr}
\end{equation}

\noindent $r_K$ is the tidal radius of the cluster (listed in Table
\ref{tbl-1}), $M_c$ is its total mass, ${\rho}_c(r')$ is its spatial
density, obtained with a \citet{K66} model, and $M_c(r')$ is the mass
of the cluster within radius $r'$. As in Paper I, we approximate the
rms velocity $v'(r')$ with the corresponding circular velocity at
that $r'$.

With $E_c \simeq -0.2GM_c/r_h$ the mean binding energy per unit mass
of the cluster, and if the cluster has a dominant maximum of the
tidal acceleration ($<${\boldmath $M$}$^2>$)$^{1/2}$ in a given
apogalactic period, bulge shock timescales in this period are defined
as \citep{GO97} 

\begin{equation} t_{bulge,1} = \left ( \frac{-E_c}{<\!(\Delta E)_b\!>}
\right )P_{orb}, \label{tb1} \end{equation}

\begin{equation} t_{bulge,2} = \left ( \frac{E_c^2}{<\!(\Delta E)_b^2\!>}
\right )P_{orb}, \label{tb2} \end{equation}

\noindent with $P_{orb}$ the apogalactic period. If the cluster has
more than one maximum of ($<${\boldmath $M$}$^2>$)$^{1/2}$ in the
given apogalactic period, as in Figure \ref{fig10}, instead of 
$P_{orb}$ we use in each main fitted peak the corresponding interval
of time taken in the fit. In each case the total destruction rate due
to bulge shocking is

\begin{equation} \frac{1}{t_{bulge}} = \frac{1}{t_{bulge,1}} +
\frac{1}{t_{bulge,2}}. \label{tbtot} \end{equation}

\section{Disk and spiral arms shocking}\label{disco}

\bf The treatment for disk shocking remains the same as in Paper I.
The corresponding expressions to Eqs. (\ref{delEtot}) and
(\ref{delE2tot}) for disk shocking are obtained averaging equations
(1) and (2) in \citet{G99b}, resulting in

\begin{equation} <\!(\Delta E)_d\!> = \frac{2g_m^2}{3v_z^2}
<\!r'^2 {\eta}_1 (x(r'))\!>, \label{delEd} \end{equation}

\begin{equation} <\!(\Delta E)_d^2\!> = \frac{4g_m^2}{9v_z^2}
<\!r'^2 v'^2 {\eta}_2 (x(r'))(1 + {\chi}_{r',v'}(r'))\!>.
\label{delE2d} \end{equation}

In both, the axisymmetric and non-axisymmetric Galactic potentials,
$|g_m|$ is the maximum acceleration produced by the corresponding
axisymmetric disk component in its perpendicular z-direction, on the
perpendicular line to the plane of the disk passing at the position
where the cluster crosses the disk, and $|v_z|$ is the z-velocity of 
the cluster at this point. Here $\tau$ in $x(r') = {\omega}(r'){\tau}$
is given by $\tau = |z_m|/|v_z|$, with $|z_m|$ the z-distance at
which $|g_m|$ is reached.

With $n$ crossings of the cluster orbit with the Galactic plane,
disk shock timescales and corresponding total destruction rate are
given by

\begin{equation} t_{disk,1} = \left ( \frac{-E_c}{<\!(\Delta E)_d\!>}
\right ) \frac{P_{orb}}{n}, \label{td1} \end{equation}

\begin{equation} t_{disk,2} = \left ( \frac{E_c^2}{<\!(\Delta E)_d^2\!>}
\right ) \frac{P_{orb}}{n}, \label{td2} \end{equation}

\begin{equation} \frac{1}{t_{disk}} = \frac{1}{t_{disk,1}} +
\frac{1}{t_{disk,2}}. \label{tdtot} \end{equation}

In the non-axisymmetric potential, the spiral arms represent a plane
mass distribution, analogous to the axisymmetric disk component, and 
also produce a shock on a cluster crossing the Galactic plane, with
corresponding averaged diffusion coefficients $<\!(\Delta E)_{arms}\!>$
and $<\!(\Delta E)_{arms}^2\!>$ . At a given crossing point, the ratios
$<\!(\Delta E)_{arms}\!>/<\!(\Delta E)_d\!>$,
$<\!(\Delta E)_{arms}^2\!>/<\!(\Delta E)_d^2\!>$ between averaged
diffusion coefficients due to the spiral arms and axisymmetric disk,
will depend on the squared ratio of corresponding maximum accelerations
$(|g_m|_{arms}/|g_m|_{disk})^2$. The velocity $|v_z|$ has the same
value in both type of diffusion coefficients, as this velocity, as well
as the orbit itself, is computed under the whole Galactic potential, 
i.e. including the axisymmetric (disk and dark halo) and
non-axisymmetric (bar and spiral arms) components. There will be also
a dependence of these ratios on the corresponding $|z_m|$ given by the
spiral arms, through the dependence of ${\eta}_1$ and ${\eta}_2$ on
$\tau$.

\begin{figure}
\plotone{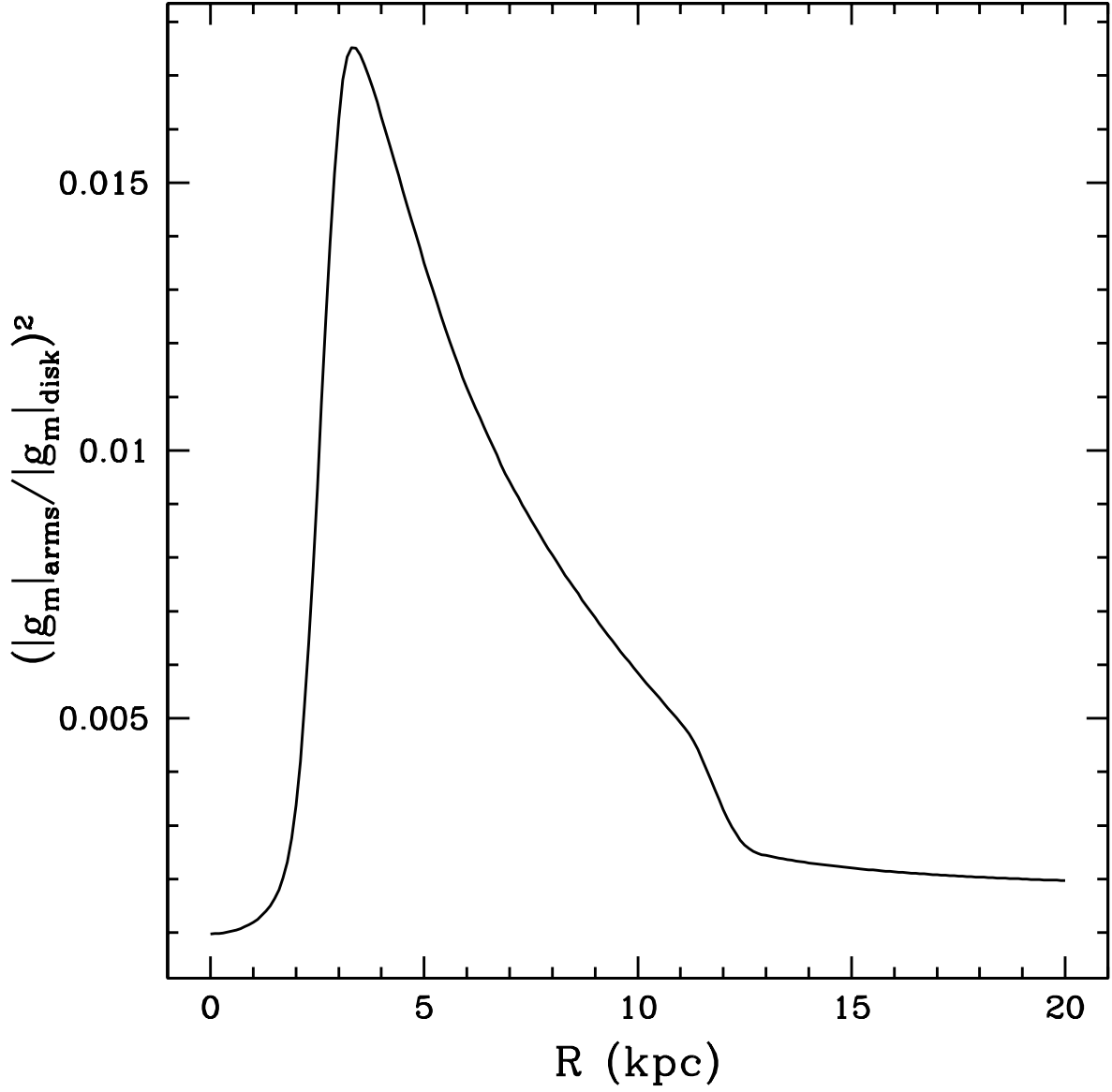}
\caption {Azimuth-averaged squared ratio of maximum z-accelerations
due to the spiral arms and axisymmetric disk component in the
non-axisymmetric potential, as a function of the distance $R$ to the
Galactic center of an orbital crossing point with the Galactic plane.}
\label{fig18}
\end{figure}

\begin{figure}
\plotone{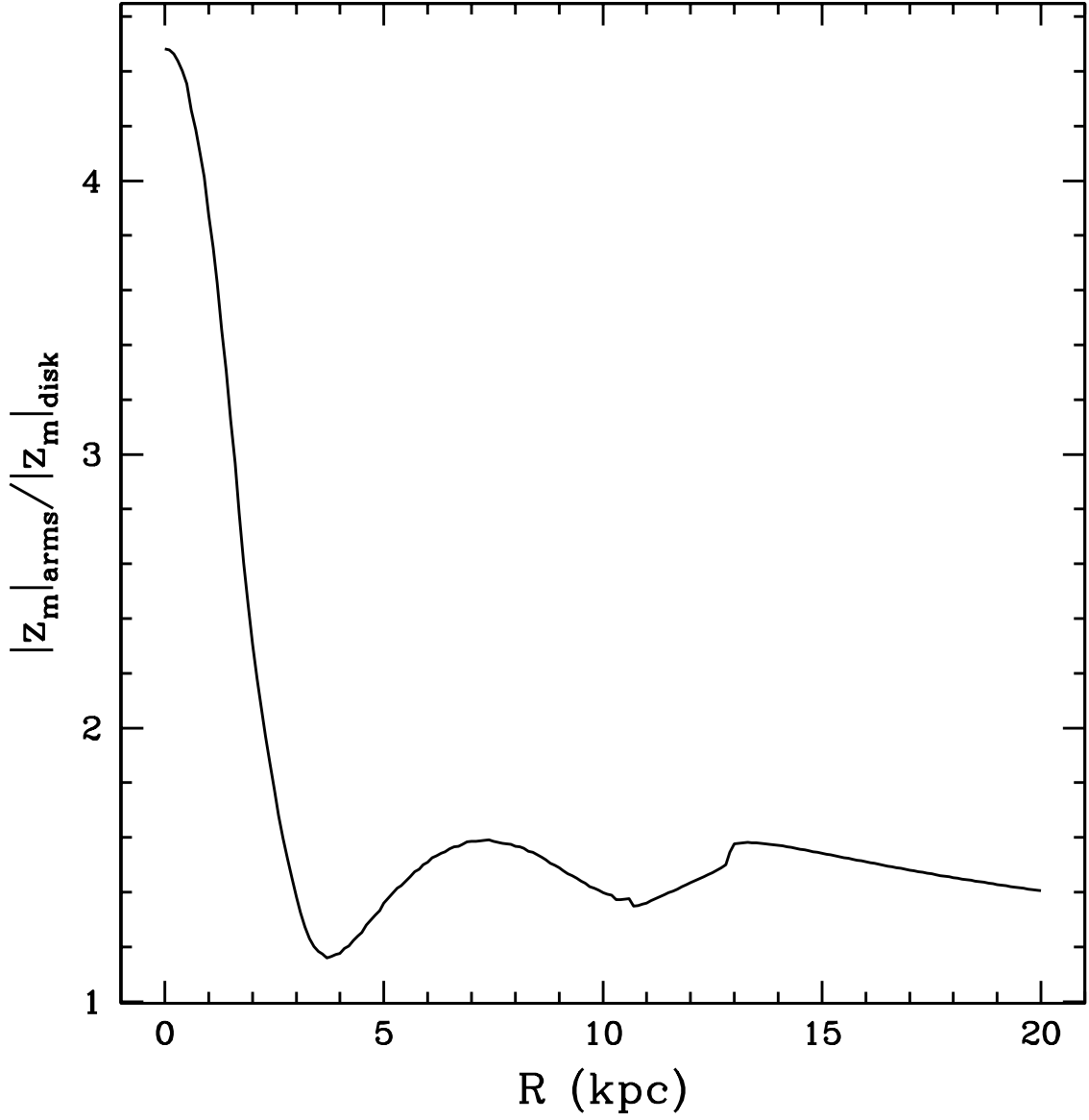}
\caption {Azimuth-averaged ratio of distances from the Galactic plane
where maximum z-accelerations of the spiral arms and axisymmetric
disk are reached, as a function of the distance $R$ to the Galactic
center of an orbital crossing point with the Galactic plane.}
\label{fig19}
\end{figure}

Figures \ref{fig18} and \ref{fig19} show the azimuth-averaged ratios
$(|g_m|_{arms}/|g_m|_{disk})^2$ and $|z_m|_{arms}/|z_m|_{disk}$ as
functions of the distance $R$ to the Galactic center of the point
where a cluster orbit crosses the Galactic plane. The squared ratio
$(|g_m|_{arms}/|g_m|_{disk})^2$ is important only in the region of the
spiral arms (2-12 kpc) and of order $10^{-2}$-$10^{-3}$. In this
region $|z_m|_{arms}/|z_m|_{disk}$ is close to unity. Thus, in our
analysis we ignore the spiral arms shocking, which compared with the
one of the disk will be two or three orders of magnitude lower. \rm

\section{Destruction rates. Results}\label{destr}

In this section we present bulge-shocking destruction rates obtained
with the formulism given in the last section, using the real
trajectories of globular clusters in the employed Galactic potentials,
and compare with corresponding values obtained with the usual linear
trajectory approximation used in Paper I. The disk-shocking destruction
rates are also computed and compared in the axisymmetric and
non-axisymmetric potentials. All the computations are done with Monte
Carlo simulations.

Table \ref{tbl-4} shows our results. Bulge-shocking destruction rates
averaged over the last $10^9$ yr in a cluster's orbit (this time is
increased in some clusters) and their lower, ${\sigma}_{-}$, and upper,
${\sigma}_{+}$, uncertainties, are listed according to the linear
(columns 2-4) or real (columns 5-7) employed cluster's trajectory.
Columns 8-10 give the disk-shocking destruction rates.
For each cluster, the first line gives values in the non-axisymmetric
potential, and the second line in the axisymmetric potential.

\begin{figure}
\plotone{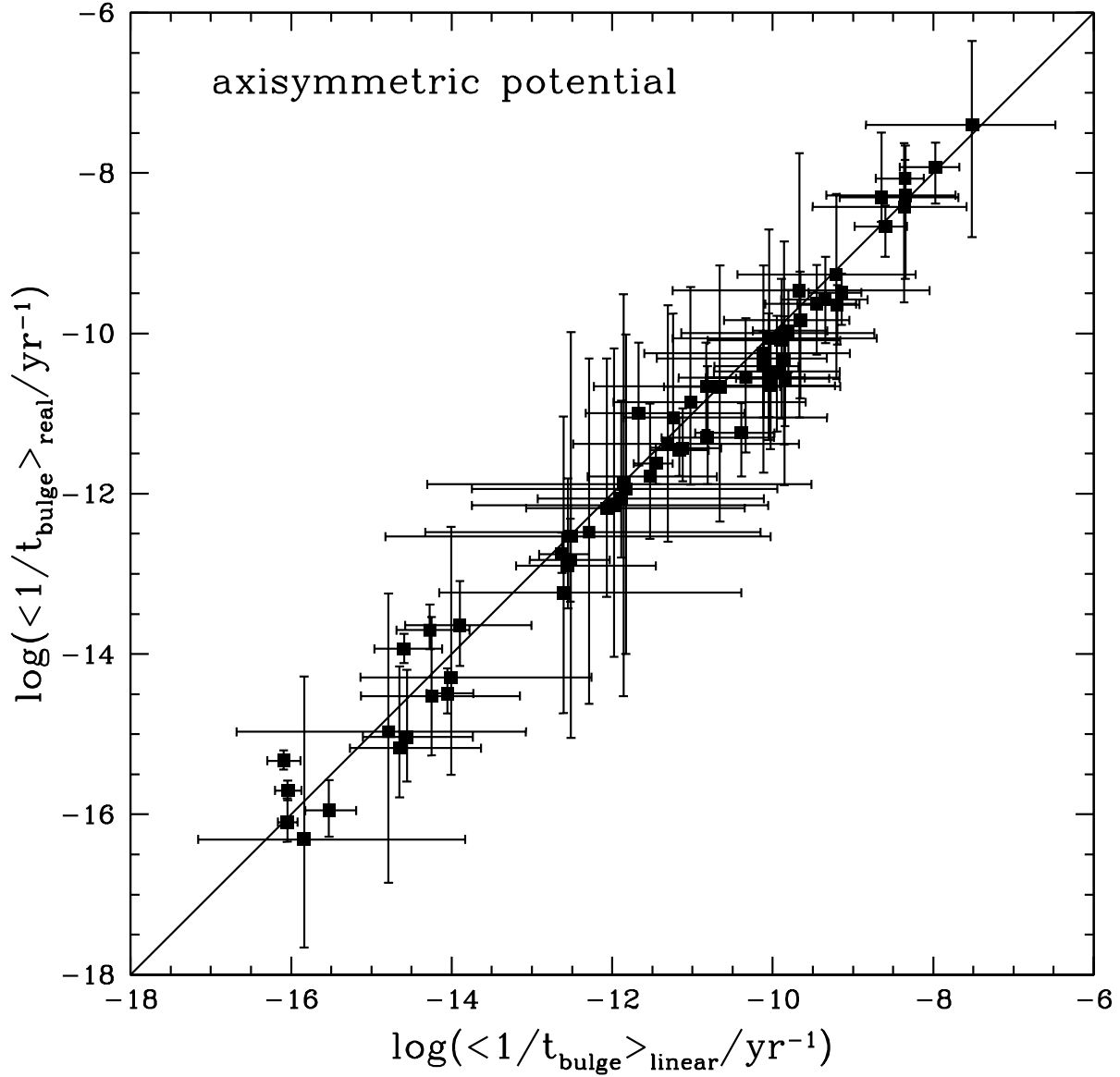}
\caption {Comparison of bulge-shocking total destruction rates in the
axisymmetric potential, employing the real and linear trajectories in
each cluster. Corresponding values are shown in the vertical and
horizontal axes. The plotted line is the line of coincidence.}
\label{fig20}
\end{figure}

\begin{figure}
\plotone{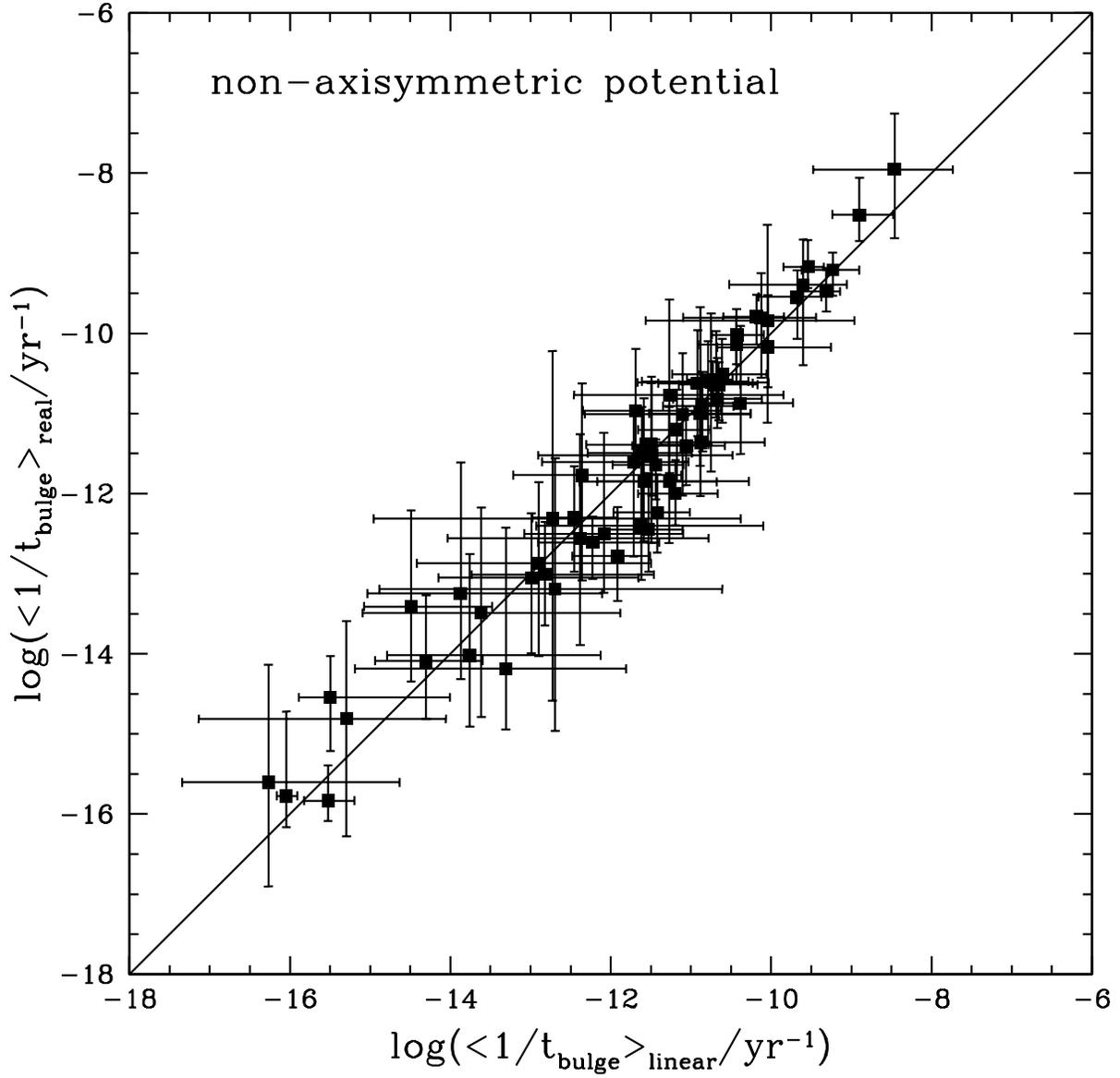}
\caption {As in Figure \ref{fig20}, here the comparison is made in
the non-axisymmetric potential.} 
\label{fig21}
\end{figure}

With the data given in Table \ref{tbl-4}, Figures \ref{fig20} and
\ref{fig21} show separately the comparison of bulge-shocking total
destruction rates in the axisymmetric and non-axisymmetric Galactic
potentials. Values obtained with the real trajectory of the cluster are
shown in the vertical axis, and in the horizontal axis those with the
linear trajectory. The error bars in these and following figures are
given by the ${\sigma}_{-}$ and ${\sigma}_{+}$ values in Table
\ref{tbl-4}. The conclusion from these two figures is that the use of
the linear trajectory, along with the associated effective interaction
time estimated as $\tau = |${\boldmath $r$}$_p|/|${\boldmath $v$}$_p|$,
gives a good approximation to compute destruction rates due to the
bulge, in the axisymmetric and non-axisymmetric Galactic potentials.

\begin{figure}
\plotone{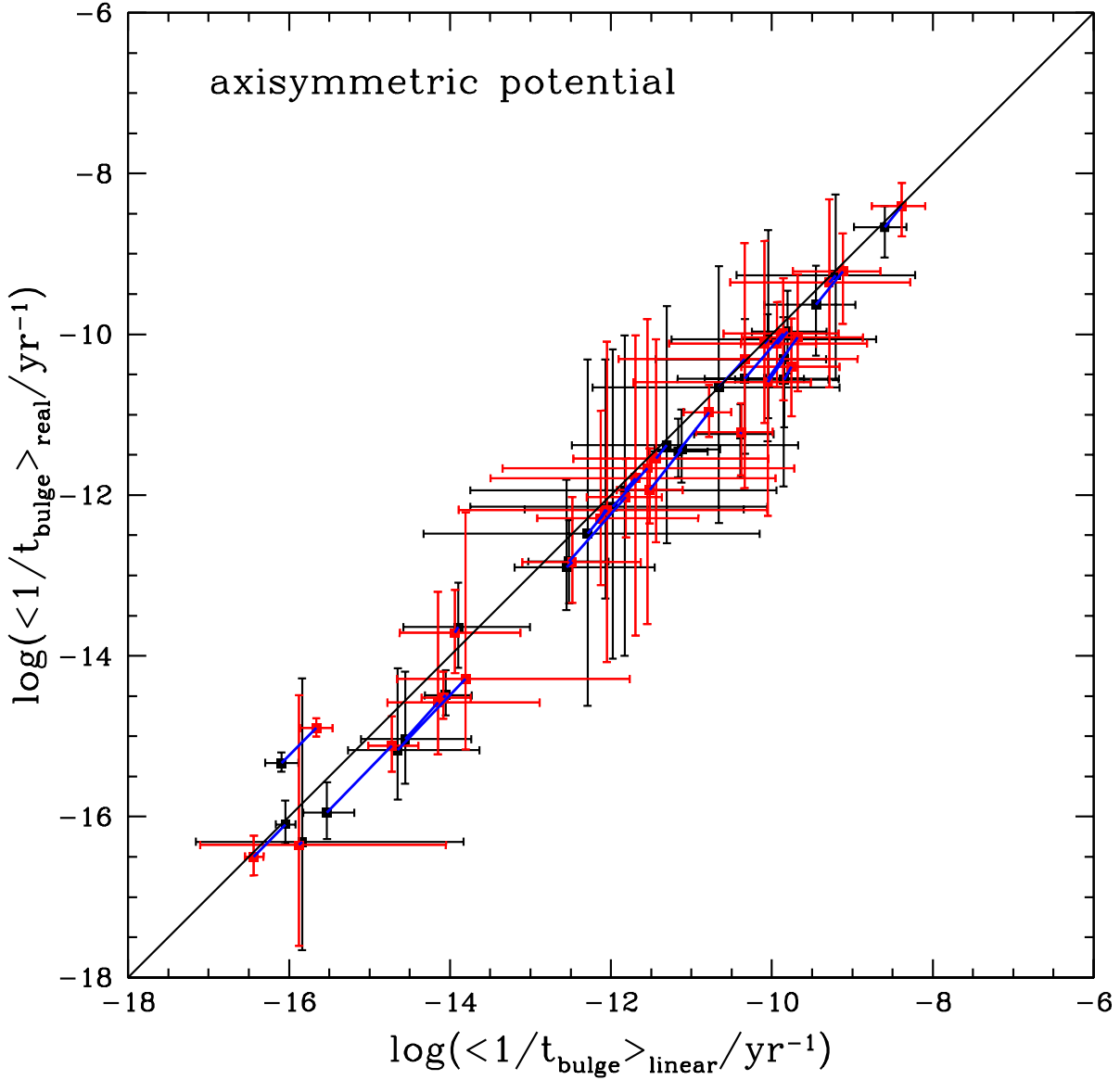}
\caption {Comparison of bulge-shocking total destruction rates in the
axisymmetric potential, employing the real and linear trajectories in
each cluster. Here the cluster mass $M_c$ is computed with dynamical
mass-to-light ratios $(M/L)_V$ given by \citet{MM05}. Black points are
points from Figure \ref{fig20}, and red points are the new points
obtained with the dynamical mass-to-light ratios. Corresponding shifts
between black and red points in a cluster are shown with blue lines.}
\label{fig22}
\end{figure}

\bf To see how much the computed destruction rates can change if we
take the cluster mass-to-light ratio $(M/L)_V$ different from the
assumed $(M/L)_V$ = 2 $M_{\odot}/L_{\odot}$, we consider, as in
Figure \ref{fig12}, dynamical mass-to-light ratios $(M/L)_V$ obtained
by \citet{MM05}. For clusters in common between our sample and in
table 13 of \citet{MM05}, the new cluster mass $M_c$ is computed, and
in Figure \ref{fig22} we show the comparison of bulge-shocking total
destruction rates in particular in the axisymmetric potential,
comparing values using the real (vertical axis) and linear (horizontal
axis) trajectory. The black points with their uncertainties are points
from Figure \ref{fig20}, obtained with the assumed
$(M/L)_V$ = 2 $M_{\odot}/L_{\odot}$. The red points are obtained with
the dynamical mass-to-light ratios given by \citet{MM05}.
The correspondig shifts between black and red points in a cluster are
shown with blue lines. Thus, the destruction rates do not change too
much, specially those with high values, and the standard
$(M/L)_V$ = 2 $M_{\odot}/L_{\odot}$ is a convenient test value. \rm 

\begin{figure}
\plotone{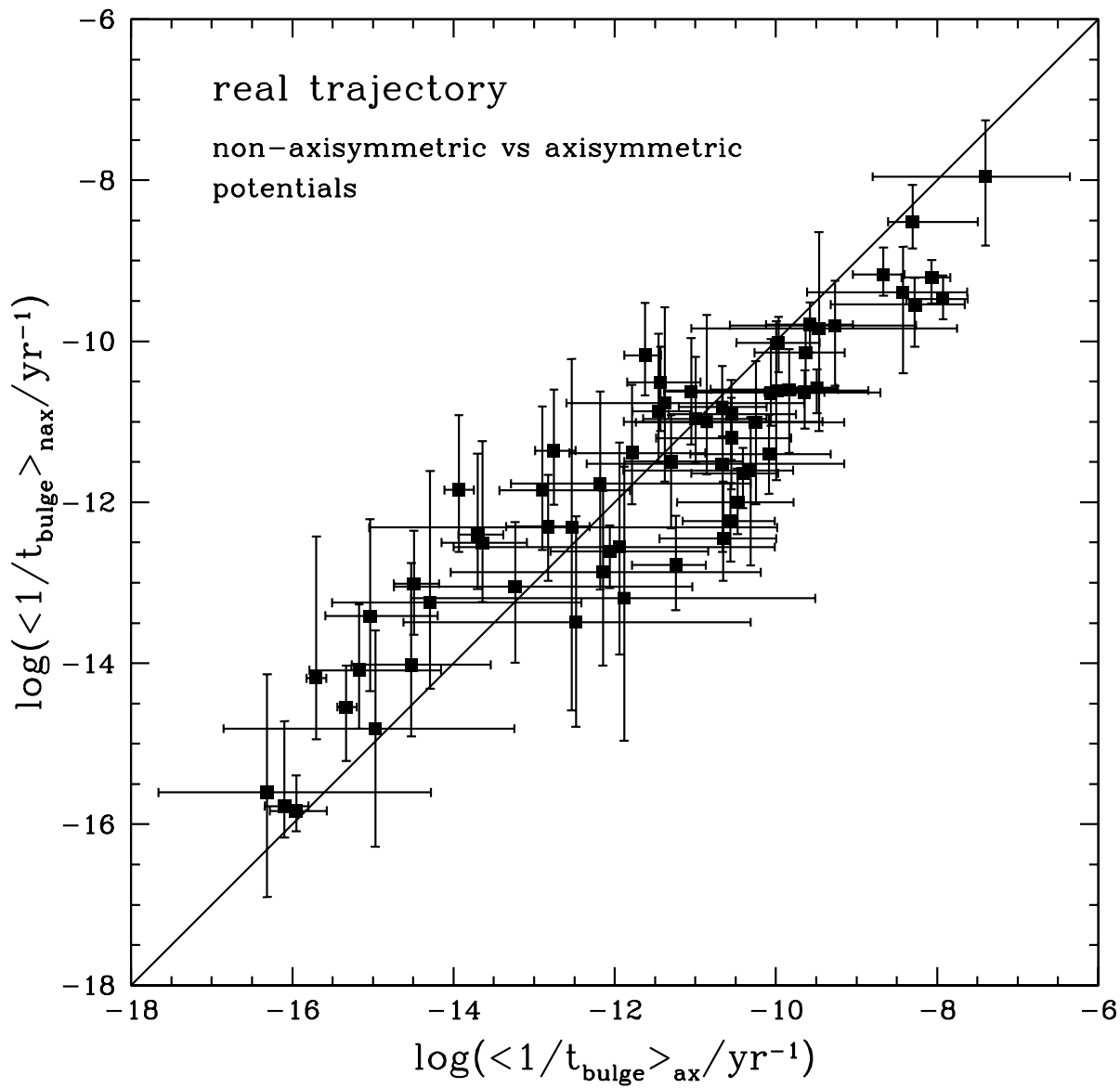}
\caption {Comparison of bulge-shocking total destruction rates
employing the real trajectory in each cluster. Values obtained in
the non-axisymmetric potential (denoted with a subindex 'nax') and
axisymmetric potential (with a subindex 'ax') are shown in the vertical
and horizontal axes, respectively. The plotted line is the line of
coincidence.}
\label{fig23}
\end{figure}

Taking the real trajectories, Figure \ref{fig23} shows the comparison
of bulge-shocking total destruction rates in the non-axisymmetric
(vertical axis) and axisymmetric (horizontal axis) Galactic potentials.
Here we note important differences between both potentials, specially
in the region of high destruction rates, where values obtained with
the non-axisymmetric potential are $smaller$ than those with the
axisymmetric potential. As noted in $\S$ \ref{gpot}, in the
non-axisymmetric potential all the bulge is represented by the Galactic
bar; thus, there is no remnant of the original concentrated spherical
bulge in the axisymmetric potential, whose mass is now distributed over
the bar, with less central concentration (see upper frame in figure 7
in Paper I) and thus less dangerous for clusters crossing its region.
This explains the behavior in the high destruction rate region in
Figure \ref{fig23}. \bf In $\S$ \ref{king} we saw a related $decrease$
of tidal radii in the inner Galactic region in the non-axisymmetric
potential; see discussion of Figure \ref{fig8} in that section. \rm 
Our conclusion is that the more appropriate non-axisymmetric Galactic
potential employed in our computations, reduces the destruction rates
due to the bulge (bar, in this case) for clusters with perigalacticons
in the inner Galactic region.

\begin{figure}
\plotone{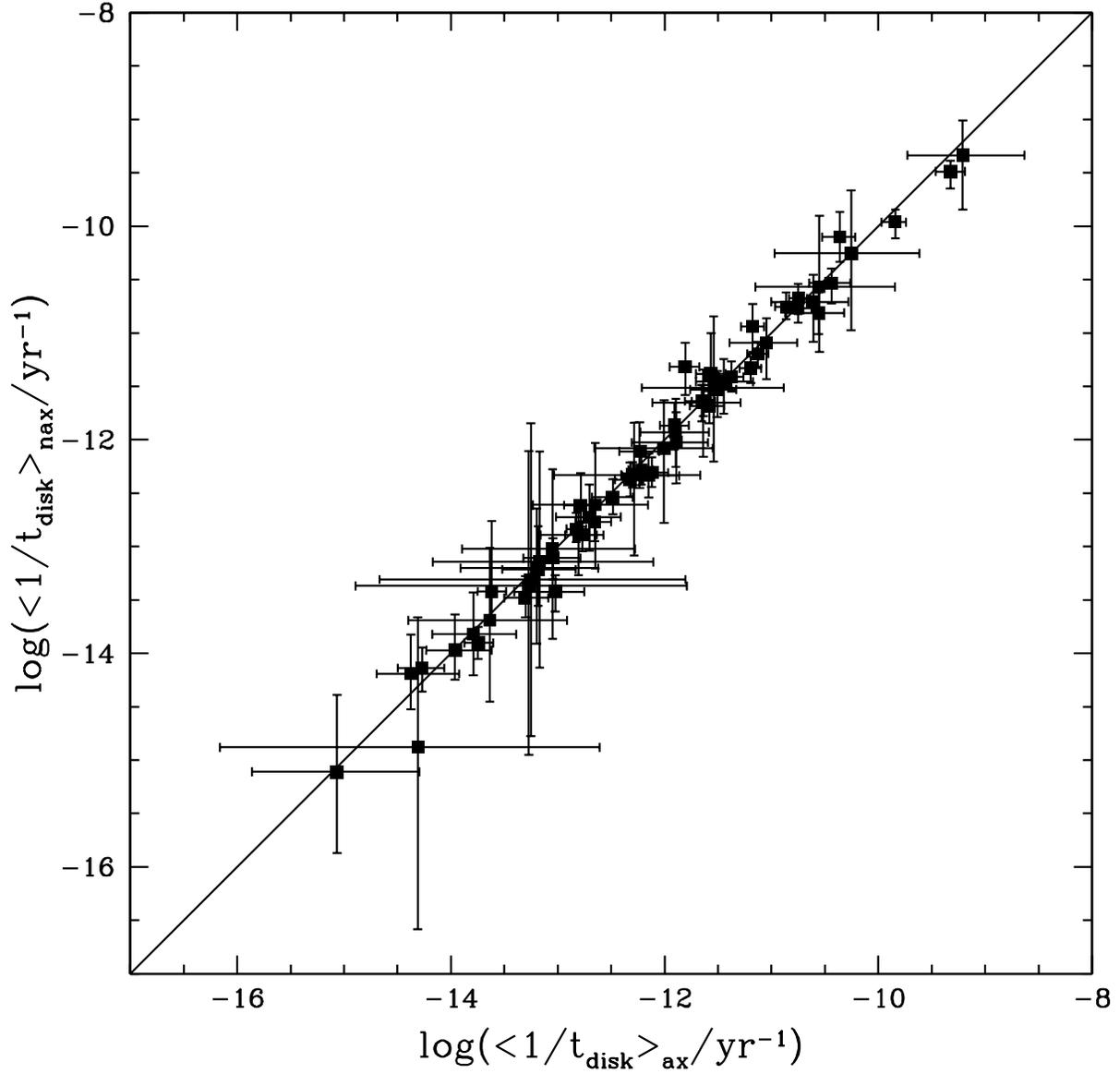}
\caption {Comparison of disk-shocking destruction rates obtained in
the non-axisymmetric potential (vertical axis) with those obtained in
the axisymmetric potential (horizontal axis). The plotted line is the
line of coincidence.}
\label{fig24}
\end{figure}

Disk-shocking destruction rates in the non-axisymmetric and
axisymmetric potentials are compared in Figure \ref{fig24}.
Practically these destruction rates are the same in both potentials.
Comparing this figure with Figure \ref{fig23}, we note that the disk
dominates the destruction rate for clusters in the low destruction
rate region, i.e. clusters with perigalacticons relatively distant
from the Galactic center.

\begin{figure}
\plotone{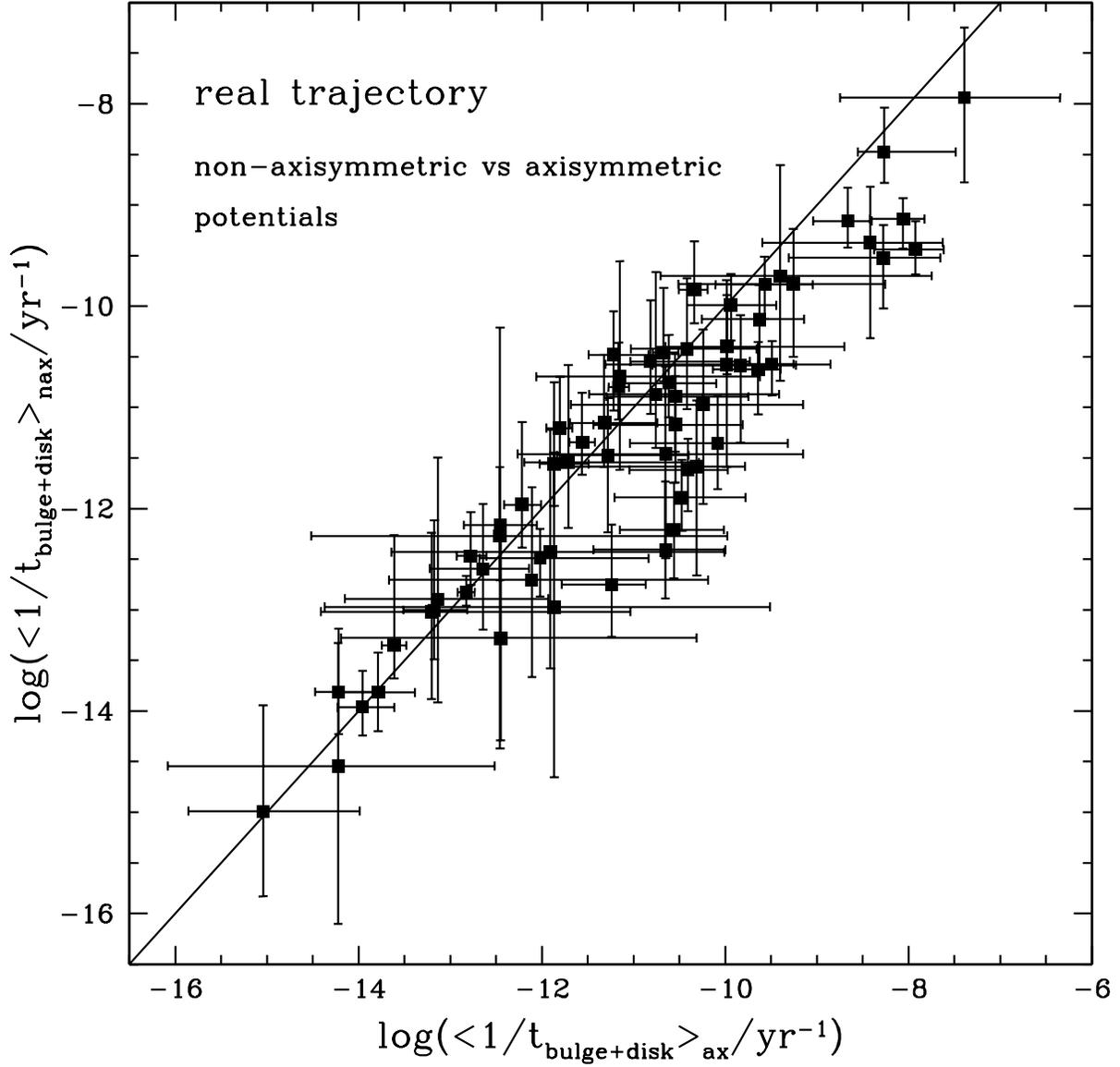}
\caption {Total bulge+disk destruction rates employing the real
trajectory of the cluster. Values obtained in the non-axisymmetric
potential and in the axisymmetric potential are shown in the vertical
and horizontal axes, respectively. The plotted line is the line of
coincidence.}
\label{fig25}
\end{figure}

\begin{figure}
\plotone{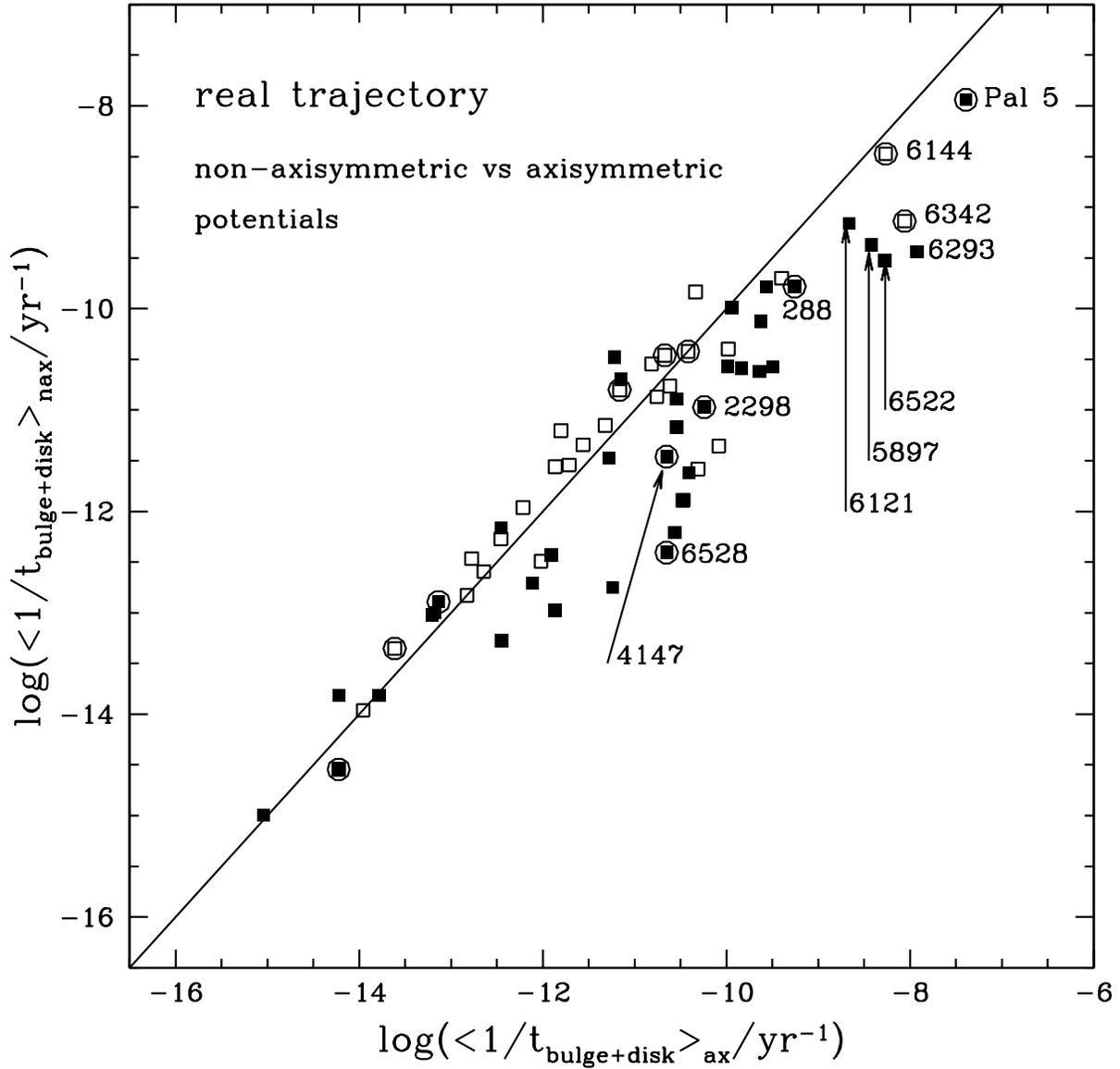}
\caption {This is Figure \ref{fig25} without the error bars. Points
shown with empty squares correspond to clusters with orbital
eccentricity $e \leq 0.5$, and those with black squares to clusters
with $e > 0.5$. The squares with a circle correspond to clusters with a
mass less than $10^5 M_{\odot}$. The position of the cluster Pal 5 is
shown, and also other clusters with their NGC numbers.}
\label{fig26}
\end{figure}

Adding the bulge-shocking total destruction rate obtained with the
real trajectory of the cluster, and the disk-shocking destruction rate,
results in the total bulge+disk destruction rate. Figure \ref{fig25}
shows the comparison of these total values in the non-axisymmetric
(vertical axis) and axisymmetric (horizontal axis) Galactic potentials.
In Figure \ref{fig26} we show the same Figure \ref{fig25} but now
without the error bars. Empty and black squares correspond to clusters
with orbital eccentricity $e \leq 0.5$ and $e > 0.5$, respectively.
The points marked with a circle show the clusters whose mass is less
than $10^5 M_{\odot}$. The position of some clusters are marked with
their NGC and Pal numbers.

As in Figure \ref{fig23}, in these last figures we note that in the
region of high destruction rates, dominated by the bulge, the total
destruction rates obtained with the non-axisymmetric potential are
$smaller$ than those resulting with the axisymmetric potential.
Figure \ref{fig26} shows that the majority of clusters with high
eccentricities ($e > 0.5$), have smaller destruction rates in the
non-axisymmetric potential.

With the non-axisymmetric Galactic potential employed in our analysis,
along with the more appropriate Monte Carlo simulations, we see from
Figure \ref{fig26} that seven clusters have particularly high
destruction rates at the present time, due to bulge and disk shocking:
Pal 5, NGC 6144, NGC 6121, NGC 6342, NGC 5897, NGC 6293, and NGC 6522. 
In Paper I, using only the Galactic bar in the non-axisymmetric
potential, we found that NGC 6528 had the greatest destruction rate;
now with the present non-axisymmetric Galactic model and the Monte
Carlo simulations, this cluster has a low destruction rate, as shown in
Figure \ref{fig26}.

\section{Conclusions}\label{concl}

We have employed the available 6-D data (positions and velocities) of
63 globular clusters in our Galaxy to analyze their Galactic orbits
and compute their tidal radii, as well as their bulge and disk
shocking destruction rates. This analysis has been made in
axisymmetric and non-axisymmetric Galactic potentials; in particular,
the used non-axisymmetric potential is a very detailed model which
includes both the Galactic bar and a 3D model for the spiral arms. Our
analysis is made using Monte Carlo simulations, to take into account
the several uncertainties in the kinematical data of the clusters. For
the computation of destruction rates due to the bulge in both Galactic
potentials, we have employed the rigorous treatment of considering the
real Galactic cluster orbit, instead of the usual linear trajectory
employed in previous studies.

Our first result is that the theoretical tidal radius computed in the
non-axisymmetric Galactic potential compares better with the observed
tidal radius than that computed in the axisymmetric potential. 
This result leaves an open question with a recent study made by
\citet{WHS13}, who propose a correction to be applied to the
theoretical tidal radius computed at perigalacticon, to have a better
comparison with the observed tidal radius. In our computations we do
not need to introduce this correction.

The first conclusion from our results of bulge-shocking destruction
rates is that the usual linear trajectory of the cluster considered
at perigalacticon, gives a good approximation to the result obtained
taking the real trajectory of the cluster. This conclusion holds in 
both the axisymmetric and non-axisymmetric potentials.

Our second conclusion is that the bulge-shocking destruction rates
for clusters with perigalacticons in the inner Galactic region,
turn out to be $smaller$ in the non-axisymmetric potential, as
compared with those in the axisymmetric one. The majority of clusters
with high orbital eccentricities ($e > 0.5$) have $smaller$ total
bulge+disk destruction rates in the non-axisymmetric potential.

We acknowledge financial support from UNAM DGAPA-PAPIIT through grant
IN114114.

%\clearpage

\clearpage
\begin{deluxetable}{crrrrrrcrr}
\tabletypesize{\scriptsize}
%\rotate
\tablecaption{Globular Clusters Data \label{tbl-1}}
\tablewidth{0pt}
\tablehead{
\colhead{Cluster} & \colhead{${\alpha}_{2000}$} & \colhead{${\delta}_{2000}$} &
\colhead{$r$} & \colhead{$v_r$} & \colhead{${\mu}_x$} &
\colhead{${\mu}_y$} & \colhead{$M_c$} & \colhead{$r_K$} &
\colhead{$r_h$} \\ 
\colhead{} & \colhead{($deg$)} & \colhead{($deg$)} &
\colhead{($kpc$)} & \colhead{($km/s$)} & \colhead{($mas$)} &
\colhead{($mas$)} & \colhead{($M_{\odot}$)} & \colhead{($pc$)} &
\colhead{($pc$)} 
}
\startdata
NGC 104 &  6.02363 & $-$72.08128 & 4.5$\pm$0.45 & $-$18.0$\pm$0.1 &
5.64$\pm$0.20 & $-$2.02$\pm$0.20 & 0.10E+07 & 55.37 & 4.15 \\   
NGC 288 & 13.18850 & $-$26.58261 & 8.9$\pm$0.89 & $-$45.4$\pm$0.2 &
4.67$\pm$0.42 & $-$5.62$\pm$0.23 & 0.86E+05 & 34.15 & 5.77 \\
NGC 362 & 15.80942 & $-$70.84878 & 8.6$\pm$0.86 & 223.5$\pm$0.5 &
5.07$\pm$0.71 & $-$2.55$\pm$0.72 & 0.40E+06 & 26.61 & 2.05 \\
NGC 1851 & 78.52817 & $-$40.04655 & 12.1$\pm$1.21 & 320.5$\pm$0.6 &
1.28$\pm$0.68 & 2.39$\pm$0.65 & 0.37E+06 & 22.95 & 1.80 \\
NGC 1904 & 81.04621 & $-$24.52472 & 12.9$\pm$1.29 & 205.8$\pm$0.4 &
2.12$\pm$0.64 & $-$0.02$\pm$0.64 & 0.24E+06 & 30.90 & 2.44 \\
NGC 2298 & 102.24754 & $-$36.00531 & 10.8$\pm$1.08 & 148.9$\pm$1.2 &
4.05$\pm$1.00 & $-$1.72$\pm$0.98 & 0.57E+05 & 23.36 & 3.08 \\
NGC 2808 & 138.01292 & $-$64.86350 & 9.6$\pm$0.96 & 101.6$\pm$0.7 &
0.58$\pm$0.45 & 2.06$\pm$0.46 & 0.97E+06 & 25.35 & 2.23 \\
Pal 3 & 151.38292 & 0.07167 & 92.5$\pm$9.25 & 83.4$\pm$8.4 &
0.33$\pm$0.23 & 0.30$\pm$0.31 & 0.32E+05 & 107.81 & 17.49 \\
NGC 3201 & 154.40342 & $-$46.41247 & 4.9$\pm$0.49 & 494.0$\pm$0.2 &
5.28$\pm$0.32 & $-$0.98$\pm$0.33 & 0.16E+06 & 36.13 & 4.42 \\
NGC 4147 & 182.52625 & 18.54264 & 19.3$\pm$1.93 & 183.2$\pm$0.7 &
$-$1.85$\pm$0.82 & $-$1.30$\pm$0.82 & 0.50E+05 & 34.16 & 2.69 \\
NGC 4372 & 186.43917 & $-$72.65900 & 5.8$\pm$0.58 & 72.3$\pm$1.2 &
$-$6.49$\pm$0.33 & 3.71$\pm$0.32 & 0.22E+06 & 58.91 & 6.60 \\
NGC 4590 & 189.86658 & $-$26.74406 & 10.3$\pm$1.03 & $-$94.7$\pm$0.2 &
$-$3.76$\pm$0.66 & 1.79$\pm$0.62 & 0.15E+06 & 44.67 & 4.52 \\
NGC 4833 & 194.89133 & $-$70.87650 & 6.6$\pm$0.66 & 200.2$\pm$1.2 &
$-$8.11$\pm$0.35 & $-$0.96$\pm$0.34 & 0.32E+06 & 34.14 & 4.63 \\
NGC 5024 & 198.23021 & 18.16817 & 17.9$\pm$1.79 & $-$62.9$\pm$0.3 &
0.50$\pm$1.00 & $-$0.10$\pm$1.00 & 0.52E+06 & 95.64 & 6.82 \\
NGC 5139 & 201.69683 & $-$47.47958 & 5.2$\pm$0.52 & 232.1$\pm$0.1 &
$-$5.08$\pm$0.35 & $-$3.57$\pm$0.34 & 0.22E+07 & 73.19 & 7.56 \\
NGC 5272 & 205.54842 & 28.37728 & 10.2$\pm$1.02 & $-$147.6$\pm$0.2 &
$-$1.10$\pm$0.51 & $-$2.30$\pm$0.54 & 0.61E+06 & 85.22 & 6.85 \\
NGC 5466 & 211.36371 & 28.53444 & 16.0$\pm$1.60 & 110.7$\pm$0.2 &
$-$4.65$\pm$0.82 & 0.80$\pm$0.82 & 0.11E+06 & 72.98 & 10.70 \\
Pal 5 & 229.02188 & $-$0.11161 & 23.2$\pm$2.32 & $-$58.7$\pm$0.2 &
$-$1.78$\pm$0.17 & $-$2.32$\pm$0.23 & 0.20E+05 & 51.17 & 18.42 \\
NGC 5897 & 229.35208 & $-$21.01028 & 12.5$\pm$1.25 & 101.5$\pm$1.0 &
$-$4.93$\pm$0.86 & $-$2.33$\pm$0.84 & 0.13E+06 & 36.88 & 7.49 \\
NGC 5904 & 229.63842 & 2.08103 & 7.5$\pm$0.75 & 53.2$\pm$0.4  &
5.07$\pm$0.68 & $-$10.70$\pm$0.56 & 0.57E+06 & 51.55 & 3.86 \\
NGC 5927 & 232.00288 & $-$50.67303 & 7.7$\pm$0.77 & $-$107.5$\pm$0.9 &
$-$5.72$\pm$0.39 & $-$2.61$\pm$0.40 & 0.23E+06 & 37.45 & 2.46 \\
NGC 5986 & 236.51250 & $-$37.78642 & 10.4$\pm$1.04 & 88.9$\pm$3.7 &
$-$3.81$\pm$0.45 & $-$2.99$\pm$0.37 & 0.41E+06 & 24.15 & 2.96 \\
NGC 6093 & 244.26004 & $-$22.97608 & 10.0$\pm$1.00 & 8.1$\pm$1.5 &
$-$3.31$\pm$0.58 & $-$7.20$\pm$0.67 & 0.33E+06 & 20.88 & 1.77 \\
NGC 6121 & 245.89675 & $-$26.52575 & 2.2$\pm$0.22 & 70.7$\pm$0.2 &
$-$12.50$\pm$0.36 & $-$19.93$\pm$0.49 & 0.13E+06 & 33.16 & 2.77 \\
NGC 6144 & 246.80775 & $-$26.02350 & 8.9$\pm$0.89 & 193.8$\pm$0.6 &
$-$3.06$\pm$0.64 & $-$5.11$\pm$0.72 & 0.94E+05 & 86.35 & 4.22 \\
NGC 6171 & 248.13275 & $-$13.05378 & 6.4$\pm$0.64 & $-$34.1$\pm$0.3 &
$-$0.70$\pm$0.90 & $-$3.10$\pm$1.00 & 0.12E+06 & 35.33 & 3.22 \\
NGC 6205 & 250.42183 & 36.45986 & 7.1$\pm$0.71 & $-$244.2$\pm$0.2 &
$-$0.90$\pm$0.71 & 5.50$\pm$1.12 & 0.45E+06 & 43.39 & 3.49 \\
NGC 6218 & 251.80908 & $-$1.94853 & 4.8$\pm$0.48 & $-$41.4$\pm$0.2 &
1.30$\pm$0.58 & $-$7.83$\pm$0.62 & 0.14E+06 & 24.13 & 2.47 \\
NGC 6254 & 254.28771 & $-$4.10031 & 4.4$\pm$0.44 & 75.2$\pm$0.7 &
$-$6.00$\pm$1.00 & $-$3.30$\pm$1.00 & 0.17E+06 & 23.64 & 2.50 \\
NGC 6266 & 255.30333 & $-$30.11372 & 6.8$\pm$0.68 & $-$70.1$\pm$1.4 &
$-$3.50$\pm$0.37 & $-$0.82$\pm$0.37 & 0.80E+06 & 23.10 & 1.82 \\
NGC 6273 & 255.65750 & $-$26.26797 & 8.8$\pm$0.88 & 135.0$\pm$4.1 &
$-$2.86$\pm$0.49 & $-$0.45$\pm$0.51 & 0.77E+06 & 37.30 & 3.38 \\
NGC 6284 & 256.11879 & $-$24.76486 & 15.3$\pm$1.53 & 27.5$\pm$1.7 &
$-$3.66$\pm$0.64 & $-$5.39$\pm$0.83 & 0.26E+06 & 102.72 & 2.94 \\
NGC 6287 & 256.28804 & $-$22.70836 & 9.4$\pm$0.94 & $-$288.7$\pm$3.5 &
$-$3.68$\pm$0.88 & $-$3.54$\pm$0.69 & 0.15E+06 & 19.02 & 2.02 \\
NGC 6293 & 257.54250 & $-$26.58208 & 9.5$\pm$0.95 & $-$146.2$\pm$1.7 &
0.26$\pm$0.85 & $-$5.14$\pm$0.71 & 0.22E+06 & 39.32 & 2.46 \\
NGC 6304 & 258.63438 & $-$29.46203 & 5.9$\pm$0.59 & $-$107.3$\pm$3.6 &
$-$2.59$\pm$0.29 & $-$1.56$\pm$0.29 & 0.14E+06 & 22.74 & 2.44 \\
NGC 6316 & 259.15542 & $-$28.14011 & 10.4$\pm$1.04 & 71.4$\pm$8.9 &
$-$2.42$\pm$0.63 & $-$1.71$\pm$0.56 & 0.37E+06 & 22.97 & 1.97 \\
NGC 6333 & 259.79692 & $-$18.51594 & 7.9$\pm$0.79 & 229.1$\pm$7.0 &
$-$0.57$\pm$0.57 & $-$3.70$\pm$0.50 & 0.26E+06 & 18.39 & 2.21 \\
NGC 6341 & 259.28079 & 43.13594 & 8.3$\pm$0.83 & $-$120.0$\pm$0.1 &
$-$3.30$\pm$0.55 & $-$0.33$\pm$0.70 & 0.33E+06 & 30.05 & 2.46 \\
NGC 6342 & 260.29200 & $-$19.58742 & 8.5$\pm$0.85 & 115.7$\pm$1.4 &
$-$2.77$\pm$0.71 & $-$5.84$\pm$0.65 & 0.63E+05 & 36.74 & 1.80 \\
NGC 6356 & 260.89554 & $-$17.81303 & 15.1$\pm$1.51 & 27.0$\pm$4.3 &
$-$3.14$\pm$0.68 & $-$3.65$\pm$0.53 & 0.43E+06 & 41.01 & 3.56 \\
NGC 6362 & 262.97913 & $-$67.04833 & 7.6$\pm$0.76 & $-$13.1$\pm$0.6 &
$-$3.09$\pm$0.46 & $-$3.83$\pm$0.46 & 0.10E+06 & 30.73 & 4.53 \\
NGC 6388 & 264.07179 & $-$44.73550 & 9.9$\pm$0.99 & 80.1$\pm$0.8 &
$-$1.90$\pm$0.45 & $-$3.83$\pm$0.51 & 0.99E+06 & 19.43 & 1.50 \\
NGC 6397 & 265.17538 & $-$53.67433 & 2.3$\pm$0.23 & 18.8$\pm$0.1 &
3.69$\pm$0.29 & $-$14.88$\pm$0.26 & 0.77E+05 & 29.79 & 1.94 \\
NGC 6441 & 267.55442 & $-$37.05144 & 11.6$\pm$1.16 & 16.5$\pm$1.0 &
$-$2.86$\pm$0.45 & $-$3.45$\pm$0.76 & 0.12E+07 & 24.11 & 1.92 \\
NGC 6522 & 270.89175 & $-$30.03397 & 7.7$\pm$0.77 & $-$21.1$\pm$3.4 &
6.08$\pm$0.20 & $-$1.83$\pm$0.20 & 0.20E+06 & 36.82 & 2.24 \\
NGC 6528 & 271.20683 & $-$30.05628 & 7.9$\pm$0.79 & 206.6$\pm$1.4 &
$-$0.35$\pm$0.23 & 0.27$\pm$0.26 & 0.73E+05 & 9.45 & 0.87 \\
NGC 6553 & 272.32333 & $-$25.90869 & 6.0$\pm$0.60 & $-$3.2$\pm$1.5 &
2.50$\pm$0.07 & 5.35$\pm$0.08 & 0.22E+06 & 13.37 & 1.80 \\
NGC 6584 & 274.65667 & $-$52.21578 & 13.5$\pm$1.35 & 222.9$\pm$15.0 &
$-$0.22$\pm$0.62 & $-$5.79$\pm$0.67 & 0.20E+06 & 30.13 & 2.87 \\
NGC 6626 & 276.13671 & $-$24.86978 & 5.5$\pm$0.55 & 17.0$\pm$1.0 &
0.63$\pm$0.67 & $-$8.46$\pm$0.67 & 0.31E+06 & 17.96 & 3.15 \\
NGC 6656 & 279.09975 & $-$23.90475 & 3.2$\pm$0.32 & $-$146.3$\pm$0.2 &
7.37$\pm$0.50 & $-$3.95$\pm$0.42 & 0.43E+06 & 29.70 & 3.13 \\
NGC 6712 & 283.26792 & $-$8.70611 & 6.9$\pm$0.69 & $-$107.6$\pm$0.5 &
4.20$\pm$0.40 & $-$2.00$\pm$0.40 & 0.17E+06 & 17.12 & 2.67 \\
NGC 6723 & 284.88813 & $-$36.63225 & 8.7$\pm$0.87 & $-$94.5$\pm$3.6 &
$-$0.17$\pm$0.45 & $-$2.16$\pm$0.50 & 0.23E+06 & 30.20 & 3.87 \\
NGC 6752 & 287.71712 & $-$59.98456 & 4.0$\pm$0.40 & $-$26.7$\pm$0.2 &
$-$0.69$\pm$0.42 & $-$2.85$\pm$0.45 & 0.21E+06 & 40.48 & 2.22 \\
NGC 6779 & 289.14821 & 30.18347 & 9.4$\pm$0.94 & $-$135.6$\pm$0.9 &
0.30$\pm$1.00 & 1.40$\pm$0.10 & 0.16E+06 & 28.86 & 3.01 \\
NGC 6809 & 294.99879 & $-$30.96475 & 5.4$\pm$0.54 & 174.7$\pm$0.3 &
$-$1.42$\pm$0.62 & $-$10.25$\pm$0.64 & 0.18E+06 & 24.07 & 4.45 \\
NGC 6838 & 298.44371 & 18.77919 & 4.0$\pm$0.40 & $-$22.8$\pm$0.2 &
$-$2.30$\pm$0.80 & $-$5.10$\pm$0.80 & 0.30E+05 & 10.35 & 1.94 \\
NGC 6934 & 308.54738 & 7.40447 & 15.6$\pm$1.56 & $-$411.4$\pm$1.6 &
1.20$\pm$1.00 & $-$5.10$\pm$1.00 & 0.16E+06 & 33.83 & 3.13 \\
NGC 7006 & 315.37242 & 16.18733 & 41.2$\pm$4.12 & $-$384.1$\pm$0.4 &
$-$0.96$\pm$0.35 & $-$1.14$\pm$0.40 & 0.20E+06 & 52.37 & 5.27 \\
NGC 7078 & 322.49304 & 12.16700 & 10.4$\pm$1.04 & $-$107.0$\pm$0.2 &
$-$0.95$\pm$0.51 & $-$5.63$\pm$0.50 & 0.81E+06 & 63.23 & 3.03 \\
NGC 7089 & 323.36258 & $-$0.82325 & 11.5$\pm$1.15 & $-$5.3$\pm$2.0 &
5.90$\pm$0.86 & $-$4.95$\pm$0.86 & 0.70E+06 & 41.65 & 3.55 \\
NGC 7099 & 325.09217 & $-$23.17986 & 8.1$\pm$0.81 & $-$184.2$\pm$0.2 &
1.42$\pm$0.69 & $-$7.71$\pm$0.65 & 0.16E+06 & 37.43 & 2.43 \\
Pal 12 & 326.66183 & $-$21.25261 & 19.0$\pm$1.90 & 27.8$\pm$1.5 &
$-$1.20$\pm$0.30 & $-$4.21$\pm$0.29 & 0.10E+05 & 105.56 & 9.51 \\
Pal 13 & 346.68517 & 12.77200 & 26.0$\pm$2.60 & 25.2$\pm$0.3 &
2.30$\pm$0.26 & 0.27$\pm$0.25 & 0.54E+04 & 16.59 & 2.72 \\
\enddata
\end{deluxetable}

\clearpage
\begin{deluxetable}{lcr}
\tablecolumns{3}
\tablewidth{0pt}
\tablecaption{Galactic parameters \label{tbl-2}} 
\tablehead{\colhead{Parameter} &\colhead{Value} & \colhead{References}}
\startdata
$R_0$               & 8.3$\pm$0.23 kpc     & 1 \\
${\Theta}_0$        & 239$\pm$7 km/s       & 1 \\
$(U,V,W)_{\odot}$   & ($-11.1\pm$1.2,12.24$\pm$2.1,7.25$\pm$0.6)
 km s$^{-1}$  &  2,1 \\
\cutinhead{Galactic Bar}
position of major axis              & 20$^{\circ}$        & 3 \\
angular velocity                    & 55$\pm$5 $\kmskpc$  & 4 \\
\cutinhead{Spiral Arms}
$M_{\rm arms}/M_{\rm disk}$         &  0.04$\pm$0.01      & 5 \\
scale length ($H$)                  & 3.9$\pm$0.6 kpc  ($R_0$ = 8.5 kpc)        & 6 \\
pitch angle                         & 15.5$\pm$3.5$^{\circ}$ & 7 \\
angular velocity                    & 24$\pm$6 $\kmskpc$  & 4 
\enddata
\tablerefs{
            1)~\citet{BRet11}.
            2)~\citet{SBD10}.
            3)~\citet{G02}.
            4)~\citet{G11}.
            5)~\citet{PMA12}.
            6)~\citet{BCHet05}.
            7)~\citet{D00}.}
\end{deluxetable}

\clearpage
\begin{deluxetable}{cccccrccc}
\tabletypesize{\scriptsize}
%\rotate
\tablecaption{Orbital parameters with the non-axisymmetric and
axisymmetric potentials \label{tbl-3}}
\tablewidth{0pt}
\tablehead{
\colhead{Cluster} &  \colhead{$<$$r_{min}$$>$} &
\colhead{$<$$r_{max}$$>$} & \colhead{$<$$|z|_{max}$$>$} & 
\colhead{$<$$e$$>$} & \colhead{$E$} & \colhead{$h$} &
\colhead{$<$$r_{K_t}$$>$} & \colhead{$<$$r_{\ast}$$>$} \\ 
\colhead{} & \colhead{($kpc$)} & \colhead{($kpc$)} & \colhead{($kpc$)} &
\colhead{} & \colhead{(10$kms^{-1})^2$} &
\colhead{(10$kms^{-1}kpc$)} & \colhead{($pc$)} & \colhead{($pc$)}  
}
\startdata
NGC 104 & 5.78 & 8.30 & 3.13 & 0.177 &     &    & 91.6 & 102.9 \\ 
        & 6.25 & 7.57 & 3.16 & 0.095 & $-1482.12$ & 134.55 & 98.3 &
112.4  \\
NGC 288 & 2.60 & 12.38 & 6.40 & 0.654 &    &    & 23.6 & 22.8 \\
        & 2.78 & 12.25 & 6.70 & 0.632 & $-1384.68$ & $-46.38$ & 24.5 &
23.5   \\
NGC 362 & 1.39 &  9.28 & 3.85 & 0.736 &    &    & 24.3 & 20.7 \\
        & 0.74 & 11.09 & 2.16 & 0.877 & $-1492.42$ & $-12.56$ & 17.1 &
15.9   \\
NGC 1851 & 6.66 & 33.28 & 7.61 & 0.667 &    &    & 70.1 & 68.2 \\ 
         & 6.73 & 31.75 & 7.59 & 0.650 & $-939.52$ & 238.99 & 71.5 &
69.5   \\
NGC 1904 & 5.25 & 18.85 & 5.54 & 0.563 &    &    & 51.4 & 51.4 \\ 
         & 5.19 & 20.52 & 5.37 & 0.596 & $-1137.89$ & 173.68 & 52.0 &
51.6   \\
NGC 2298 & 3.18 & 20.38 & 11.49 & 0.731 &    &    & 22.0 & 20.8 \\ 
         & 3.20 & 17.90 &  9.52 & 0.698 & $-1212.81$ & $-56.36$ & 22.7 &
21.3   \\
NGC 2808 & 2.27 & 10.74 &  2.39 & 0.649 &    &    & 46.0 & 43.4 \\ 
         & 2.73 & 12.74 &  2.59 & 0.647 & $-1395.39$ & 94.19 & 53.3 &
51.4   \\
NGC 3201 & 9.00 & 16.86 &  4.54 & 0.304 &    &    & 67.5 & 72.3 \\ 
         & 8.99 & 17.12 &  4.52 & 0.311 & $-1151.30$ & $-251.56$ & 67.8
 & 72.5  \\
NGC 4147 & 3.89 & 27.33 & 13.97 & 0.750 &    &    & 25.0 & 22.8 \\ 
         & 3.78 & 28.64 & 14.73 & 0.766 & $-1001.35$ & 66.83 & 25.4 &
22.8  \\
NGC 4372 & 2.39 &  5.30 &  1.57 & 0.386 &    &    & 30.4 & 33.2 \\ 
         & 3.19 &  7.41 &  1.60 & 0.397 & $-1624.75$ & 95.34 & 37.7 &
39.9  \\
NGC 4590 & 9.60 & 30.81 & 11.86 & 0.525 &    &    & 67.5 & 68.6 \\ 
         & 9.57 & 30.40 & 11.82 & 0.521 & $-932.42$ & 264.42 & 68.0 &
68.6  \\
NGC 4833 & 1.04 &  8.41 &  1.54 & 0.778 &    &    & 22.1 & 18.3 \\ 
         & 0.98 &  7.43 &  1.93 & 0.767 & $-1685.97$ & 26.11 & 18.3 &
17.3  \\
NGC 5024 & 16.43 & 36.30 & 24.34 & 0.377 &    &    & 155.4 & 161.6 \\   
         & 16.44 & 36.46 & 24.44 & 0.379 & $-811.52$ & 143.90 & 155.7 &
161.9  \\
NGC 5139 &  1.49 &  5.81 &  1.69 & 0.592 &    &    & 47.5 & 47.4 \\ 
         &  0.98 &  6.45 &  1.16 & 0.737 & $-1770.34$ & $-34.25$ & 36.7
 & 35.5  \\
NGC 5272 &  5.61 & 13.28 &  8.77 & 0.404 &    &    & 76.3 & 80.2 \\ 
         &  5.60 & 14.22 &  8.99 & 0.435 & $-1267.90$ & 79.36 & 76.7 &
79.4  \\
NGC 5466 &  6.81 & 60.45 & 36.45 & 0.797 &    &    & 48.8 & 44.1 \\ 
         &  6.85 & 60.20 & 36.33 & 0.796 & $-663.05$ & $-32.56$ & 49.1 &
44.2  \\
Pal 5    &  3.80 & 18.74 & 10.93 & 0.663 &    &    & 19.2 & 18.5 \\   
         &  3.97 & 18.88 & 10.92 & 0.653 & $-1179.26$ & 54.78 & 19.6 &
18.7  \\
NGC 5897 &  1.89 &  7.95 &  5.09 & 0.621 &    &    & 23.4 & 22.5 \\ 
         &  1.48 &  8.94 &  4.49 & 0.719 & $-1552.41$ & 23.18 & 17.5 &
16.6  \\
NGC 5904 &  2.65 & 36.80 & 17.89 & 0.866 &    &    & 46.5 & 40.4 \\ 
         &  2.76 & 37.35 & 18.11 & 0.863 & $-888.65$ & 40.12 & 46.6 &
40.5  \\
NGC 5927 &  3.44 &  4.64 &  0.80 & 0.150 &    &    & 41.4 & 48.8 \\ 
         &  4.50 &  5.45 &  0.79 & 0.095 & $-1693.39$ & 110.15 & 48.7 &
57.3  \\
NGC 5986 &  1.05 &  3.97 &  1.33 & 0.562 &    &    & 24.7 & 24.6 \\ 
         &  0.46 &  4.90 &  1.31 & 0.831 & $-1904.82$ & 1.51 & 12.9 &
12.4  \\
NGC 6093 &  2.07 &  3.11 &  3.02 & 0.201 &    &    & 34.1 & 37.0 \\ 
         &  2.01 &  3.78 &  3.14 & 0.311 & $-1867.55$ & 10.18 & 31.7 &
32.8  \\
\\
NGC 6121 &  0.39 &  5.95 &  0.49 & 0.874 &    &    & 11.6 & 7.7 \\ 
         &  0.55 &  5.47 &  2.11 & 0.827 & $-1824.59$ & $-1.58$ & 9.8 &
8.7   \\
NGC 6144 &  2.14 &  2.99 &  2.64 & 0.166 &    &    & 22.2 & 24.9 \\ 
         &  2.08 &  2.66 &  2.33 & 0.123 & $-1981.75$ & $-20.51$ & 22.1
 & 23.3  \\
NGC 6171 &  2.28 &  3.14 &  2.34 & 0.157 &    &    & 25.1 & 28.4 \\ 
         &  2.70 &  3.31 &  2.41 & 0.104 & $-1886.70$ & 39.42 & 28.3 &
31.2  \\
NGC 6205 &  5.35 & 21.93 & 13.90 & 0.609 &    &    & 67.7 & 67.2 \\ 
         &  5.30 & 22.61 & 14.22 & 0.621 & $-1087.59$ & $-30.31$ & 67.1
 & 66.4  \\
NGC 6218 &  2.76 &  5.94 &  2.21 & 0.363 &    &    & 30.0 & 32.1 \\ 
         &  2.73 &  5.32 &  2.56 & 0.323 & $-1744.69$ & 59.13 & 29.5 &
31.2  \\
NGC 6254 &  3.86 &  5.84 &  2.43 & 0.204 &    &    & 38.8 & 43.5 \\ 
         &  3.46 &  4.93 &  2.40 & 0.175 & $-1737.04$ & 71.23 & 37.3 &
41.9  \\
NGC 6266 &  1.52 &  2.63 &  0.83 & 0.276 &    &    & 37.2 & 43.6 \\ 
         &  1.41 &  2.22 &  0.85 & 0.223 & $-2167.78$ & 33.20 & 32.8 &
34.9  \\
NGC 6273 &  1.28 &  2.40 &  1.28 & 0.304 &    &    & 34.4 & 38.4 \\ 
         &  1.35 &  1.83 &  1.60 & 0.153 & $-2169.73$ & $-11.81$ & 32.0
 & 33.2  \\
NGC 6284 &  6.34 &  8.52 &  2.78 & 0.147 &    &    & 62.2 & 69.7 \\ 
         &  6.40 &  8.10 &  2.68 & 0.117 & $-1461.17$ & 148.72 & 64.3 &
73.3  \\
NGC 6287 &  0.91 &  5.03 &  2.82 & 0.707 &    &    & 16.3 & 14.2 \\ 
         &  0.87 &  4.30 &  2.44 & 0.671 & $-1895.79$ & $-3.07$ & 8.7 &
7.7  \\
NGC 6293 &  0.32 &  3.34 &  0.46 & 0.826 &    &    & 14.2 & 8.9 \\ 
         &  0.37 &  2.67 &  1.19 & 0.756 & $-2168.38$ & $-3.58$ & 8.6 &
7.7  \\
NGC 6304 &  1.90 &  3.25 &  0.53 & 0.276 &    &    & 23.7 & 27.4 \\ 
         &  1.84 &  3.09 &  0.57 & 0.253 & $-2054.85$ & 48.88 & 22.8 &
24.3  \\
NGC 6316 &  0.72 &  3.07 &  1.18 & 0.626 &    &    & 19.9 & 19.2 \\ 
         &  0.96 &  2.59 &  0.83 & 0.460 & $-2170.75$ & $-26.10$ & 18.2
 & 19.0  \\
NGC 6333 &  1.44 &  5.32 &  1.53 & 0.582 &    &    & 21.5 & 21.5 \\ 
         &  1.02 &  4.37 &  1.34 & 0.623 & $-1937.01$ & 27.83 & 15.3 &
15.1  \\
NGC 6341 &  1.20 & 10.43 &  2.43 & 0.793 &    &    & 22.9 & 20.6 \\ 
         &  1.30 & 10.86 &  2.59 & 0.786 & $-1496.54$ & 30.29 & 21.6 &
20.1  \\
NGC 6342 &  1.29 &  2.09 &  1.37 & 0.245 &    &    & 14.9 & 17.1 \\ 
         &  0.73 &  1.68 &  1.13 & 0.401 & $-2304.73$ & 10.92 & 8.3 &
8.6   \\
NGC 6356 &  2.45 &  7.94 &  2.08 & 0.528 &    &    & 38.2 & 39.6 \\ 
         &  2.30 &  7.74 &  1.98 & 0.542 & $-1637.01$ & 67.45 & 37.3 &
37.8  \\
NGC 6362 &  2.31 &  5.05 &  1.98 & 0.374 &    &    & 22.4 & 24.0 \\ 
         &  2.28 &  5.80 &  1.70 & 0.436 & $-1760.08$ & 62.23 & 23.4 &
24.4  \\
NGC 6388 &  0.70 &  3.03 &  1.10 & 0.627 &    &    & 27.6 & 26.1 \\ 
         &  0.53 &  2.95 &  0.88 & 0.696 & $-2148.65$ & $-13.57$ & 16.2
 & 16.2  \\
NGC 6397 &  2.53 &  5.12 &  1.46 & 0.344 &    &    & 23.1 & 25.4 \\ 
         &  3.33 &  6.42 &  1.66 & 0.317 & $-1672.42$ & 91.10 & 27.4 &
29.7  \\
NGC 6441 &  0.57 &  3.97 &  0.87 & 0.751 &    &    & 27.6 & 22.5 \\ 
         &  0.48 &  3.15 &  1.45 & 0.742 & $-2082.24$ & $-2.77$ & 16.9
 & 15.7  \\
NGC 6522 &  0.37 &  3.87 &  0.57 & 0.831 &    &    & 13.4 & 8.9 \\ 
         &  0.81 &  2.23 &  1.16 & 0.471 & $-2199.97$ & 18.26 & 13.2 &
13.4  \\
\\
NGC 6528 &  0.81 &  2.85 &  1.20 & 0.571 &    &    & 12.5 & 12.0 \\ 
         &  0.57 &  1.51 &  0.77 & 0.454 & $-2399.18$ & 14.31 & 7.3 &
7.6  \\
NGC 6553 &  2.09 &  8.75 &  0.33 & 0.615 &    &    & 28.9 & 28.1 \\ 
         &  2.31 & 12.02 &  0.52 & 0.677 & $-1445.87$ & 97.14 & 30.2 &
28.0  \\
NGC 6584 &  1.38 & 12.43 &  4.43 & 0.804 &    &    & 21.7 & 19.4 \\ 
         &  1.06 & 12.20 &  3.11 & 0.843 & $-1434.63$ & 24.27 & 15.3 &
14.5  \\
NGC 6626 &  1.03 &  2.82 &  0.85 & 0.472 &    &    & 21.7 & 24.0 \\ 
         &  0.74 &  3.09 &  0.88 & 0.613 & $-2111.49$ & 21.96 & 14.2 &
14.3  \\
NGC 6656 &  2.85 &  7.96 &  1.18 & 0.472 &    &    & 42.9 & 45.1 \\ 
         &  3.10 &  9.18 &  1.28 & 0.495 & $-1542.68$ & 105.56 & 45.7 &
46.8  \\
NGC 6712 &  0.60 &  5.56 &  1.14 & 0.809 &    &    & 15.9 & 13.6 \\ 
         &  0.91 &  6.34 &  1.86 & 0.749 & $-1767.49$ & 13.60 & 12.9 &
12.5  \\
NGC 6723 &  2.00 &  3.25 &  3.03 & 0.242 &    &    & 29.1 & 31.4 \\ 
         &  2.06 &  2.66 &  2.65 & 0.127 & $-1969.12$ & $-0.19$ & 30.1
 & 30.8  \\
NGC 6752 &  4.65 &  6.71 &  1.75 & 0.180 &    &    & 45.8 & 52.7 \\ 
         &  4.74 &  5.81 &  1.72 & 0.102 & $-1640.42$ & 109.14 & 48.9 &
56.8  \\
NGC 6779 &  0.62 & 12.52 &  0.77 & 0.906 &    &    & 15.2 & 10.7 \\ 
         &  0.84 & 12.49 &  2.44 & 0.875 & $-1428.17$ & $-24.15$ & 11.7
 & 10.5  \\
NGC 6809 &  1.98 &  5.95 &  3.87 & 0.508 &    &    & 25.5 & 26.1 \\ 
         &  1.78 &  5.61 &  3.59 & 0.526 & $-1741.44$ & 19.74 & 22.9 &
22.7  \\
NGC 6838 &  5.01 &  6.56 &  0.33 & 0.131 &    &    & 25.2 & 29.4 \\ 
         &  4.88 &  6.98 &  0.29 & 0.177 & $-1599.94$ & 134.92 & 25.8 &
29.6  \\
NGC 6934 &  6.88 & 34.87 & 20.35 & 0.670 &    &    & 54.6 & 52.3 \\ 
         &  6.83 & 35.12 & 20.60 & 0.674 & $-892.95$ & $-56.25$ & 54.6
 & 52.4  \\
NGC 7006 & 17.89 & 79.15 & 26.93 & 0.631 &    &    & 115.4 & 111.1 \\ 
         & 17.90 & 79.32 & 26.97 & 0.632 & $-514.26$ & 572.52 & 115.7 &
111.4  \\
NGC 7078 &  6.12 & 10.89 &  5.25 & 0.281 &    &    & 91.5 & 98.5 \\ 
         &  6.48 & 11.00 &  5.45 & 0.259 & $-1347.69$ & 139.60 & 94.4 &
103.1  \\
NGC 7089 &  6.10 & 33.12 & 18.03 & 0.689 &    &    & 83.4 & 78.9 \\ 
         &  6.14 & 34.19 & 18.77 & 0.695 & $-908.90$ & $-67.44$ & 84.1
 & 79.3  \\
NGC 7099 &  3.16 &  6.91 &  4.20 & 0.373 &    &    & 32.6 & 35.0 \\ 
         &  3.09 &  7.38 &  4.70 & 0.412 & $-1584.65$ & $-46.62$ & 32.9
 & 34.0  \\
Pal 12   & 15.19 & 19.64 & 15.76 & 0.128 &    &    & 40.3 & 44.7 \\ 
         & 15.25 & 19.86 & 15.90 & 0.131 & $-1010.08$ & 165.80 & 40.4 &
44.8  \\
Pal 13   & 11.84 & 88.01 & 38.03 & 0.763 &    &    & 25.4 & 23.5 \\ 
         & 11.86 & 88.13 & 38.24 & 0.763 & $-485.90$ & $-329.53$ & 25.5
 & 23.6  \\
\enddata
\end{deluxetable}

\clearpage
\begin{deluxetable}{cccccccccc}
\tabletypesize{\scriptsize}
%\rotate
\tablecaption{Destruction Rates of globular clusters in the
non-axisymmetric and axisymmetric potentials, obtained with
Monte Carlo simulations \label{tbl-4}}
\tablewidth{0pt}
\tablehead{
\colhead{} &
\multicolumn{3}{c}{----------- Linear Trajectory -----------} &
\multicolumn{3}{c}{------------ Real Trajectory ------------} &
\colhead{} & \colhead{} & \colhead{} \\
\colhead{Cluster} & \colhead{$<$$1/t_{bulge}$$>$} &
\colhead{${\sigma}_{-}$} & \colhead{${\sigma}_{+}$} &
\colhead{$<$$1/t_{bulge}$$>$} & \colhead{${\sigma}_{-}$} &
\colhead{${\sigma}_{+}$} & \colhead{$<$$1/t_{disk}$$>$} &
\colhead{${\sigma}_{-}$} & \colhead{${\sigma}_{+}$} \\
\colhead{} & \colhead{($yr^{-1}$)} & \colhead{($yr^{-1}$)} &
\colhead{($yr^{-1}$)} & \colhead{($yr^{-1}$)} & \colhead{($yr^{-1}$)} & 
\colhead{($yr^{-1}$)} & \colhead{($yr^{-1}$)} & \colhead{($yr^{-1}$)} &
\colhead{($yr^{-1}$)} 
}
\startdata
NGC 104 & 0.321E-15 & 0.191E-15 & 0.962E-14 & 0.286E-14 & 0.225E-14 &
0.650E-14 & 0.146E-12 & 0.359E-13 & 0.616E-13 \\      
 & 0.806E-16 & 0.304E-16 & 0.497E-16 & 0.466E-15 & 0.104E-15 &
0.165E-15 & 0.148E-12 & 0.281E-13 & 0.352E-13 \\
NGC 288 & 0.750E-10 & 0.669E-10 & 0.289E-09 & 0.157E-09 & 0.129E-09 &
0.408E-09 & 0.809E-11 & 0.440E-11 & 0.563E-11 \\ 
 & 0.621E-09 & 0.585E-09 & 0.538E-08 & 0.542E-09 & 0.516E-09 &
0.496E-08 & 0.899E-11 & 0.493E-11 & 0.842E-11  \\
NGC 362 & 0.365E-11 & 0.259E-11 & 0.445E-11 & 0.228E-11 & 0.144E-11 &
0.244E-11 & 0.128E-12 & 0.383E-13 & 0.439E-13 \\ 
 & 0.767E-10 & 0.580E-10 & 0.134E-09 & 0.390E-10 & 0.301E-10 &
0.672E-10 & 0.170E-12 & 0.528E-13 & 0.614E-13 \\
NGC 1851 & 0.540E-16 & 0.495E-16 & 0.227E-14 & 0.248E-15 & 0.236E-15 &
0.703E-14 & 0.777E-15 & 0.642E-15 & 0.332E-14 \\ 
 & 0.145E-15 & 0.138E-15 & 0.146E-13 & 0.488E-16 & 0.466E-16 &
0.523E-14 & 0.860E-15 & 0.723E-15 & 0.419E-14 \\
NGC 1904 & 0.127E-12 & 0.123E-12 & 0.306E-11 & 0.135E-12 & 0.126E-12 &
0.126E-11 & 0.628E-13 & 0.506E-13 & 0.164E-12 \\ 
 & 0.105E-11 & 0.104E-11 & 0.866E-10 & 0.714E-12 & 0.704E-12 &
0.639E-10 & 0.635E-13 & 0.512E-13 & 0.176E-12 \\
NGC 2298 & 0.789E-11 & 0.741E-11 & 0.469E-10 & 0.983E-11 & 0.889E-11 &
0.465E-10 & 0.835E-12 & 0.668E-12 & 0.151E-11 \\ 
 & 0.769E-10 & 0.743E-10 & 0.834E-09 & 0.562E-10 & 0.544E-10 &
0.645E-09 & 0.989E-12 & 0.768E-12 & 0.182E-11 \\
NGC 2808 & 0.498E-14 & 0.383E-14 & 0.201E-13 & 0.819E-14 & 0.666E-14 &
0.459E-13 & 0.727E-14 & 0.289E-14 & 0.406E-14 \\ 
 & 0.225E-14 & 0.171E-14 & 0.210E-13 & 0.672E-15 & 0.510E-15 &
0.636E-14 & 0.537E-14 & 0.220E-14 & 0.328E-14 \\
Pal 3 & 0.506E-15 & 0.499E-15 & 0.830E-14 & 0.154E-14 & 0.149E-14 &
0.239E-13 & 0.132E-14 & 0.129E-14 & 0.203E-13 \\ 
 & 0.163E-14 & 0.161E-14 & 0.826E-13 & 0.107E-14 & 0.106E-14 &
0.556E-13 & 0.490E-14 & 0.483E-14 & 0.241E-12 \\
NGC 3201 & 0.903E-16 & 0.221E-16 & 0.331E-16 & 0.168E-15 & 0.999E-16 &
0.174E-14 & 0.107E-13 & 0.504E-14 & 0.125E-13 \\
 & 0.895E-16 & 0.212E-16 & 0.302E-16 & 0.801E-16 & 0.345E-16 &
0.779E-16 & 0.109E-13 & 0.505E-14 & 0.133E-13 \\
NGC 4147 & 0.320E-11 & 0.308E-11 & 0.300E-10 & 0.300E-11 & 0.276E-11 &
0.209E-10 & 0.466E-12 & 0.384E-12 & 0.984E-12 \\
 & 0.219E-10 & 0.213E-10 & 0.669E-09 & 0.217E-10 & 0.213E-10 &
0.685E-09 & 0.519E-12 & 0.427E-12 & 0.164E-11 \\
NGC 4372 & 0.910E-10 & 0.697E-10 & 0.463E-09 & 0.670E-10 & 0.458E-10 &
0.234E-09 & 0.792E-10 & 0.327E-10 & 0.567E-10 \\
 & 0.352E-11 & 0.167E-11 & 0.213E-11 & 0.238E-11 & 0.106E-11 &
0.136E-11 & 0.436E-10 & 0.137E-10 & 0.171E-10 \\
NGC 4590 & 0.299E-15 & 0.149E-15 & 0.337E-15 & 0.147E-15 & 0.655E-16 &
0.258E-15 & 0.151E-13 & 0.887E-14 & 0.223E-13 \\
 & 0.295E-15 & 0.145E-15 & 0.351E-15 & 0.112E-15 & 0.590E-16 &
0.154E-15 & 0.162E-13 & 0.951E-14 & 0.247E-13 \\
NGC 4833 & 0.661E-10 & 0.408E-10 & 0.792E-10 & 0.160E-09 & 0.878E-10 &
0.144E-09 & 0.389E-11 & 0.107E-11 & 0.152E-11 \\
 & 0.454E-09 & 0.326E-09 & 0.106E-08 & 0.264E-09 & 0.189E-09 &
0.633E-09 & 0.422E-11 & 0.100E-11 & 0.126E-11 \\
NGC 5024 & 0.188E-12 & 0.187E-12 & 0.415E-10 & 0.487E-12 & 0.484E-12 &
0.597E-10 & 0.491E-13 & 0.475E-13 & 0.138E-11 \\
 & 0.307E-12 & 0.305E-12 & 0.937E-10 & 0.292E-12 & 0.291E-12 &
0.104E-09 & 0.561E-13 & 0.539E-13 & 0.151E-11 \\
NGC 5139 & 0.371E-10 & 0.187E-10 & 0.443E-10 & 0.958E-10 & 0.546E-10 &
0.105E-09 & 0.639E-11 & 0.157E-11 & 0.184E-11 \\
 & 0.156E-09 & 0.992E-10 & 0.320E-09 & 0.108E-09 & 0.754E-10 &
0.241E-09 & 0.746E-11 & 0.151E-11 & 0.189E-11 \\
NGC 5272 & 0.438E-12 & 0.377E-12 & 0.718E-11 & 0.170E-11 & 0.161E-11 &
0.221E-10 & 0.118E-11 & 0.618E-12 & 0.124E-11 \\
 & 0.854E-12 & 0.770E-12 & 0.441E-10 & 0.661E-12 & 0.609E-12 &
0.481E-10 & 0.127E-11 & 0.681E-12 & 0.132E-11 \\
NGC 5466 & 0.538E-11 & 0.503E-11 & 0.136E-09 & 0.171E-10 & 0.153E-10 &
0.247E-09 & 0.307E-11 & 0.245E-11 & 0.113E-10 \\
 & 0.495E-11 & 0.462E-11 & 0.208E-09 & 0.417E-11 & 0.392E-11 &
0.221E-09 & 0.288E-11 & 0.227E-11 & 0.101E-10 \\
Pal 5 & 0.345E-08 & 0.312E-08 & 0.149E-07 & 0.110E-07 & 0.950E-08 &
0.445E-07 & 0.461E-09 & 0.317E-09 & 0.516E-09 \\
 & 0.304E-07 & 0.289E-07 & 0.300E-06 & 0.399E-07 & 0.383E-07 &
0.406E-06 & 0.616E-09 & 0.428E-09 & 0.172E-08 \\
NGC 5897 & 0.250E-09 & 0.220E-09 & 0.625E-09 & 0.406E-09 & 0.366E-09 &
0.108E-08 & 0.196E-10 & 0.113E-10 & 0.154E-10 \\
 & 0.434E-08 & 0.402E-08 & 0.218E-07 & 0.378E-08 & 0.354E-08 &
0.198E-07 & 0.246E-10 & 0.146E-10 & 0.280E-10 \\
NGC 5904 & 0.350E-12 & 0.244E-12 & 0.853E-12 & 0.499E-12 & 0.393E-12 &
0.170E-11 & 0.189E-12 & 0.976E-13 & 0.190E-12 \\
 & 0.302E-12 & 0.208E-12 & 0.631E-12 & 0.149E-12 & 0.104E-12 &
0.342E-12 & 0.198E-12 & 0.102E-12 & 0.192E-12 \\
NGC 5927 & 0.535E-11 & 0.467E-11 & 0.473E-10 & 0.142E-11 & 0.118E-11 &
0.106E-10 & 0.483E-11 & 0.220E-11 & 0.324E-11 \\
 & 0.255E-14 & 0.146E-14 & 0.506E-14 & 0.116E-13 & 0.395E-14 &
0.620E-14 & 0.156E-11 & 0.445E-12 & 0.564E-12 \\
NGC 5986 & 0.222E-10 & 0.121E-10 & 0.187E-10 & 0.233E-10 & 0.151E-10 &
0.204E-10 & 0.523E-12 & 0.144E-12 & 0.172E-12 \\
 & 0.640E-09 & 0.437E-09 & 0.564E-09 & 0.228E-09 & 0.155E-09 &
0.174E-09 & 0.605E-12 & 0.147E-12 & 0.171E-12 \\
NGC 6093 & 0.194E-11 & 0.180E-11 & 0.735E-11 & 0.248E-11 & 0.232E-11 &
0.900E-11 & 0.128E-12 & 0.743E-13 & 0.872E-13 \\
 & 0.140E-09 & 0.136E-09 & 0.332E-09 & 0.487E-10 & 0.474E-10 &
0.116E-09 & 0.156E-12 & 0.873E-13 & 0.112E-12 \\
NGC 6121 & 0.286E-09 & 0.144E-09 & 0.167E-09 & 0.676E-09 & 0.310E-09 &
0.786E-09 & 0.176E-10 & 0.416E-11 & 0.651E-11 \\
 & 0.252E-08 & 0.147E-08 & 0.219E-08 & 0.215E-08 & 0.125E-08 &
0.180E-08 & 0.137E-10 & 0.288E-11 & 0.372E-11 \\
NGC 6144 & 0.126E-08 & 0.678E-09 & 0.206E-08 & 0.304E-08 & 0.161E-08 &
0.575E-08 & 0.325E-09 & 0.986E-10 & 0.862E-10 \\
 & 0.226E-08 & 0.157E-08 & 0.182E-07 & 0.495E-08 & 0.249E-08 &
0.272E-07 & 0.472E-09 & 0.128E-09 & 0.176E-09 \\
NGC 6171 & 0.206E-10 & 0.135E-10 & 0.468E-10 & 0.229E-10 & 0.140E-10 &
0.840E-10 & 0.170E-10 & 0.450E-11 & 0.486E-11 \\
 & 0.897E-10 & 0.840E-10 & 0.188E-08 & 0.866E-10 & 0.775E-10 &
0.189E-08 & 0.177E-10 & 0.470E-11 & 0.604E-11 \\
NGC 6205 & 0.325E-14 & 0.241E-14 & 0.300E-13 & 0.388E-13 & 0.342E-13 &
0.577E-12 & 0.611E-13 & 0.331E-13 & 0.931E-13 \\
 & 0.277E-14 & 0.198E-14 & 0.156E-13 & 0.920E-15 & 0.662E-15 &
0.542E-14 & 0.657E-13 & 0.354E-13 & 0.813E-13 \\
NGC 6218 & 0.260E-11 & 0.219E-11 & 0.183E-10 & 0.140E-11 & 0.115E-11 &
0.141E-10 & 0.136E-11 & 0.549E-12 & 0.870E-12 \\
 & 0.280E-12 & 0.216E-12 & 0.323E-11 & 0.126E-12 & 0.889E-13 &
0.142E-11 & 0.123E-11 & 0.332E-12 & 0.454E-12 \\
NGC 6254 & 0.821E-12 & 0.737E-12 & 0.714E-11 & 0.315E-12 & 0.258E-12 &
0.544E-11 & 0.777E-12 & 0.424E-12 & 0.679E-12 \\
 & 0.127E-13 & 0.100E-13 & 0.859E-13 & 0.228E-13 & 0.157E-13 &
0.587E-13 & 0.584E-12 & 0.206E-12 & 0.308E-12 \\
NGC 6266 & 0.594E-12 & 0.470E-12 & 0.341E-11 & 0.245E-12 & 0.158E-12 &
0.268E-12 & 0.781E-13 & 0.290E-13 & 0.418E-13 \\
 & 0.130E-11 & 0.118E-11 & 0.760E-10 & 0.865E-12 & 0.705E-12 &
0.136E-10 & 0.900E-13 & 0.421E-13 & 0.726E-13 \\
NGC 6273 & 0.214E-10 & 0.154E-10 & 0.553E-10 & 0.152E-10 & 0.863E-11 &
0.339E-10 & 0.208E-11 & 0.656E-12 & 0.765E-12 \\
 & 0.148E-10 & 0.104E-10 & 0.108E-09 & 0.216E-10 & 0.153E-10 &
0.546E-10 & 0.262E-11 & 0.892E-12 & 0.994E-12 \\
NGC 6284 & 0.898E-10 & 0.871E-10 & 0.100E-08 & 0.144E-09 & 0.136E-09 &
0.212E-08 & 0.557E-10 & 0.451E-10 & 0.162E-09 \\
 & 0.215E-09 & 0.209E-09 & 0.883E-08 & 0.342E-09 & 0.333E-09 &
0.172E-07 & 0.556E-10 & 0.450E-10 & 0.188E-09 \\
NGC 6287 & 0.188E-10 & 0.981E-11 & 0.141E-10 & 0.264E-10 & 0.136E-10 &
0.181E-10 & 0.493E-12 & 0.132E-12 & 0.195E-12 \\
 & 0.721E-09 & 0.443E-09 & 0.553E-09 & 0.323E-09 & 0.195E-09 &
0.234E-09 & 0.758E-12 & 0.266E-12 & 0.325E-12 \\
NGC 6293 & 0.487E-09 & 0.242E-09 & 0.240E-09 & 0.334E-09 & 0.146E-09 &
0.319E-09 & 0.295E-10 & 0.106E-10 & 0.107E-10 \\
 & 0.107E-07 & 0.684E-08 & 0.105E-07 & 0.118E-07 & 0.769E-08 &
0.122E-07 & 0.364E-10 & 0.140E-10 & 0.183E-10 \\
NGC 6304 & 0.130E-10 & 0.100E-10 & 0.428E-10 & 0.100E-10 & 0.779E-11 &
0.203E-09 & 0.349E-11 & 0.173E-11 & 0.220E-11 \\
 & 0.945E-11 & 0.842E-11 & 0.247E-09 & 0.139E-10 & 0.126E-10 &
0.364E-09 & 0.358E-11 & 0.162E-11 & 0.316E-11 \\
NGC 6316 & 0.872E-11 & 0.687E-11 & 0.178E-10 & 0.396E-11 & 0.268E-11 &
0.826E-11 & 0.461E-12 & 0.174E-12 & 0.281E-12 \\
 & 0.128E-09 & 0.113E-09 & 0.566E-09 & 0.818E-10 & 0.732E-10 &
0.399E-09 & 0.710E-12 & 0.309E-12 & 0.659E-12 \\
NGC 6333 & 0.243E-11 & 0.192E-11 & 0.785E-11 & 0.321E-11 & 0.273E-11 &
0.895E-11 & 0.170E-12 & 0.577E-13 & 0.931E-13 \\
 & 0.154E-10 & 0.113E-10 & 0.878E-10 & 0.501E-11 & 0.370E-11 &
0.341E-10 & 0.222E-12 & 0.757E-13 & 0.929E-13 \\
NGC 6341 & 0.668E-11 & 0.449E-11 & 0.108E-10 & 0.627E-11 & 0.482E-11 &
0.135E-10 & 0.472E-12 & 0.114E-12 & 0.144E-12 \\
 & 0.462E-10 & 0.395E-10 & 0.203E-09 & 0.280E-10 & 0.247E-10 &
0.126E-09 & 0.501E-12 & 0.107E-12 & 0.133E-12 \\
NGC 6342 & 0.586E-09 & 0.347E-09 & 0.675E-09 & 0.620E-09 & 0.324E-09 &
0.406E-09 & 0.109E-09 & 0.325E-10 & 0.344E-10 \\
 & 0.445E-08 & 0.252E-08 & 0.321E-08 & 0.856E-08 & 0.491E-08 &
0.608E-08 & 0.144E-09 & 0.372E-10 & 0.363E-10 \\
NGC 6356 & 0.179E-10 & 0.157E-10 & 0.747E-10 & 0.244E-10 & 0.225E-10 &
0.153E-09 & 0.223E-11 & 0.154E-11 & 0.232E-11 \\
 & 0.139E-09 & 0.132E-09 & 0.171E-08 & 0.101E-09 & 0.971E-10 &
0.130E-08 & 0.231E-11 & 0.154E-11 & 0.283E-11 \\
NGC 6362 & 0.417E-10 & 0.281E-10 & 0.145E-09 & 0.135E-10 & 0.104E-10 &
0.110E-09 & 0.212E-10 & 0.611E-11 & 0.782E-11 \\
 & 0.683E-11 & 0.385E-11 & 0.906E-11 & 0.349E-11 & 0.183E-11 &
0.543E-11 & 0.178E-10 & 0.320E-11 & 0.396E-11 \\
NGC 6388 & 0.122E-11 & 0.883E-12 & 0.186E-11 & 0.166E-12 & 0.120E-12 &
0.512E-12 & 0.125E-13 & 0.366E-14 & 0.507E-14 \\
 & 0.408E-10 & 0.299E-10 & 0.639E-10 & 0.572E-11 & 0.409E-11 &
0.780E-11 & 0.179E-13 & 0.457E-14 & 0.684E-14 \\
NGC 6397 & 0.131E-10 & 0.108E-10 & 0.699E-10 & 0.438E-11 & 0.345E-11 &
0.206E-10 & 0.115E-10 & 0.479E-11 & 0.714E-11 \\
 & 0.237E-12 & 0.114E-12 & 0.279E-12 & 0.176E-12 & 0.724E-13 &
0.153E-12 & 0.667E-11 & 0.145E-11 & 0.185E-11 \\
NGC 6441 & 0.379E-11 & 0.270E-11 & 0.586E-11 & 0.584E-12 & 0.401E-12 &
0.115E-11 & 0.331E-13 & 0.113E-13 & 0.196E-13 \\
 & 0.142E-09 & 0.107E-09 & 0.366E-09 & 0.274E-10 & 0.204E-10 &
0.694E-10 & 0.500E-13 & 0.185E-13 & 0.314E-13 \\
NGC 6522 & 0.211E-09 & 0.142E-09 & 0.213E-09 & 0.285E-09 & 0.200E-09 &
0.328E-09 & 0.154E-10 & 0.569E-11 & 0.842E-11 \\
 & 0.450E-08 & 0.403E-08 & 0.145E-07 & 0.528E-08 & 0.480E-08 &
0.168E-07 & 0.277E-10 & 0.143E-10 & 0.203E-10 \\
NGC 6528 & 0.296E-11 & 0.213E-11 & 0.492E-11 & 0.355E-12 & 0.249E-12 &
0.144E-11 & 0.377E-13 & 0.130E-13 & 0.161E-13 \\
 & 0.935E-10 & 0.800E-10 & 0.506E-09 & 0.221E-10 & 0.185E-10 &
0.782E-10 & 0.948E-13 & 0.557E-13 & 0.827E-13 \\
NGC 6553 & 0.103E-12 & 0.954E-13 & 0.210E-11 & 0.894E-13 & 0.792E-13 &
0.475E-12 & 0.643E-14 & 0.345E-14 & 0.853E-14 \\
 & 0.247E-12 & 0.240E-12 & 0.405E-10 & 0.582E-13 & 0.563E-13 &
0.909E-11 & 0.422E-14 & 0.220E-14 & 0.776E-14 \\
NGC 6584 & 0.163E-10 & 0.139E-10 & 0.356E-10 & 0.252E-10 & 0.211E-10 &
0.545E-10 & 0.944E-12 & 0.554E-12 & 0.865E-12 \\
 & 0.220E-09 & 0.195E-09 & 0.685E-09 & 0.145E-09 & 0.130E-09 &
0.440E-09 & 0.128E-11 & 0.793E-12 & 0.126E-11 \\
NGC 6626 & 0.640E-11 & 0.422E-11 & 0.152E-10 & 0.999E-12 & 0.594E-12 &
0.160E-11 & 0.290E-12 & 0.896E-13 & 0.139E-12 \\
 & 0.113E-09 & 0.943E-10 & 0.569E-09 & 0.333E-10 & 0.273E-10 &
0.132E-09 & 0.328E-12 & 0.118E-12 & 0.173E-12 \\
NGC 6656 & 0.151E-12 & 0.133E-12 & 0.328E-11 & 0.972E-13 & 0.746E-13 &
0.344E-12 & 0.244E-12 & 0.906E-13 & 0.241E-12 \\
 & 0.885E-14 & 0.399E-14 & 0.990E-14 & 0.322E-14 & 0.140E-14 &
0.345E-14 & 0.164E-12 & 0.488E-13 & 0.767E-13 \\
NGC 6712 & 0.140E-10 & 0.950E-11 & 0.199E-10 & 0.124E-10 & 0.904E-11 &
0.208E-10 & 0.423E-12 & 0.130E-12 & 0.191E-12 \\
 & 0.886E-10 & 0.740E-10 & 0.591E-09 & 0.279E-10 & 0.232E-10 &
0.150E-09 & 0.475E-12 & 0.131E-12 & 0.155E-12 \\
NGC 6723 & 0.121E-10 & 0.823E-11 & 0.470E-10 & 0.238E-10 & 0.186E-10 &
0.853E-10 & 0.466E-11 & 0.126E-11 & 0.134E-11 \\
 & 0.572E-11 & 0.433E-11 & 0.470E-09 & 0.890E-11 & 0.474E-11 &
0.167E-09 & 0.640E-11 & 0.137E-11 & 0.162E-11 \\
NGC 6752 & 0.241E-11 & 0.229E-11 & 0.776E-10 & 0.394E-12 & 0.311E-12 &
0.361E-11 & 0.415E-11 & 0.207E-11 & 0.581E-11 \\
 & 0.535E-14 & 0.330E-14 & 0.113E-13 & 0.198E-13 & 0.834E-14 &
0.215E-13 & 0.272E-11 & 0.749E-12 & 0.101E-11 \\
NGC 6779 & 0.367E-10 & 0.244E-10 & 0.536E-10 & 0.722E-10 & 0.473E-10 &
0.868E-10 & 0.229E-11 & 0.809E-12 & 0.958E-12 \\
 & 0.353E-09 & 0.272E-09 & 0.746E-09 & 0.233E-09 & 0.179E-09 &
0.485E-09 & 0.222E-11 & 0.666E-12 & 0.750E-12 \\
NGC 6809 & 0.245E-10 & 0.187E-10 & 0.638E-10 & 0.309E-10 & 0.232E-10 &
0.551E-10 & 0.228E-11 & 0.612E-12 & 0.969E-12 \\
 & 0.756E-11 & 0.405E-11 & 0.153E-10 & 0.371E-11 & 0.228E-11 &
0.795E-11 & 0.231E-11 & 0.515E-12 & 0.642E-12 \\
NGC 6838 & 0.493E-13 & 0.487E-13 & 0.150E-11 & 0.653E-14 & 0.539E-14 &
0.367E-12 & 0.380E-13 & 0.182E-13 & 0.135E-12 \\
 & 0.905E-16 & 0.276E-16 & 0.436E-16 & 0.198E-15 & 0.489E-16 &
0.662E-16 & 0.241E-13 & 0.633E-14 & 0.870E-14 \\
NGC 6934 & 0.415E-12 & 0.405E-12 & 0.161E-10 & 0.277E-12 & 0.264E-12 &
0.524E-11 & 0.948E-13 & 0.812E-13 & 0.436E-12 \\
 & 0.148E-11 & 0.146E-11 & 0.112E-09 & 0.115E-11 & 0.114E-11 &
0.961E-10 & 0.895E-13 & 0.769E-13 & 0.439E-12 \\
NGC 7006 & 0.203E-12 & 0.201E-12 & 0.244E-10 & 0.641E-13 & 0.631E-13 &
0.271E-11 & 0.429E-13 & 0.418E-13 & 0.740E-12 \\
 & 0.138E-11 & 0.138E-11 & 0.303E-09 & 0.131E-11 & 0.130E-11 &
0.305E-09 & 0.536E-13 & 0.523E-13 & 0.156E-11 \\
NGC 7078 & 0.175E-13 & 0.159E-13 & 0.734E-12 & 0.956E-14 & 0.832E-14 &
0.168E-12 & 0.246E-12 & 0.183E-12 & 0.689E-12 \\
 & 0.562E-14 & 0.488E-14 & 0.656E-13 & 0.299E-14 & 0.245E-14 &
0.260E-13 & 0.225E-12 & 0.166E-12 & 0.472E-12 \\
NGC 7089 & 0.243E-13 & 0.235E-13 & 0.130E-11 & 0.324E-13 & 0.308E-13 &
0.636E-12 & 0.204E-13 & 0.168E-13 & 0.772E-13 \\
 & 0.515E-12 & 0.510E-12 & 0.697E-10 & 0.330E-12 & 0.328E-12 &
0.484E-10 & 0.231E-13 & 0.191E-13 & 0.989E-13 \\
NGC 7099 & 0.322E-11 & 0.272E-11 & 0.133E-10 & 0.409E-11 & 0.315E-11 &
0.246E-10 & 0.296E-11 & 0.132E-11 & 0.144E-11 \\
 & 0.296E-11 & 0.246E-11 & 0.172E-10 & 0.163E-11 & 0.136E-11 &
0.117E-10 & 0.312E-11 & 0.138E-11 & 0.155E-11 \\
Pal 12 & 0.205E-11 & 0.160E-11 & 0.218E-10 & 0.109E-10 & 0.782E-11 &
0.528E-10 & 0.270E-10 & 0.204E-10 & 0.982E-10 \\
 & 0.214E-11 & 0.167E-11 & 0.423E-10 & 0.101E-10 & 0.785E-11 &
0.668E-10 & 0.281E-10 & 0.210E-10 & 0.115E-09 \\
Pal 13 & 0.135E-13 & 0.126E-13 & 0.764E-12 & 0.568E-13 & 0.519E-13 &
0.238E-11 & 0.718E-13 & 0.645E-13 & 0.705E-12 \\
 & 0.982E-14 & 0.910E-14 & 0.542E-12 & 0.512E-14 & 0.481E-14 &
0.381E-12 & 0.679E-13 & 0.611E-13 & 0.714E-12 \\
\enddata
\end{deluxetable}

\clearpage

\clearpage

\clearpage

\clearpage

\clearpage

\clearpage

\clearpage

\clearpage

\clearpage

\clearpage

\clearpage

\clearpage

\clearpage

\clearpage

\clearpage

\clearpage

\clearpage

\clearpage

\clearpage

\clearpage

\clearpage

\clearpage

\clearpage

\clearpage

\clearpage

\clearpage

\end{document}